\documentclass[onecolumn]{aastex631}

\begin{document}

\shorttitle{Multi-Epoch Observations of NGC 3938 with {\it 
Chandra}}
\shortauthors{Raut et al.}

\title{Multi-Epoch Observations of the Nearby Spiral Galaxy NGC
3938 with the {\it Chandra} X-ray Observatory} 

\author{Siddhi Raut}
\affiliation{Ronald Reagan High School, 
  19000 Ronald Reagan, San Antonio, TX 78258, USA\footnote{Currently an
  undergraduate at the University of Chicago}}
\email{siddhi.raut.missionartemis@gmail.com}

\author[0000-0002-4162-8190]{Eric M. Schlegel}
\affiliation{Department of Physics and Astronomy,
University of Texas at San Antonio,
One UTSA Circle, San Antonio, TX 78249 USA}
\email{eric.schlegel@utsa.edu}

\author[0000-0001-7658-1065]{Thomas G. Pannuti}
\affiliation{Department of Engineering Science, 123 Lappin Hall,
 Morehead State University, 150 University Drive,
 Morehead, KY 40351, USA}
 \email{t.pannuti@moreheadstate.edu}

\author{Brannon W. Jones}
\affiliation{Department of Engineering Science, 123 Lappin Hall,
 Morehead State University, 150 University Drive,
 Morehead, KY 40351, USA}

\author{Jacobo Matallana}
\affiliation{Department of Engineering Science, 123 Lappin Hall,
 Morehead State University, 150 University Drive,
 Morehead, KY 40351, USA}

\begin{abstract}

We present an analysis of two epochs of ACIS observations of the
SA(s)c spiral galaxy NGC 3938 with the {\it Chandra} X-ray
Observatory.  The total exposure time of the observations was 95
ksecs with a limiting unabsorbed luminosity of ${\approx
}10^{38}$ ergs/sec assuming a distance of 22 Mpc. A
total of 47 discrete merged sources from both epochs were
detected at the ${\approx}3{\sigma}$
level or greater with the D25 radius.  We demonstrate that at
the time of the {\it Chandra} observations, the nucleus was {\it
not} detected.  We connect the detected sources to counterparts
in other wavebands to the degree possible.  Based on the two
epochs, we identify three variable sources and an additional two
that may have varied between the two observations.  We do not
formally detect any of the five historical supernovae that have
occurred in NGC 3938.  The luminosity function of NGC 3938
is compared to a recent compilation of 38 galaxies and we 
identify a potentially significant problem with the `known' 
distance to NGC 3938.   Star formation rate and metallicity values 
are also computed; the star formation rate is highly dependent upon 
the adopted distance.  The metallicity appears to lie in the range
of 8.2-9.2, consistent with values from other work.  We include 
in an appendix a short discussion of the sources that lie in {\it
Chandra}'s field-of-view but lie outside of NGC 3938.
 
\end{abstract}

\keywords{X-ray sources (1822) --- Spiral galaxies (1560) 
-- X-ray astronomy (1810)}

\section{Introduction} \label{sec:intro}

The {\it Chandra} X-ray Observatory's sharp point spread
function has yielded advances of our understanding of the
properties of discrete X-ray sources (e.g., X-ray binaries (XRBs
)(e.g., \cite{Lehmer2021}; \cite{Binder2017}; \cite{Mineo2012});
supernova remnants (SNRs) (e.g., \cite{Sasaki2020}) in nearby
galaxies.  The as-yet unsurpassed angular resolution of {\it
Chandra} of ${\sim}$1'' (\cite{vanSpeybroeck1997},
\cite{Gaetz2000})has led to an increase in the number of
discrete X-ray sources for which variability analyses
and positional accuracy aid in the identification of possible
counterparts at other wavelengths (particularly the infrared and
radio wavelengths).

The X-ray sources within a galaxy are often connected to stellar
evolution, either as SNRs or XRBs.  By observing the XRBs and SNRs in
other galaxies, greater insight can be provided about our own, the
Milky Way.  Interstellar dust within the Milky Way prevents linking
many sources across the electromagnetic spectrum whereas nearby, face-on
spirals permit a uniform survey of X-ray-emitting sources.

The study of galaxies in the X-ray band typically breaks in two
directions: the study of X-ray-emitting objects in the galaxy or the
study of the nucleus.  Much of that break is driven by the distances
to individual galaxies.  The {\it Einstein} generation of detectors
could resolve a few very nearby galaxies into {\it any} 
individual sources; all others were essentially an unresolved 
blur \citep{Fabbiano1992}.  {\it ROSAT},  with its improved 
spatial resolution, permitted a larger volume to be investigated 
before sources became too confused,
allowing the first identifications of individual sources, 
e.g., ultra-luminous X-ray sources (ULX; \cite{Zampieri2006}). 
{\it Chandra} allows the study of individual sources in
galaxies to ${\approx}$5 Mpc.  That resolution also permits
pushing the outer boundary outward to ${\approx}$20-25 Mpc
at which point most sources are confused or blurred together.  
Within that volume lies the face-on unbarred SA(s)c galaxy 
galaxy NGC 3938.

NGC 3938 has a low-luminosity nucleus hence the LINER classification (LINER = Low Ionization
Nuclear Emission Region; \cite{Pellegrini2000}, \cite{Ho1997}).
It is one of the brightest in the Ursa Major South galaxy group at
an inclination angle of {\it i} ${\sim}14^{\circ}$
\citep{JimenezVicente1999}.  A previous, 50 ksec study of NGC 3938
using {\it Chandra} detected ${\sim}$45 sources within the D25
radius\footnote{The diameter at which the surface brightness of the
galaxy in an optical band has fallen to mag 25 / arcsec$^2$.}
\citep{Buhidar2017}.  Regions of star formation in NGC 3938 have
also been observed through the near-ultraviolet (NUV) from 
GALEX\footnote{Galaxy Evolution Explorer, \cite{Bianchi2014}}, 
H$\alpha$ from JKT\footnote{Jacobus Kapteyn Telescope, La Palma} and
KPNO\footnote{Kitt Peak National Observatory}, 8 and 
24 $\mu$m from {\it Spitzer}, and 
CO from BIMA\footnote{Berkeley-Illinois-Maryland Array,
\cite{Thornley2004}} which resulted in an
obtained power-law relationship between the emission 
region volume and luminosity \citep{CalduPrimo2009}. 
Five supernovae have erupted in NGC 3938: SN 2017ein is 
a progenitor candidate for the Type Ic supernova
\citep{VanDyk2017} along with four other supernovae: SN 1961U
(SN~IIL), SN 1964L (SN~Ib), SN 2005ay (SN~IIP), and SN 2022xlp
(SN~Iax).

NGC 3938 visibly lacks nearby neighbors, suggesting a 
relatively isolated galaxy.  However, it is useful to note 
that NGC 3938 is part of the Ursa Major South galaxy group, 
so it is isolated to a certain extent.  NGC 3938 lies at a 
``preferred'' distance, according to the NASA Extragalactic
Database\footnote{https://ned.ipac.caltech.edu}(hereafter, NED),
of 15.3 Mpc.  However, that distance is based essentially on 
a mean value of measured distances that differ substantially.
Recent measures fall in the 17 \citep{Poznanski2009} to 22
\citep{Rodriguez2014} Mpc range.  We initially adopted 22 Mpc 
as the distance to NGC 3938.  That places the galaxy at 
essentially the outer edge of a volume in which {\it Chandra} 
will return a starting census of the X-ray-emitting objects: 
an object emitting of order 10$^{38}$ ergs s$^{-1}$ will 
produce a few counts in a 50-ksec exposure at that distance.  
However, we return to a short discussion of the distance in
\S\ref{lumFunc} and its possible `incorrectness' when 
describing the distribution of
source counts with luminosity.

We describe and characterize two observations of
the NGC 3938 obtained by the {\it Chandra} X-ray Observatory.  
The data from the two epochs was also
examined for counterparts at other wavelengths.

The paper is organized in this matter: Section 2 describes the
observations and data reductions.  Section 3 delves into the
specific sources and their detection. X-ray properties derived 
from color-color
hardness are then discussed in Section 4, and Sections 5 and 6 talk
about time-variable emission and multi-wavelength source comparison
respectively. Section 7 will detail the star formation rate and
metallicity of NGC 3938, and Section 8 will summarize our results
and conclusions.

\begin{figure}[h!]
 \centering
 \caption{Field-of-view from {\it Chandra} for NGC 3938 with
 detected sources identified. Each observation was made using 
 ACIS-I, so the
   field-of-view of the four CCDs is ${\sim}$16' on a side.  (left)
   sources outside the D25 circle; (right) expanded view of the
   sources inside of the D25 circle.  The image scale is set by the
   D25 radius of ${\sim}$190$''$.  North is up and East is left for
   both images.  Ellipses are determined (size, orientation) by the
   source detection algorithm (\S\ref{ChanObs}); ellipse colors are
   defined in Figure~\ref{FoVopt} (to better see their meaning), refer
   to the observation epochs, and are used throughout the paper.}
 \scalebox{0.38}{\includegraphics{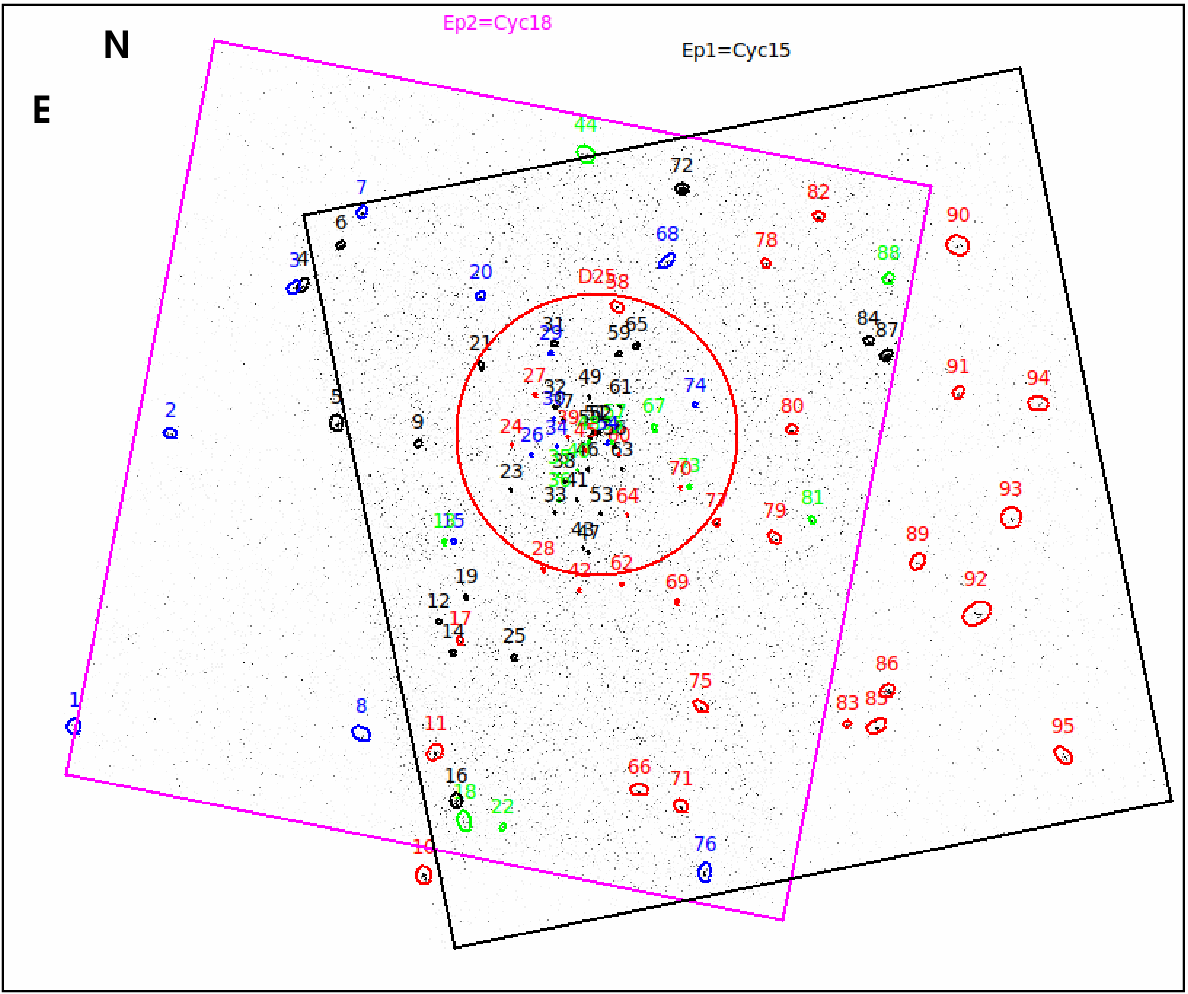}}
 \scalebox{0.40}{\includegraphics{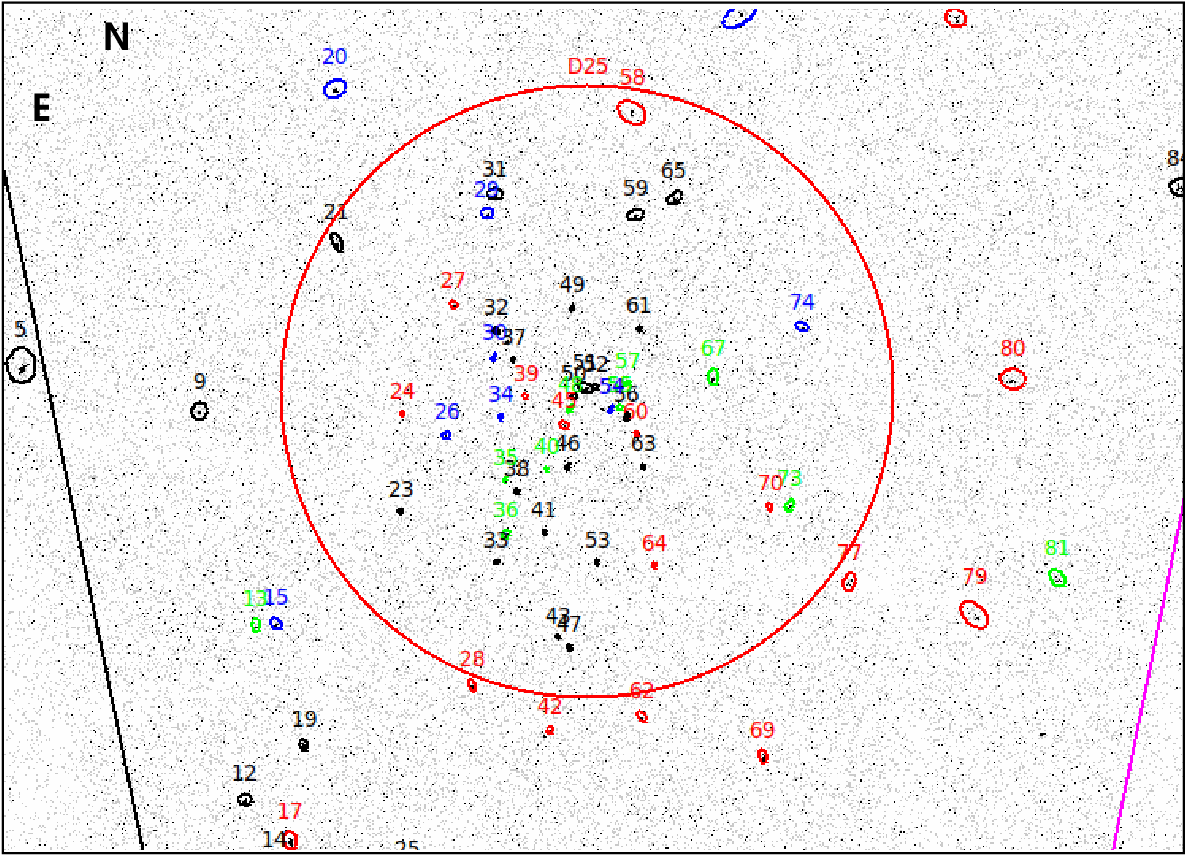}}
 \label{FoVfig}
\end{figure}

\section{Observations and Data Reduction}

\subsection{\it {Chandra} Observations}\label{ChanObs}

\begin{figure}[h!]
 \centering
 \caption{(left) Detected X-ray sources within the D25 radius (red
   circle) overlaid on an IRAC 3.6${\mu}m$ image (left) of NGC 3938
   and on a SDSS g (DR7) image (right).  North is up and East is
   left. A 1$'$ scale bar is in the lower left corner of both images.
   As stated previously, the ellipses are determined by the source
   detection routine.  Colors are assigned as follows: black (or
   white/yellow if a dark background): detected source is present
   in both epochs; red: present only in Epoch 1; blue: only present
   in Epoch 2; green: only present in the merged data.  This color
   coding is used whenever X-ray detections are overlaid on an
   image of NGC 3938.}
 \scalebox{0.5}{\includegraphics{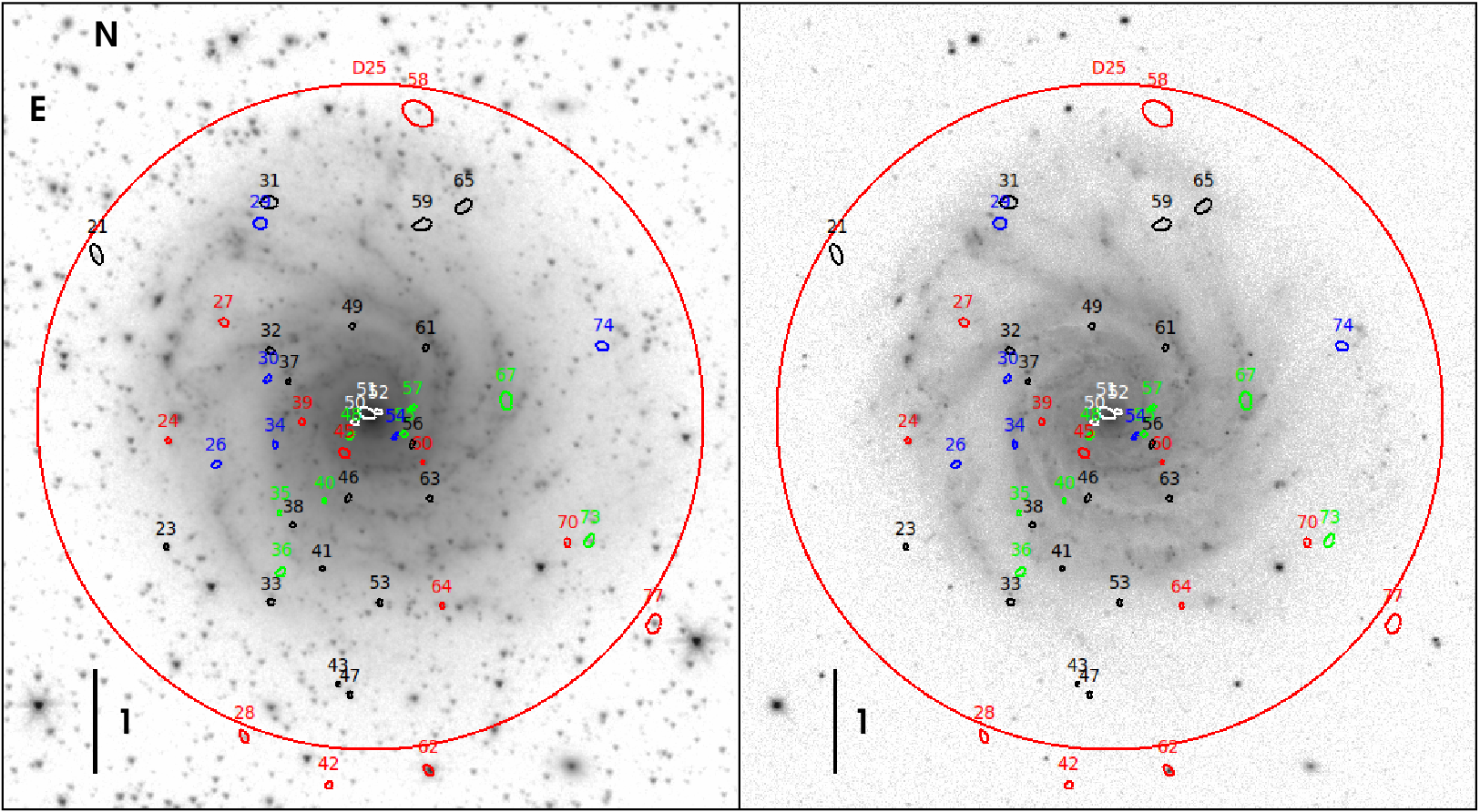}}
 \label{FoVopt}
\end{figure}

NGC 3938 was the target of two pointed observations made with the
Advanced Charge-Coupled Device (CCD) Imaging Spectrometer (ACIS)
\citep{2003SPIE.4851...28G} aboard {\it Chandra}
\citep{2002PASP..114....1W}: both of these observations were
guaranteed time to G. Garmire (Pennsylvania State University). The
first observation occurred on 2013 June 28 (ObsID 15389) and the
second occurred on 2016 Oct 1 (ObsID 18456): the exposure times of
these observations were 40 ksec and 45 ksec, respectively. The aim
point for both observations was off the nucleus: the first by
${\sim}2.7$ arcmin SSW and the second by ${\sim}1$ arcmin SSE.  In
both cases, the entire galaxy was covered essentially by the I3 CCD.
Both observations were obtained using the front-illuminated I0-I3 CCDs
in Very Faint mode.  After correcting for the deadtime, the two
exposures were 49.3 and 45.5 ksec.  Table~\ref{table1} lists some
basic properties of NGC 3938 and the {\it Chandra} observations.
Figure~\ref{FoVfig} shows the detected sources (described next)
overlaid on the {\it Chandra} images.  Figure~\ref{FoVopt} overlays
the {\it Chandra} sources on IRAC 3.6${\mu}$ and Digital Sky Survey~2
blue images of NGC 3938.

\begin{table}
  \centering
  \caption{Basic Properties of NGC 3938}
  \label{table1}
  \begin{tabular}{ll}
    Quantity   &   Value \\  \hline
    Position: RA, Dec   & 11:52:49.4, +44:07:14.6 (NED) \\
    Distance            &  22 Mpc (see \S\ref{sec:intro} and \S\ref{lumFunc}) \\
    Column Density      & $1.77{\times}10^{20}$ cm$^{-2}$ (HI4PI; \cite{Bekhti2016}) \\
    D25 radius          & 190.3$''$ (NED) \\
    {\it Chandra} exposures & obsID 15389, 49.3 ksec = Epoch 1 \\
         & obsID 18456, 45.5 ksec = Epoch 2 \\
    Pointing center & 11:52:46.1, +44:05:26.4 (Ep1) \\
                & 11:52:59.8, +44:06:38.4 (Ep2) \\
    \hline
  \end{tabular}
\end{table}

The Chandra Interactive Analysis of Observations (CIAO) software
(version 4.13) and the associated calibration files (version 4.10.3)
were used \citep{2006SPIE.6270E..60F}.  We accumulated source-free
background areas away from the galaxy -- broadly described as
originating in the corners of the CCDs.  We extracted a light curve
using 50-second bins to test for the presence of soft background
flares; no flares were detected.

\begin{table}[h!]
 \centering
 \caption{Number of Detected Sources by Epoch, Band, and Position}
 \label{NmbrDetSrcs}
 \begin{tabular}{lrrrrrr}
  Data set   &  B   &   S   &   M   &   H & $<$D25 & $>$D25 \\ \hline
 Epoch 1     & 66   &  14   &  57   &  45 & 33 & 33 \\
 Epoch 2     & 48   &  ~2   &  37   &  34 & 27 & 21 \\
 merged      & 95   &  19   &  44   &  70 & 47 & 48 \\ \hline
\end{tabular}
\end{table}

The epoch data were merged using CIAO task {\tt merge\_obs} that
yielded a single data set with an exposure time of $\sim$94 ksec.  The
source detection routine {\tt wavdetect} was run on each epoch
individually, on the merged data set, and on four bands: B = Broad:
0.3-8 keV; S = Soft: 0.3-1.0 keV; M = Medium: 1.0-2.0 keV; H = Hard:
2-8 keV (Figure~\ref{FoVfig})\footnote{The words Broad, Soft, Medium,
and Hard will be used throughout the paper in place of the energy
ranges.}.

Point sources were detected using {\tt wavdetect} at 1$''$,
2$''$, 4$''$ and 8$''$ scales \citep{2002ApJS..138..185F} using the
recommended $10^{-6}$ false-source threshold (1 false source per CCD).
The X-ray background for each source was determined from merging
`blank-sky' images from the {\it Chandra} archive and re-projecting
the merged data to the position of NGC 3938.  For each detected
source a corresponding region of the merged blank sky was extracted
to determine the background counts.   The reader should also note
that we do not consider variability at this stage -- a variable 
source could lie above our detection threshold for a limited portion
of the observation, but over the entire observation it falls below
our threshold.

The detected sources were merged into a final source list after
eliminating duplicate detections.  Source and background counts were
extracted by {\tt wavdetect} using apertures centered on the {\tt
  wavdetect} positions and of sufficient radius to enclose 95\% of
the detected X-rays.  The J2000 epoch is used for any and all
positions throughout this paper.   We also checked the
astrometry for sources in common between {\it Chandra} and {\it
GAIA} and {\it Hubble} and {\it GAIA}.  Positional differences
were less than ${\sim}0.08$ arcsec.

\begin{table}[h!]
\centering
\caption{NGC 3938: Merged Source List: Descending RA Order}
\label{MrgdSrc}
\tiny
\begin{tabular}{rrrrrrrrrrrrll}
 N  &       RA     &       Dec     &   Counts &    Unc  &  BkgCts &  BkUnc &   SrcRate &  SrcRUnc &  BkgRate &   BkRUnc & Signif & Epoch$^a$ & Cntrpt \\ \hline
\multicolumn{14}{c}{Within the D25 Circle} \\
21  &  11:53:03.92 &   44:08:47.04 &     45.6 &   7.48  &    10.4 &   0.07 &  4.79e-04 & 7.86e-05 & 1.09e-04 & 7.55e-07 &   10.5 &  1,2  &   \\
23  &  11:53:00.25 &   44:06:00.11 &    136.7 &  11.87  &     4.3 &   0.05 &  1.44e-03 & 1.25e-04 & 4.54e-05 & 4.76e-07 &   42.0 &  1,2  &   \\
24  &  11:53:00.13 &   44:07:00.77 &     25.3 &   5.29  &     2.7 &   0.04 &  2.66e-04 & 5.56e-05 & 2.80e-05 & 3.75e-07 &    8.9 &  1    & IR? \\
26  &  11:52:57.59 &   44:06:47.27 &     12.0 &   3.87  &     3.0 &   0.04 &  1.26e-04 & 4.07e-05 & 3.15e-05 & 3.88e-07 &    4.1 &  2    &   \\
27  &  11:52:57.17 &   44:08:08.38 &     65.5 &   8.43  &     5.5 &   0.05 &  6.88e-04 & 8.85e-05 & 5.74e-05 & 5.28e-07 &   18.8 &  1    &   \\
28  &  11:52:56.10 &   44:04:11.97 &     17.2 &   4.58  &     3.8 &   0.04 &  1.80e-04 & 4.81e-05 & 4.01e-05 & 4.46e-07 &    5.5 &  1    &   \\
29  &  11:52:55.24 &   44:09:04.97 &     15.2 &   4.58  &     5.8 &   0.05 &  1.60e-04 & 4.81e-05 & 6.10e-05 & 5.67e-07 &    4.3 &  2    &   \\
30  &  11:52:54.85 &   44:07:35.93 &     13.4 &   4.00  &     2.6 &   0.03 &  1.41e-04 & 4.20e-05 & 2.75e-05 & 3.66e-07 &    4.7 &  2    &   \\
31  &  11:52:54.79 &   44:09:16.98 &     19.4 &   5.29  &     8.6 &   0.07 &  2.04e-04 & 5.56e-05 & 9.02e-05 & 6.87e-07 &    4.8 &  1,2  &   \\
32  &  11:52:54.71 &   44:07:52.04 &    163.3 &  13.04  &     6.7 &   0.06 &  1.72e-03 & 1.37e-04 & 7.03e-05 & 5.97e-07 &   43.8 &  1,2  &   \\
33  &  11:52:54.69 &   44:05:28.88 &     39.9 &   6.56  &     3.1 &   0.04 &  4.19e-04 & 6.89e-05 & 3.25e-05 & 4.00e-07 &   13.5 &  1,2  & HII? IR? \\
34  &  11:52:54.46 &   44:06:58.43 &     15.4 &   4.24  &     2.6 &   0.03 &  1.62e-04 & 4.46e-05 & 2.75e-05 & 3.66e-07 &    5.4 &  2    &   \\
35  &  11:52:54.22 &   44:06:19.57 &      8.4 &   3.16  &     1.6 &   0.03 &  8.80e-05 & 3.32e-05 & 1.70e-05 & 2.91e-07 &    3.3 &  mrgd &   \\
36  &  11:52:54.16 &   44:05:45.85 &      9.8 &   3.61  &     3.2 &   0.04 &  1.03e-04 & 3.79e-05 & 3.35e-05 & 4.08e-07 &    3.3 &  mrgd &   \\
37  &  11:52:53.74 &   44:07:34.48 &     40.0 &   6.63  &     4.0 &   0.04 &  4.20e-04 & 6.97e-05 & 4.20e-05 & 4.58e-07 &   12.6 &  1,2  & IR?  \\
38  &  11:52:53.51 &   44:06:12.51 &     70.2 &   8.60  &     3.8 &   0.04 &  7.38e-04 & 9.03e-05 & 3.96e-05 & 4.41e-07 &   22.5 &  1,2  &   \\
39  &  11:52:53.03 &   44:07:11.47 &     25.6 &   5.39  &     3.4 &   0.04 &  2.69e-04 & 5.66e-05 & 3.58e-05 & 4.26e-07 &    8.4 &  1    &   \\
40  &  11:52:51.84 &   44:06:26.45 &     11.9 &   3.74  &     2.1 &   0.03 &  1.25e-04 & 3.93e-05 & 2.17e-05 & 3.29e-07 &    4.5 &  mrgd &   \\
41  &  11:52:51.94 &   44:05:47.59 &     89.5 &   9.64  &     3.5 &   0.04 &  9.40e-04 & 1.01e-04 & 3.67e-05 & 4.27e-07 &   29.3 &  1,2  & IR?  \\
42  &  11:52:51.60 &   44:03:44.10 &     22.3 &   5.10  &     3.7 &   0.04 &  2.34e-04 & 5.36e-05 & 3.91e-05 & 4.44e-07 &    7.2 &  1    &   \\
43  &  11:52:51.15 &   44:04:42.17 &     48.6 &   7.28  &     4.4 &   0.05 &  5.11e-04 & 7.65e-05 & 4.57e-05 & 4.77e-07 &   14.9 &  1,2  & IR?  \\
45  &  11:52:50.79 &   44:06:53.63 &     14.2 &   4.36  &     4.8 &   0.05 &  1.50e-04 & 4.58e-05 & 5.00e-05 & 4.95e-07 &    4.3 &  1    & HII? \\
46  &  11:52:50.59 &   44:06:27.83 &     27.7 &   5.57  &     3.3 &   0.04 &  2.90e-04 & 5.85e-05 & 3.51e-05 & 4.17e-07 &    9.1 &  1,2  &   \\
47  &  11:52:50.52 &   44:04:35.93 &    185.5 &  13.86  &     6.5 &   0.06 &  1.95e-03 & 1.46e-04 & 6.84e-05 & 5.85e-07 &   50.2 &  1,2  &   \\
48  &  11:52:50.46 &   44:07:03.76 &     12.7 &   4.00  &     3.3 &   0.04 &  1.33e-04 & 4.20e-05 & 3.50e-05 & 4.15e-07 &    4.2 &  mrgd &   \\
49  &  11:52:50.36 &   44:08:06.29 &    167.2 &  13.19  &     6.8 &   0.06 &  1.76e-03 & 1.39e-04 & 7.19e-05 & 6.03e-07 &   44.5 &  1,2  &   \\
50  &  11:52:50.23 &   44:07:11.41 &     53.0 &   7.68  &     6.0 &   0.05 &  5.57e-04 & 8.07e-05 & 6.25e-05 & 5.70e-07 &   14.8 &  1,2  &   \\
51  &  11:52:49.59 &   44:07:16.56 &     57.9 &   8.31  &    11.1 &   0.07 &  6.08e-04 & 8.72e-05 & 1.17e-04 & 7.73e-07 &   13.0 &  1,2  &   \\
52  &  11:52:49.00 &   44:07:17.52 &     28.6 &   5.74  &     4.4 &   0.05 &  3.00e-04 & 6.03e-05 & 4.66e-05 & 4.86e-07 &    8.7 &  1,2  &   \\
53  &  11:52:48.91 &   44:05:28.55 &     36.4 &   6.32  &     3.6 &   0.04 &  3.83e-04 & 6.64e-05 & 3.74e-05 & 4.29e-07 &   11.8 &  1,2  & IR? \\
54  &  11:52:48.09 &   44:07:03.51 &     14.0 &   4.24  &     4.0 &   0.04 &  1.47e-04 & 4.46e-05 & 4.20e-05 & 4.52e-07 &    4.4 &  2    &   \\
55  &  11:52:47.62 &   44:07:04.56 &      8.9 &   3.46  &     3.1 &   0.04 &  9.32e-05 & 3.64e-05 & 3.28e-05 & 4.13e-07 &    3.0 &  mrgd &   \\
56  &  11:52:47.19 &   44:06:58.76 &     22.6 &   5.10  &     3.4 &   0.04 &  2.37e-04 & 5.36e-05 & 3.57e-05 & 4.26e-07 &    7.4 &  1,2  &   \\
57  &  11:52:47.17 &   44:07:19.48 &      9.5 &   3.46  &     2.5 &   0.03 &  9.99e-05 & 3.64e-05 & 2.61e-05 & 3.59e-07 &    3.4 &  mrgd &   \\
58  &  11:52:46.89 &   44:10:07.67 &     23.8 &   6.56  &    19.2 &   0.11 &  2.50e-04 & 6.89e-05 & 2.02e-04 & 1.13e-06 &    4.3 &  1    &   \\
59  &  11:52:46.67 &   44:09:04.11 &     25.7 &   5.92  &     9.3 &   0.07 &  2.70e-04 & 6.21e-05 & 9.79e-05 & 7.15e-07 &    6.2 &  1,2  &   \\
60  &  11:52:46.65 &   44:06:48.46 &    363.2 &  19.26  &     7.8 &   0.06 &  3.81e-03 & 2.02e-04 & 8.24e-05 & 6.35e-07 &   92.4 &  1    &   \\
61  &  11:52:46.48 &   44:07:53.67 &    104.2 &  10.49  &     5.8 &   0.05 &  1.09e-03 & 1.10e-04 & 6.09e-05 & 5.52e-07 &   29.3 &  1,2  & HII? \\
62  &  11:52:46.32 &   44:03:52.55 &     16.3 &   4.69  &     5.7 &   0.05 &  1.71e-04 & 4.93e-05 & 5.99e-05 & 5.56e-07 &    4.6 &  1    &   \\
63  &  11:52:46.25 &   44:06:27.85 &    145.8 &  12.25  &     4.2 &   0.04 &  1.53e-03 & 1.29e-04 & 4.43e-05 & 4.71e-07 &   45.2 &  1,2  &   \\ 
64  &  11:52:45.63 &   44:05:26.79 &     16.9 &   4.36  &     2.1 &   0.03 &  1.78e-04 & 4.58e-05 & 2.20e-05 & 3.34e-07 &    6.3 &  1    &   \\
65  &  11:52:44.45 &   44:09:14.60 &     45.3 &   7.48  &    10.7 &   0.07 &  4.76e-04 & 7.86e-05 & 1.12e-04 & 7.65e-07 &   10.3 &  1,2  &   \\
67  &  11:52:42.22 &   44:07:23.77 &     17.0 &   5.00  &     8.0 &   0.06 &  1.79e-04 & 5.25e-05 & 8.40e-05 & 6.68e-07 &    4.3 &  mrgd &   \\
70  &  11:52:38.96 &   44:06:02.52 &     23.1 &   5.29  &     4.9 &   0.05 &  2.43e-04 & 5.56e-05 & 5.15e-05 & 5.21e-07 &    6.8 &  1    & IR? \\
73  &  11:52:37.80 &   44:06:03.89 &     10.5 &   4.00  &     5.5 &   0.05 &  1.11e-04 & 4.20e-05 & 5.73e-05 & 5.44e-07 &    3.0 &  mrgd &   \\
74  &  11:52:37.09 &   44:07:54.61 &     11.4 &   4.24  &     6.6 &   0.06 &  1.20e-04 & 4.46e-05 & 6.93e-05 & 6.09e-07 &    3.1 &  2    &   \\
77  &  11:52:34.40 &   44:05:16.25 &     17.2 &   4.80  &     5.8 &   0.05 &  1.81e-04 & 5.04e-05 & 6.05e-05 & 5.51e-07 &    4.9 &  2    &   \\  \hline
\end{tabular}

Notes: $^a$Epoch: sources are listed as `1' or `2' from detection in
Epoch 1 or Epoch 2, respectively.  `mrged' means that the source was
detected {\it only} in the merged data.  That does not preclude it
from being present in Epochs 1 or 2, but such sources did not exceed
the statistical cutoff.

Table heading abbreviations: `Unc' = UNCertainty; `BkgCts' = BacKGound
CounTS; `BkUnc' = BacKground count UNCertainty; `SrcRate' = SouRCe
count RATE; `SrcRUnc' = SouRCe Rate UNCertainty; `BkgRate' =
BacKGround RATE; `BkRUnc' = BacKground Rate UNCertainty; `Signif' =
statistical SIGNIFicance; `Cntrpt' = Counterpart (see \S2.3; Inside D25).

\end{table}

\setcounter{table}{2}
\begin{table}[h!]
\centering
\caption{NGC 3938: Merged Source List (continued)}
\tiny
\begin{tabular}{rrrrrrrrrrrrll}
 N  &       RA     &       Dec     &   Counts &    Unc  &  BkgCts &  BkUnc &   SrcRate &  SrcRUnc &  BkgRate &   BkRUnc & Signif & Epoch$^a$ & Cntrpt  \\ \hline
\multicolumn{14}{c}{Outside the D25 Circle}  \\
 1  &  11:53:54.95 &   44:00:39.37 &     17.9 &   5.57  &    13.1 &   0.09 &  1.88e-04 & 5.85e-05 & 1.38e-04 & 9.69e-07 &    3.8 &  2    & (2) \\
 2  &  11:53:42.95 &   44:07:15.76 &     42.9 &   7.94  &    20.1 &   0.11 &  4.50e-04 & 8.34e-05 & 2.11e-04 & 1.19e-06 &    7.7 &  2    & star \\
 3  &  11:53:27.49 &   44:10:33.65 &     22.7 &   6.25  &    16.3 &   0.10 &  2.38e-04 & 6.56e-05 & 1.72e-04 & 1.07e-06 &    4.4 &  2    & (2) \\
 4  &  11:53:26.38 &   44:10:35.92 &     25.9 &   7.07  &    24.1 &   0.12 &  2.72e-04 & 7.43e-05 & 2.53e-04 & 1.30e-06 &    4.3 &  1,2  & galaxy \\
 5  &  11:53:22.11 &   44:07:30.51 &     78.4 &  10.68  &    35.6 &   0.15 &  8.23e-04 & 1.12e-04 & 3.74e-04 & 1.57e-06 &   11.1 &  1,2  & star \\
 6  &  11:53:21.60 &   44:11:31.22 &     78.2 &  11.66  &    57.8 &   0.19 &  8.21e-04 & 1.22e-04 & 6.07e-04 & 2.01e-06 &    9.0 &  1,2  & IR src \\ 
 7  &  11:53:18.95 &   44:12:15.47 &     31.3 &   7.35  &    22.7 &   0.12 &  3.28e-04 & 7.72e-05 & 2.39e-04 & 1.26e-06 &    5.3 &  2    & galaxy \\
 8  &  11:53:18.94 &   44:00:30.10 &     22.1 &   6.17  &    15.9 &   0.10 &  2.32e-04 & 6.48e-05 & 1.67e-04 & 1.05e-06 &    4.4 &  2    & (3?) \\ 
 9  &  11:53:11.83 &   44:07:02.15 &     36.5 &   7.07  &    13.5 &   0.08 &  3.83e-04 & 7.43e-05 & 1.42e-04 & 8.61e-07 &    7.6 &  1,2  & galaxy \\
10  &  11:53:11.11 &   43:57:19.53 &     72.9 &   9.85  &    24.1 &   0.12 &  7.65e-04 & 1.03e-04 & 2.53e-04 & 1.31e-06 &   12.2 &  1    & (2?) \\
11  &  11:53:09.72 &   44:00:05.07 &     49.8 &   9.22  &    35.2 &   0.15 &  5.23e-04 & 9.68e-05 & 3.70e-04 & 1.57e-06 &    7.1 &  1    & (3?) \\
12  &  11:53:09.19 &   44:03:00.93 &     31.2 &   6.32  &     8.8 &   0.07 &  3.28e-04 & 6.64e-05 & 9.20e-05 & 6.92e-07 &    7.6 &  1,2  & (1) \\
13  &  11:53:08.57 &   44:04:49.19 &     10.1 &   3.87  &     4.9 &   0.05 &  1.06e-04 & 4.07e-05 & 5.15e-05 & 5.22e-07 &    3.0 &  mrgd & galaxy \\ 
14  &  11:53:07.46 &   44:02:19.35 &    119.7 &  12.17  &    28.3 &   0.13 &  1.26e-03 & 1.28e-04 & 2.97e-04 & 1.39e-06 &   18.8 &  1,2  & QSO \\
15  &  11:53:07.40 &   44:04:50.38 &     16.4 &   4.90  &     7.6 &   0.06 &  1.72e-04 & 5.15e-05 & 7.98e-05 & 6.43e-07 &    4.2 &  2    & no cntrprt?? \\
16  &  11:53:07.01 &   43:58:59.61 &     53.8 &   9.27  &    32.2 &   0.14 &  5.65e-04 & 9.74e-05 & 3.38e-04 & 1.49e-06 &    8.0 &  1,2  & star \\
17  &  11:53:06.57 &   44:02:35.85 &     40.0 &   7.00  &     9.0 &   0.07 &  4.20e-04 & 7.35e-05 & 9.48e-05 & 7.27e-07 &    9.7 &  1    & galaxy \\
18  &  11:53:06.03 &   43:58:32.44 &     24.4 &   7.62  &    33.6 &   0.14 &  2.56e-04 & 8.00e-05 & 3.53e-04 & 1.52e-06 &    3.6 &  mrgd & galaxy \\
19  &  11:53:05.80 &   44:03:34.85 &    116.7 &  11.18  &     8.3 &   0.06 &  1.23e-03 & 1.17e-04 & 8.68e-05 & 6.77e-07 &   29.2 &  1,2  & galaxy \\
20  &  11:53:04.01 &   44:10:22.41 &     29.9 &   7.00  &    19.1 &   0.11 &  3.14e-04 & 7.35e-05 & 2.00e-04 & 1.14e-06 &    5.5 &  2    & IR src \\
22  &  11:53:01.18 &   43:58:23.28 &     29.5 &   7.94  &    33.5 &   0.15 &  3.10e-04 & 8.34e-05 & 3.52e-04 & 1.53e-06 &    4.3 &  mrgd & galaxy \\
25  &  11:52:59.80 &   44:02:13.48 &     38.1 &   6.93  &     9.9 &   0.07 &  4.01e-04 & 7.28e-05 & 1.04e-04 & 7.36e-07 &    9.0 &  1,2  & galaxy \\
44  &  11:52:50.79 &   44:13:33.43 &     27.4 &   7.62  &    30.6 &   0.14 &  2.87e-04 & 8.00e-05 & 3.22e-04 & 1.45e-06 &    4.1 &  mrgd & IR src \\
66  &  11:52:44.06 &   43:59:14.79 &     20.0 &   6.78  &    26.0 &   0.13 &  2.10e-04 & 7.12e-05 & 2.73e-04 & 1.34e-06 &    3.2 &  1    & galaxy \\
68  &  11:52:40.63 &   44:11:09.48 &     24.9 &   7.07  &    25.1 &   0.13 &  2.62e-04 & 7.43e-05 & 2.64e-04 & 1.31e-06 &    4.1 &  2    & (3?) \\
69  &  11:52:39.36 &   44:03:28.05 &     37.4 &   6.71  &     7.6 &   0.06 &  3.93e-04 & 7.05e-05 & 7.97e-05 & 6.43e-07 &    9.6 &  1    & IR src \\
71  &  11:52:38.90 &   43:58:52.62 &     41.8 &   8.66  &    33.2 &   0.14 &  4.39e-04 & 9.10e-05 & 3.49e-04 & 1.51e-06 &    6.1 &  1    & galaxy \\
72  &  11:52:38.72 &   44:12:47.03 &    612.8 &  26.59  &    94.2 &   0.25 &  6.44e-03 & 2.79e-04 & 9.89e-04 & 2.61e-06 &   57.0 &  1,2  & galaxy \\
75  &  11:52:36.43 &   44:01:07.52 &     22.3 &   6.40  &    18.7 &   0.11 &  2.34e-04 & 6.73e-05 & 1.97e-04 & 1.13e-06 &    4.1 &  1    & IR src \\
76  &  11:52:35.95 &   43:57:23.44 &     37.4 &   8.89  &    41.6 &   0.16 &  3.93e-04 & 9.34e-05 & 4.37e-04 & 1.70e-06 &    5.0 &  2    & (1 ft) \\
78  &  11:52:28.20 &   44:11:06.12 &     35.8 &   8.60  &    38.2 &   0.15 &  3.75e-04 & 9.04e-05 & 4.02e-04 & 1.62e-06 &    4.9 &  1    & (3?) \\
79  &  11:52:27.17 &   44:04:55.72 &     27.2 &   6.71  &    17.8 &   0.11 &  2.86e-04 & 7.05e-05 & 1.87e-04 & 1.10e-06 &    5.1 &  1    & (1?) \\
80  &  11:52:24.95 &   44:07:22.06 &     43.1 &   9.59  &    48.9 &   0.18 &  4.53e-04 & 1.01e-04 & 5.13e-04 & 1.84e-06 &    5.4 &  1    & (1 br, 2 ft) \\
81  &  11:52:22.39 &   44:05:18.34 &     20.3 &   6.56  &    22.7 &   0.12 &  2.13e-04 & 6.89e-05 & 2.38e-04 & 1.25e-06 &    3.5 &  mrgd & (1) \\
82  &  11:52:21.55 &   44:12:09.61 &     51.7 &  10.05  &    49.3 &   0.18 &  5.43e-04 & 1.06e-04 & 5.18e-04 & 1.85e-06 &    6.4 &  1    & (1) \\
83  &  11:52:18.05 &   44:00:42.85 &     26.1 &   6.71  &    18.9 &   0.11 &  2.74e-04 & 7.05e-05 & 1.99e-04 & 1.14e-06 &    4.8 &  1    & (1) \\
84  &  11:52:15.28 &   44:09:21.04 &     43.7 &   8.95  &    36.3 &   0.15 &  4.59e-04 & 9.40e-05 & 3.81e-04 & 1.58e-06 &    6.2 &  1,2  & star \\
85  &  11:52:14.41 &   44:00:40.71 &     47.5 &   8.37  &    22.5 &   0.12 &  4.99e-04 & 8.79e-05 & 2.37e-04 & 1.25e-06 &    8.1 &  1    & (3) \\
86  &  11:52:13.09 &   44:01:28.32 &     62.8 &   9.27  &    23.2 &   0.12 &  6.59e-04 & 9.74e-05 & 2.44e-04 & 1.27e-06 &   10.6 &  1    & IR src \\
87  &  11:52:13.07 &   44:09:01.97 &    155.5 &  15.84  &    95.5 &   0.25 &  1.63e-03 & 1.66e-04 & 1.00e-03 & 2.59e-06 &   14.4 &  1,2  & radio src \\
88  &  11:52:12.80 &   44:10:45.61 &     35.4 &   8.55  &    37.6 &   0.15 &  3.72e-04 & 8.97e-05 & 3.95e-04 & 1.62e-06 &    4.9 &  mrgd & (2?) \\
89  &  11:52:09.23 &   44:04:22.41 &     29.5 &   6.86  &    17.5 &   0.10 &  3.10e-04 & 7.20e-05 & 1.83e-04 & 1.10e-06 &    5.6 &  1    & (4?) \\
90  &  11:52:04.10 &   44:08:11.45 &     26.5 &   7.14  &    24.5 &   0.12 &  2.78e-04 & 7.50e-05 & 2.58e-04 & 1.31e-06 &    4.4 &  1    & (1 star?) \\
91  &  11:52:04.05 &   44:11:29.82 &     45.8 &   9.44  &    43.2 &   0.17 &  4.81e-04 & 9.91e-05 & 4.54e-04 & 1.74e-06 &    6.0 &  1    & galaxy \\
92  &  11:52:01.84 &   44:03:12.23 &     87.6 &  12.57  &    70.4 &   0.21 &  9.20e-04 & 1.32e-04 & 7.39e-04 & 2.22e-06 &    9.3 &  1    & (1 br, 6 ft) \\
93  &  11:51:57.46 &   44:05:21.37 &     40.0 &   8.78  &    37.0 &   0.15 &  4.20e-04 & 9.22e-05 & 3.89e-04 & 1.61e-06 &    5.6 &  1    & star \\
94  &  11:51:54.02 &   44:07:55.79 &     23.7 &   7.00  &    25.3 &   0.13 &  2.49e-04 & 7.35e-05 & 2.65e-04 & 1.33e-06 &    3.9 &  1    & (2) \\
95  &  11:51:51.00 &   43:59:59.95 &     32.3 &   7.75  &    27.7 &   0.13 &  3.39e-04 & 8.14e-05 & 2.91e-04 & 1.41e-06 &    5.1 &  1    & (3?) \\  \hline
\end{tabular}

Notes: See the Appendix for a description of the objects in this
$>$D25 table.  $^a$Epoch: sources are listed as `1' or `2' from
detection in Epoch 1 or Epoch 2, respectively.  `mrged' means that the
source was detected {\it only} in the merged data.  That does not
preclude it from being present in Epochs 1 or 2, but such sources did
not exceed the statistical cutoff.  Under `Cntrprts': numbers in ()
indicate potential counterparts visible in {\it Spitzer} IRAC data.
`br' = bright; `ft' = faint.

\end{table}

Table~\ref{NmbrDetSrcs} displays the numbers of detected sources both
merged, by band, and by observation epoch.  Sources in the broad band
were counted as detected if the statistical significance was
${>}3{\sigma}$.  Sources in the sub-bands were also counted with that
significance.  Sources in the broad band were matched into the sub-bands
regardless of the statistical significance in the sub-band.  This
approach permitted
extracting color information about each detected source even when some
of the color information led to upper limits.  

All detected sources were also compared to the limiting {\it count}
sensitivity of the CCDs as determined by the CIAO routine {\tt
  lim\_sens}.  The limiting sensitivity task convolves the exposure
map across the field-of-view with the effective area that any given
X-ray would sense.  Consequently, sources well off the optical axis
must necessarily be brighter to be detected.  Such sources all fell
below 10 counts.  Detections with extracted counts fewer than the
limiting sensitivity value at the location of the detected source were
dropped.  That limit is also the point at which the source count
distributions of the {\it ChaMP} project peaked (their Figure 16;
\citealt{2007ApJS..169..401K}).  We followed the limiting sensitivity
values across the field-of-view.  However, for NGC 3938, the
field-of-view of ACIS-I is sufficiently large and the galaxy
sufficiently distant and small that sources near the edge of the {\it
  galaxy} are essentially unaffected by the limiting sensitivity map.
This is our primary motivation for splitting the detected sources into
`inside' and `outside' the D25 circle.  Table~\ref{MrgdSrc}
lists {\it all} detected sources in descending Right Ascension order and further
separated into `inside' and `outside' the D25 circle.  The sources {\it
outside} the D25 circle are described in the Appendix for organization sake.

The merged, detected sources were {\it numbered} in descending 
Right Ascension (RA) order regardless of their position within 
the ACIS 
field-of-view.  The merged source list was then separated 
into `inside' and `outside' the D25 circle as mentioned above
(Table~\ref{MrgdSrc}).  The detected source tables for 
Epochs 1 and 2 are contained in the Appendix and follow the 
same numbering approach.  We count 47 sources within or on
the D25 circle: sources 21, 23, 24, 26 to 43, 45 to 65, 67, 
70, 73, 74, and 77.  All other detected objects lie outside 
of the D25 circle.

\subsection{Detected Sources {\it versus} the {\it Chandra}
Source Catalog}

We must address differences between our detected source list 
and the list from the Chandra Source Catalog \citep{Evans2010}
(hereafter, CSC).  We here focus on the source differences
within the D25 circle and leave the sources outside the D25
circle to Appendix~\ref{SrcOutD25}.

There are 36 sources in common between the two source lists.
However, there are six sources detected within D25 on the CSC
list that are not on our list and five sources on our list 
that are not in the CSC.  For both source lists, the detected
objects are essentially all on axis or sufficiently close 
that changes with off-axis angle are not at all important.

The detected source list differences could fall into two
reasonable categories: below our minimum signal-to-noise 
value or only detected from the merged Epoch 1 and Epoch 2 
data sets.  There are sources, however, for which we do 
not have a reasonable explanation.  The details follow.

Five of the six CSC sources we fail to detect fall below 
our minimum signal-to-noise threshold (2CXO J115242.0+440532;
2CXO J115256.7+440902; 2CXO J115258.6+440528; 2CXO
J115256.7+440918; 2CXO 115247.1+440434).  The sixth one
technically would be above our threshold, but the detected
events are distributed across a region larger than the 
PSF (2CXO J115248.5+440704), puzzling for an essentially 
on-axis source.  That may suggest the source is extended
-- but if it is, it then falls below our detection threshold.

For the five sources on our list not detected by the CSC, 
we only have an explanation for one: it is on our {\it merged}
list (number 55).  The CSC only merges observations if the
pointing directions lie within 1$'$ of each other.  The
two observations of NGC 3938 lie ${\approx}$3' apart.
The other sources (numbers 26, 29, 30, 34, and 74) all lie 
above our detection threshold (ranging from ${\sim}$8 to 15
counts).  An examination of each shows a small `pile' of 
events consistent with {\it Chandra}'s PSF.  We consequently 
do not have an explanation for their lack of detection in
the CSC.  Source 29 is particularly puzzling as it lies 
within ${\sim}15''$ of a source with a similar count rate 
that is present on both our and the CSC detected source
lists.

\subsection{MultiWavelength Source Comparison}

\begin{figure}[h!]
 \centering
 \caption{{\it Spitzer} images of NGC 3938 in the (left to right) IRAC
   3.6, 4.5, 5.8, 8.0, and IRS 24 ${\mu}$m bands.  X-ray sources are
   circled in each band but only identified in the 3.6 ${\mu}$m band.
   The color coding follows that of Figure~\ref{FoVopt}.  X-ray sources do
   not appear to correlate with any infrared features.  North is up,
   East left.  The fields are ${\sim}5'$ on a side.}
 \scalebox{0.75}{\includegraphics[angle=90.0]{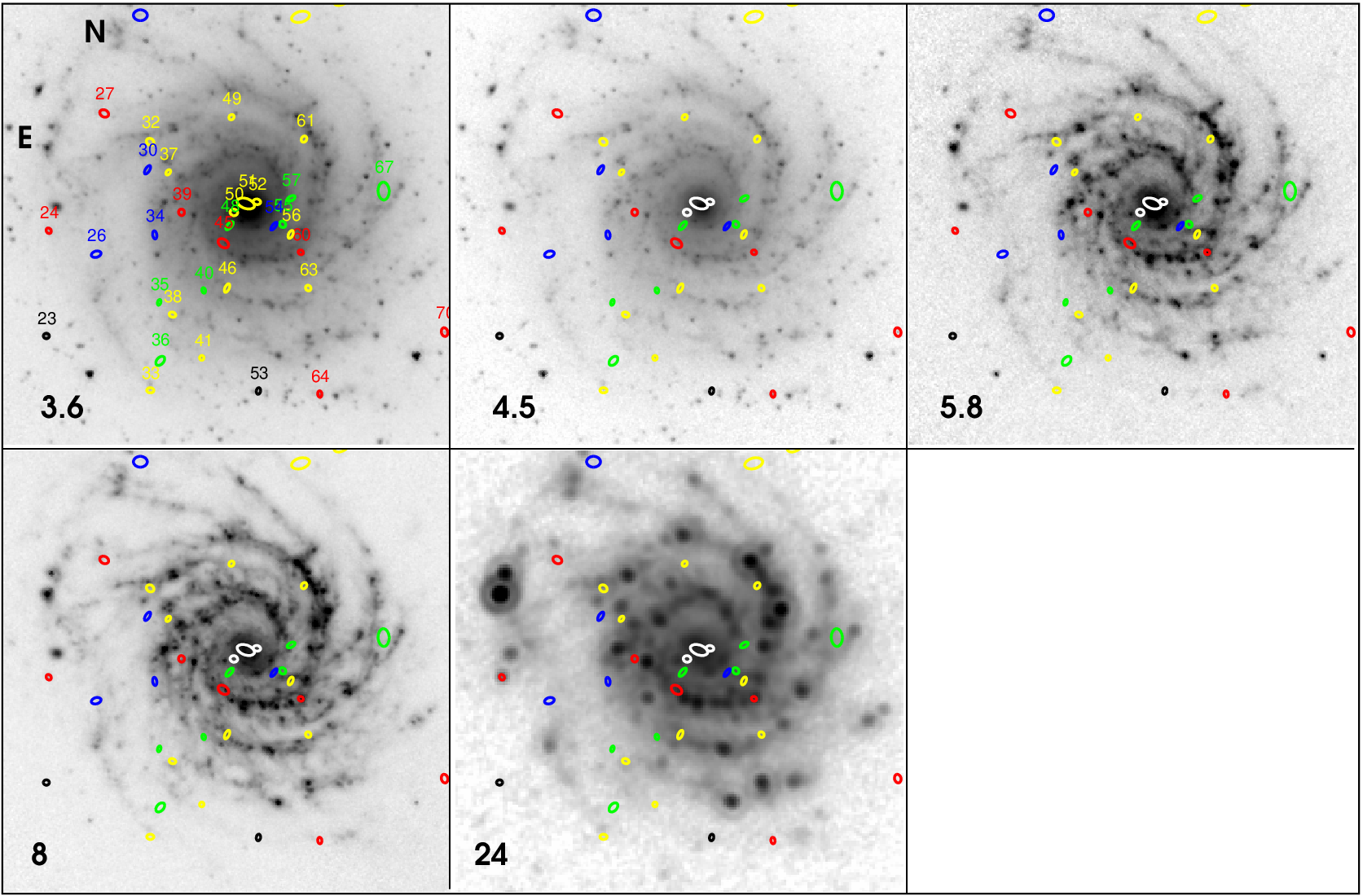}}
 \label{IRimg}
\end{figure}

One motivation to conduct high angular resolution X-ray
observations of nearby spiral galaxies with face-on 
orientations is the search for counterparts at
multiple wavebands to the detected discrete X-ray sources.
The ability to see sources in other bands enhances the multi-
wavelength connections of types of sources that may not
be easily visible in the Milky Way.  Consequently, we attempted to
correlate detected {\it Chandra} sources with sources from other
bands.  We understand that NGC 3938 is sufficiently distant that
matching sources is difficult.

Normally, we would separate the very soft sources from the harder
sources with the overall assumption that the soft sources are
supernova remnants (SNRs).  However, given that both epochs of
observation of NGC 3938 were obtained relatively late in {\it
Chandra}'s life, the build-up of the contaminant on the ACIS 
optical blocking window has dramatically reduced the instrument's
effective area sharply for photons with energies of less than 1 keV
(below ${\sim}$0.7-0.8 keV). This essentially eliminates SNRs as 
a possible category, particularly for a relatively distant galaxy 
since SNRs typically produce a majority of their X-ray photons with
energies in this range. Furthermore, the limiting luminosity 
is ${\approx}10^{38}$ erg s$^{-1}$, about 10-100 times larger than
the typical integrated X-ray luminosity of SNRs (\cite{HS1984};
\cite{CS1995}; \cite{Long2017}).   Consequently, we ignore
the possibility that any X-ray sources in NGC 3938 are SNRs.

The NED archive also lists a number of H~II regions within 
the D25 circle of NGC 3938.  In general, we would not expect H~II
regions to be X-ray sources, particularly in more distant 
galaxies, as they are often low-luminosity X-ray-emitting 
star formation regions (${\sim}10^{33}$ erg s$^{-1}$ e.g., 
\cite{Townsley2003}) unless powered by a central
black hole or some equivalent high-energy source.

We extracted {\it Spitzer} IRAC sources from the IRAC
archive\footnote{ipac.caltech.edu} to learn whether any detected
{\it Chandra} sources matched.  The archival data were 
obtained using the
{\it Spitzer} Space Telescope's Infrared Array Camera (IRAC -- see
\citet{Fazio2004}) at 3.6, 4.5, 5.8 and 8.0 $\mu$m.
Figure~\ref{IRimg} shows the images overlaid with {\it Chandra}
detected sources.  The sources listed in Table~\ref{MrgdSrc} show
potentially seven infrared sources matching X-ray positions within
the uncertainties (`IR?' in the far right column of
Table~\ref{MrgdSrc}).
None of the other X-ray sources or infrared sources revealed a
matching counterpart.

\begin{figure}[h!]
 \centering
 \caption{{\it Swift} UVOT images of NGC 3938 in the (left to right)
  UW2, UM2, UW1, U, B, and V bands.  X-ray sources are circled in each
  band but only identified in the UW2 band.  The color coding is
  identical to the previous images (Fig.~\ref{FoVopt}).  X-ray
  sources outside
  of the nucleus but within the D25 circle essentially fall into two
  groups: on or close to an arm or UV-emitting region, or in a
  UV-deficient region.  North is up, East left.  The fields are
  ${\sim}$5' on a side.}
 \scalebox{0.85}{\includegraphics[angle=90.0]{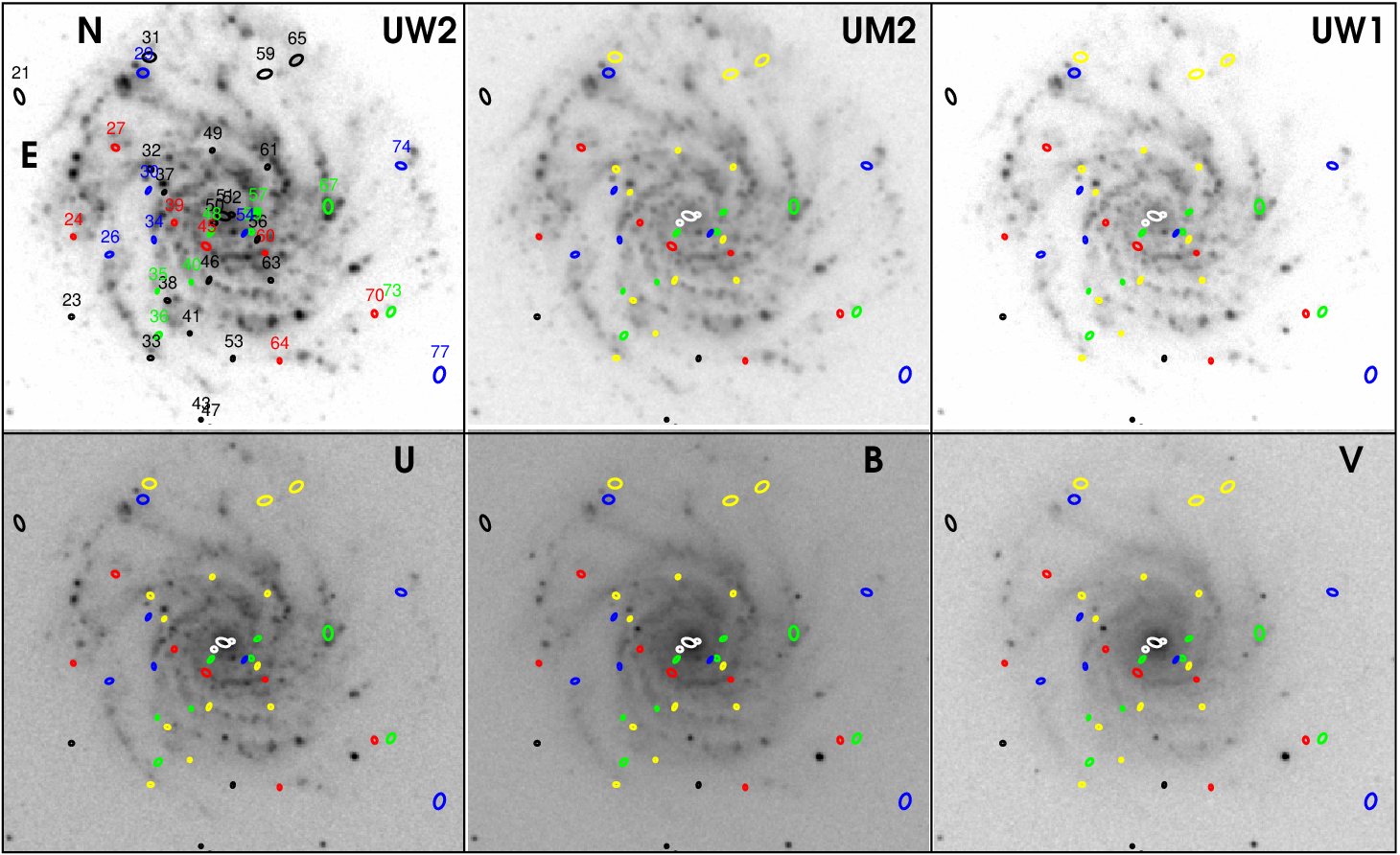}}
 \label{UVimg}
\end{figure}

Figure~\ref{UVimg} shows the full range of images of NGC 3938
obtained
from summing frames obtained with {\it Swift}'s UVOT.  In
comparing the optical with the UV images, one should see that the V
band is relatively devoid of specific sources.  In contrast, 
the UV bands exhibit significant emission variations within 
a spiral arm as the wavelength decreases.  That strong UV behavior
may mean that high-mass X-ray binaries (HMXBs) could be 
identified and separated from low-mass
XRBs (LMXBs) using the {\it HST} images (see \S\ref{SpecSrcs}) 
to build color diagrams from which masses of the main sequence
stars in the XRBs are inferred \citep{Chandar2020}.  It would 
be interesting to see the reliability of the approach as a 
function of distance -- to date,
this approach has been applied to M83 (\cite{Chandar2020};
\cite{Hunt2021}) and M81 \citep{Hunt2023a} individually and 
six other
galaxies (NGC 628, NGC 3351, NGC 3627, NGC 4321, NGC 4569, and NGC
4826) in a collective study \citep{Hunt2023b}.  Either approach,
however, lies outside the scope of this paper.

\begin{figure}[h!]
 \centering
 \caption{{\it HST} images (left = F555W; right = F814W) of the
  nucleus and eastern portions of NGC 3938.  X-ray sources are circled
  in each band.  The color coding is identical to that in
  Figure~\ref{FoVopt}.  Note that even at the resolution of HST's WFC
  instrument and outside of the nucleus only source 24 (near-center
  left) can be identified to a particular source
  (q.v. Fig.~\ref{SrcHST}). This significantly contrasts to the
  situation for sources outside the D25 circle where nearly every
  source has a known counterpart or a few candidate counterparts (see
  Table~\ref{MrgdSrc}).  North up, East left.  The fields are
  ${\sim}$2' on a side.}
 \scalebox{0.35}{\includegraphics{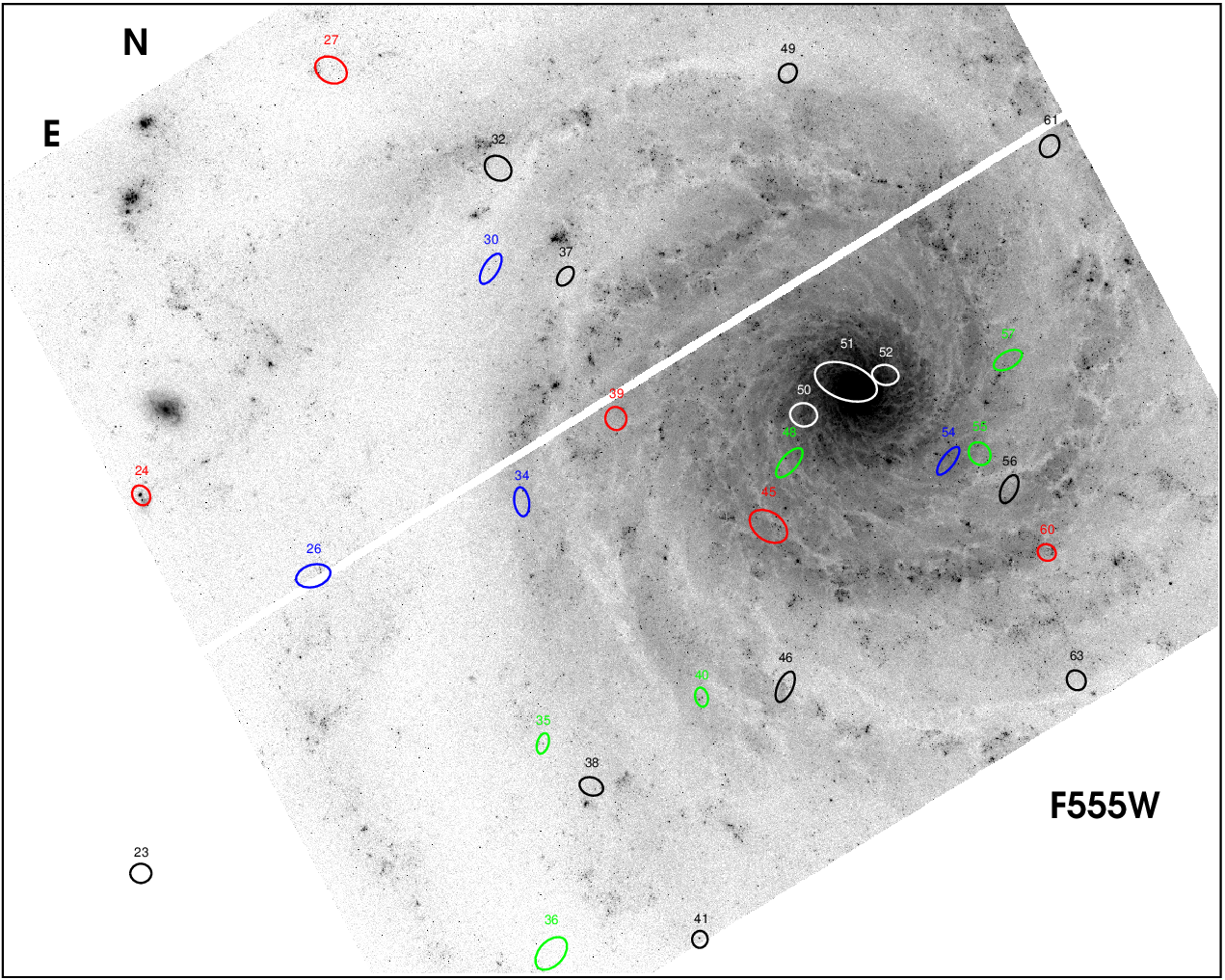}}
 \scalebox{0.35}{\includegraphics{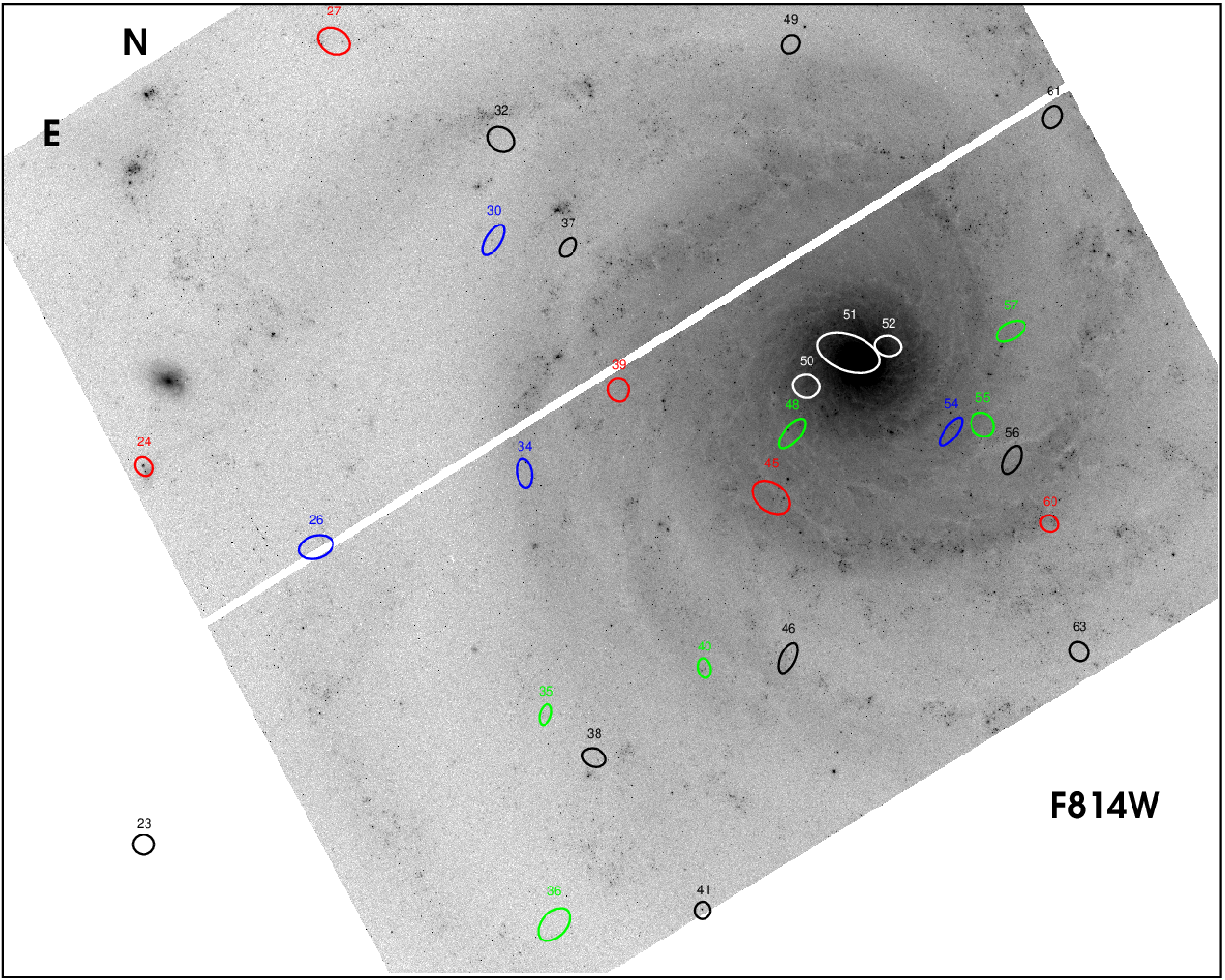}}
 \label{HSTimgs}
\end{figure}

\section{Specific Sources -- Detected or Not?}\label{SpecSrcs}

Five supernovae (SNe) have occurred in NGC 3938 over the 
historical record:
SN1961U (RA 11:52:56.81, Dec +44:09:01.1), SN1964L (11:52:49.09,
+44:07:45.4), SN 2005ay (11:52:48.07, +44:06:18.4), SN2017ein
(11:52:53.25, +44:07:26.20), and SN2022xlp (11:52:49.58, +44:06:03.5).
None of the SNe are detected with statistical significance in
our analysis.  No X-ray-emitting progenitor is found at the position 
of SN2022xlp nor pre-/post-X-ray emission at the position of 
SN2017ein.

However, SN1961U and SN1964L reveal a hint of emission -- both are
detected at ${\sim}$2 - 2.5 ${\sigma}$ which, properly, is consistent
with the background.  SN1961U and SN1964L both show detected X-rays --
just not enough to merit the `significantly detected' label.  The
counts are consistent with an X-ray flux upper limit of
${\sim}3{\times}10^{37}$ erg~s$^{-1}$.  It is quite possible that a
future, larger telescope with a PSF at least equivalent to {\it
Chandra}
will demonstrate that these objects are X-ray-emitting SNRs.
If that is nearly possible with {\it Chandra} looking at SNRs at
${\sim}$20 Mpc, then future researchers should be able to study 
many historical SNe in nearby galaxies at relatively early stages of
evolution as SNRs.

\begin{figure}[h!]
 \centering
 \caption{Expanded view of Sources 24 (left half) and 60 (right half)
   using the {\it Hubble} WFC3 images (filters F555W and F814W).  The
   frames are ${\sim}8'$ on a side.  The color coding matches that of
   Figure~\ref{FoVopt}.}
  \scalebox{0.5}{\includegraphics{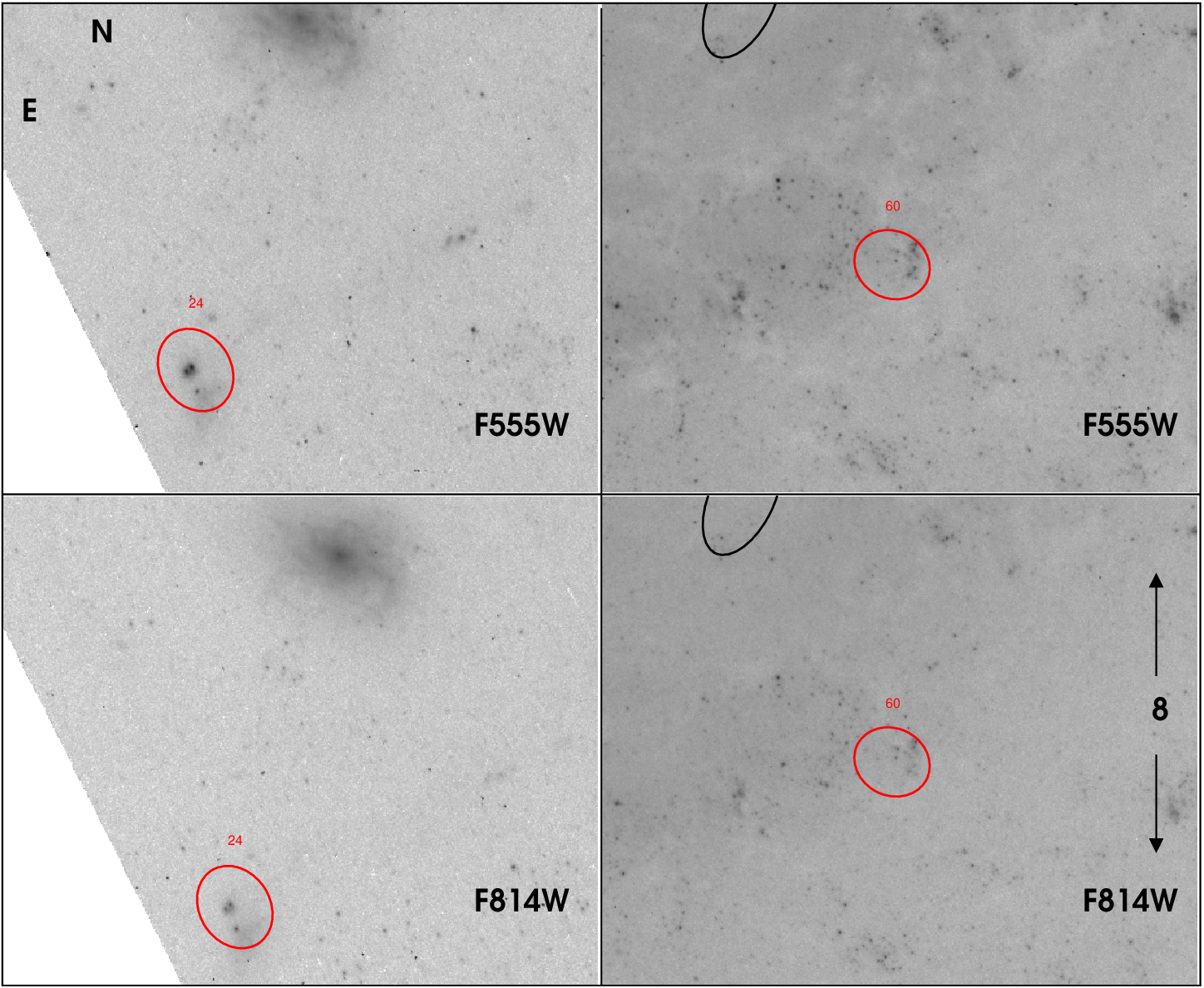}}
 \label{SrcHST}
\end{figure}

In addition to SNe, there are two {\it Hubble} images available
for the nucleus and immediate surroundings of NGC 3938 (WFC3 555W and
814W; Figure~\ref{HSTimgs})\footnote{These images were obtained from
MAST \citep{MAST2021} using {\it HST}.}.  One X-ray source matches
within 0.2 arcsec  with a corresponding optical counterpart in each
of the images: source 24.  Figure~\ref{SrcHST} (left) expands the
view.  The source is not a bright UV source (Fig.~\ref{UVimg}) but is
visible in the IR (Fig.~\ref{IRimg}).  In addition, source 60 matches
with a small group of stars (right) but we are unable to pin down a
specific candidate.

\begin{figure}[h!]
 \centering
 \caption{The nuclear region of NGC 3938.  From left to right, the
   nucleus is shown for (left) Epoch 1; (center) the merged data; 
   and (right) Epoch 2.  The ellipses are detected sources within
   the D25
   circle of the galaxy.  The ellipse for the {\it merged} data
   changes orientation slightly likely because the event
   distribution changes between the two epochs. The box is 
   centered on the nucleus and is 2$''$ on a side.  The 
   entire field-of-view is ${\sim}50''$ 
   on a side.  The color coding matches that of Fig.~\ref{FoVopt}.}
 \scalebox{0.5}{\includegraphics{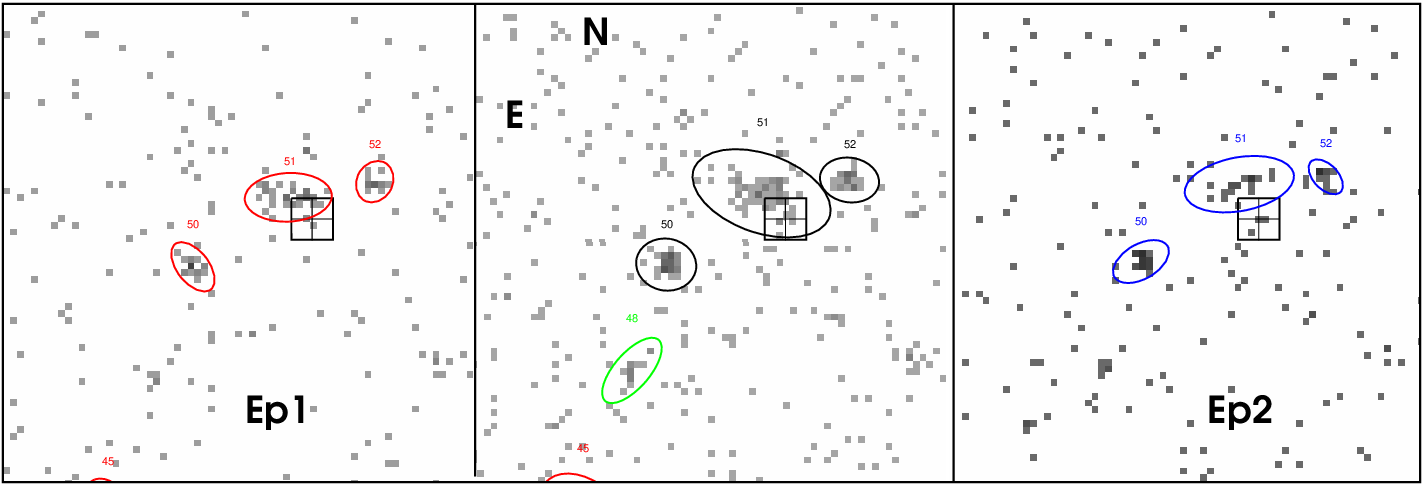}}
 \label{nucFigXray}
\end{figure}

Finally, we mention the possibility that the nucleus of NGC 3938 is
not X-ray bright -- perhaps not surprising given its status as a
LINER AGN.  Figure~\ref{nucFigXray} shows the situation:  there
appears to be a moderately bright X-ray source ${\approx}3''$ to 
the North (number 51; 36${\pm}$6.4 and 17.6${\pm}$4.6 counts in
Epochs 1 and 2, respectively), as shown in Figure~\ref{nucFigXray}. 
We immediately considered whether the ``known'' position of the
nucleus is slightly in error.  Alternatively, the sources are 
low-count statistical variations.  We deem the first explanation
unlikely given the {\it HST} WFC3 UVIS F555W image that shows a 
well-defined `circulation' pattern around the nucleus and 
{\it centered} on the nucleus using the coordinates as listed in
Table~\ref{NmbrDetSrcs}.  Those coordinates were verified 
in the GAIA reference frame.  The Epoch 1 detection list shows
a weak, {\it Soft} source at the nucleus (5.3${\pm}2.4$ counts
equivalent to a $2.4{\sigma}$ detection.  No corresponding object
exists in the Epoch 2 detection list regardless of the significance.

Furthermore, source 51 is extended E-W, suggesting either that a
source is being ripped apart {\bf or an extended `cloud' illuminated
  by the nucleus} {\it or} that it is a background extended source
{\it or} that there are multiple sources that are confused at {\it
  Chandra}'s resolution.  Since NGC 3938's nucleus produces emission
$>5''$ over two epochs of on-axis observations, the possibility that
{\it Chandra} is not resolving the nucleus seems unlikely {\it if} the
`source' is two separate objects.  If the `source' is a {\it string}
of X-ray-emitting objects that are blurred together, {\it Chandra}
likely could not resolve them at the distance of NGC 3938.  That a
background galaxy's light is getting through the nuclear region also
seems unlikely.  The {\it HST} image appears uniform in azimuth about
the nucleus, supporting the statement that a background galaxy's light
could not penetrate the nuclear region.  {\bf That leaves a limited
  set of possibilities: ripped-apart source or a blurred string of
  X-ray-emitting objects or something equivalent to an illuminated
  cloud, for example, the X-ray counterpart of Hanny's Voorwerp in IC
  2497 \citep{Fabbiano2019} or the polarized cloud near Sgr A$^*$
  \citep{Marin2023}.  None of these is an immediately-sensible
  possibility: a shredded source would likely need to be lit up by the
  nucleus, but the nucleus is not significantly brighter to do so; a
  string of X-ray-emitting objects is possible but immediately
  generates origination questions.  An X-ray gas cloud similar to the
  Voorwerp or Sgr A$^*$'s gas cloud would also appear implausible given
  the {\it HST} image.  For IC 2497, the X-ray `cloud' lies
  ${\sim}20''$ south of the nucleus, considerably more distant from
  the nucleus given the ${\sim}$225 Mpc distance to IC 2497.  For the
  Sgr A$^*$ cloud, the ${\approx}4''$ distance translates to ${\sim}250$
  pc at NGC 3938, i.e., ${\sim}$10 times farther or an ${\sim}$100
  times lower irradiation.  A sufficiently detailed investigation of
  any of these possibilities lies beyond the scope of this paper.}
Clearly the understanding of the nucleus must await a
higher-resolution and higher-sensitivity X-ray telescope, e.g., AXIS
\cite{Marchesi2020}.

\begin{figure}[h!]
 \centering
 \caption{The nuclear region of NGC 3938 as observed with the WFC3 
 UVIS
   instrument (filter F555W) on {\it HST}.  The ellipses all lie
   within ${\sim}10''$ of the nucleus but are {\it not} centered on
   the nucleus.  The nucleus lies at the intersection of the cross
   hairs within the box.  The box is 2$''$ on a side.  The
   `circulation' is visible in the image.  The entire
   field-of-view is ${\sim}22''$ on a side.  The color coding matches
   that of Fig.~\ref{FoVopt}.}
 \scalebox{0.5}{\includegraphics{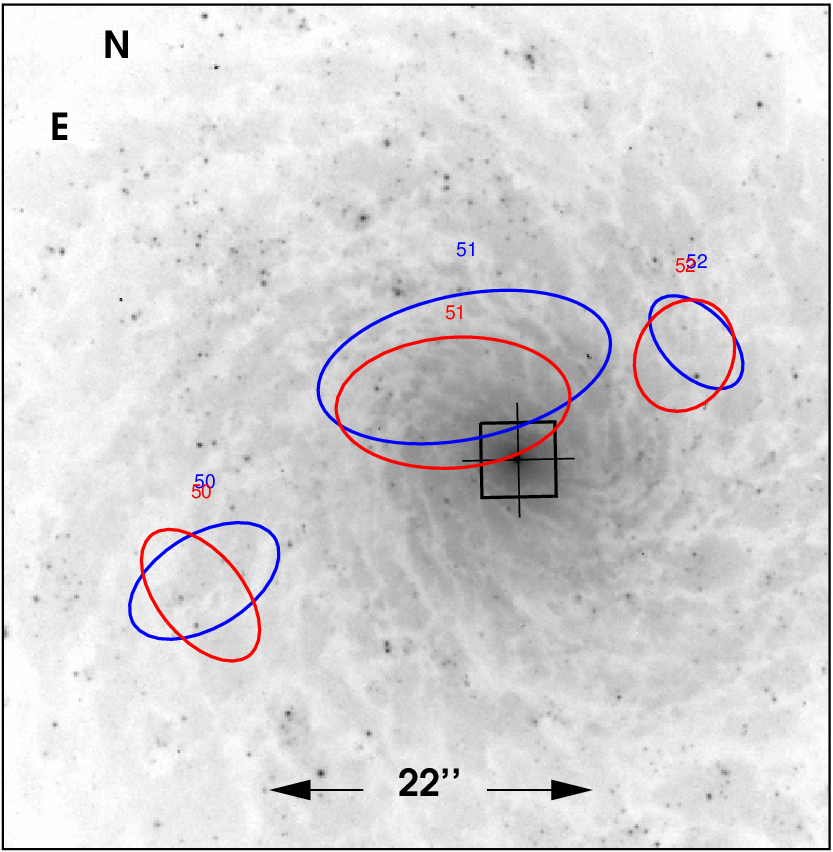}}
 \label{nucFigHST}
\end{figure}

\section{X-ray Properties via Color-Color Hardness}\label{SecColCol}

\citet{2003ApJ...595..719P} describe an X-ray color-color analysis
of the discrete sources based on separating each source's 
counts into
soft, medium, and hard bands.  If we follow that approach, defining
the bands to be `soft' (S), equal to 0.3-1.0 keV, `medium' (M) as
1.0-2.0 keV, and `hard' (H) as 2.0-8.0 keV which then leads to 
two colors: M~-~S and H~-~M, each normalized by the total counts.  
The color information is listed in Tables~\ref{Ep1ColTab} 
(Epoch 1) and \ref{Ep2ColTab} (Epoch 2), both located in the 
paper's appendix.  The contaminant is expected to alter the
positions of the ellipses -- we choose to retain the original
definitions for comparison with prior publications using this
approach.

\begin{figure}[h!]
 \centering
 \caption{Color-color plots for each epoch (left) Epoch 1; (right)
   Epoch 2 for objects {\it in}side the D25 circle.  The left arrow 
   in the Epoch 2 plot indicates the approximate change in each 
   data point solely
   because of the increase of the optical blocking window absorption
   between the two epochs.  To avoid an unreadable graph, we
   eliminated uncertainties from some of the data points, leaving a
   representative sample.  For data points without uncertainties,
   assume the uncertainties are at least as large as those plotted. 
   The `powerlaw' model describes the arc of changes to the colors 
   as the power law index changes (`dPL') from its starting point 
   at 3.0.  `UL' = Upper Limit. The region ellipses are defined
   in \cite{2003ApJ...595..719P}.}
 \scalebox{0.35}{\rotatebox{-90}{\includegraphics{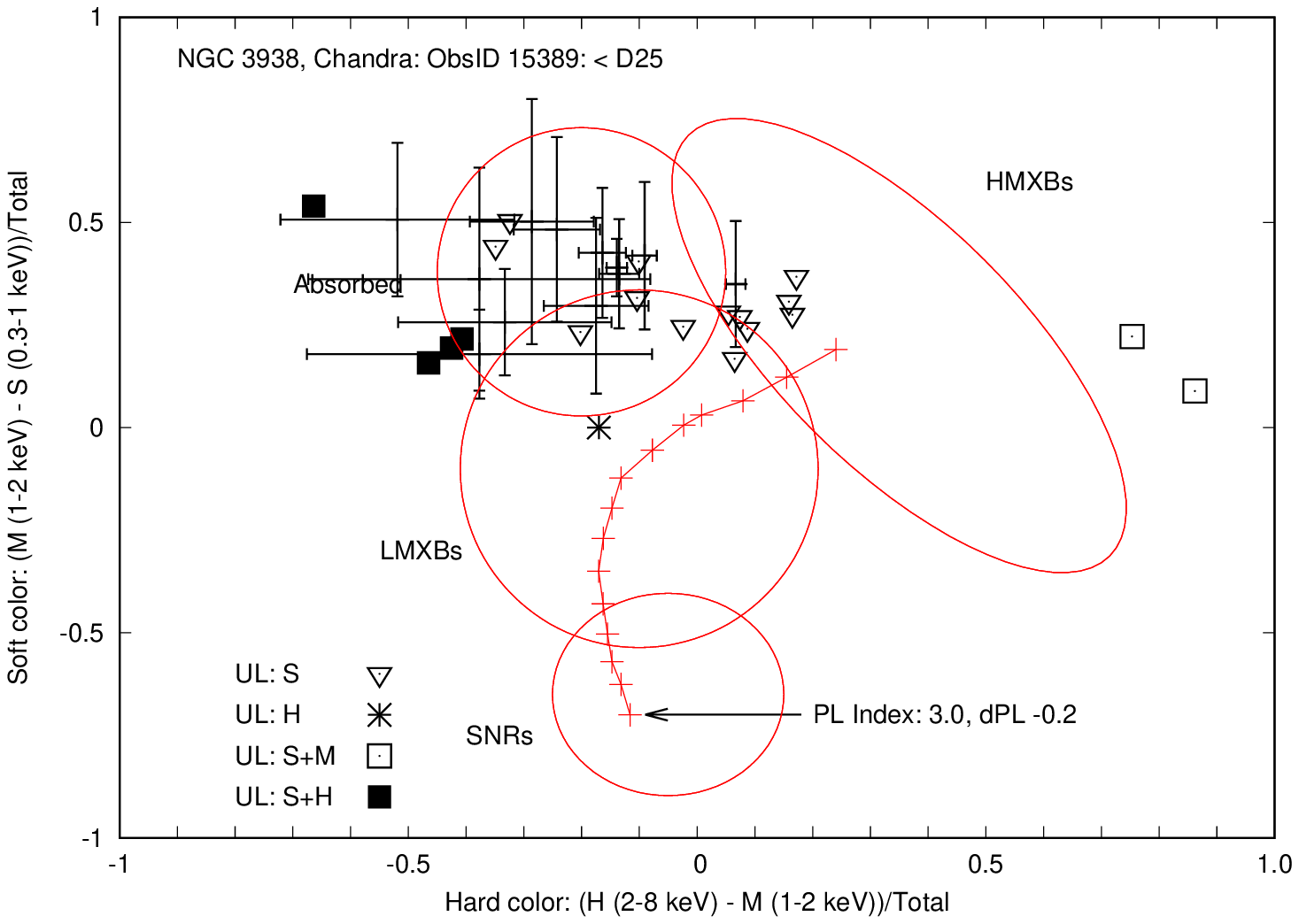}}}
 \scalebox{0.35}{\rotatebox{-90}{\includegraphics{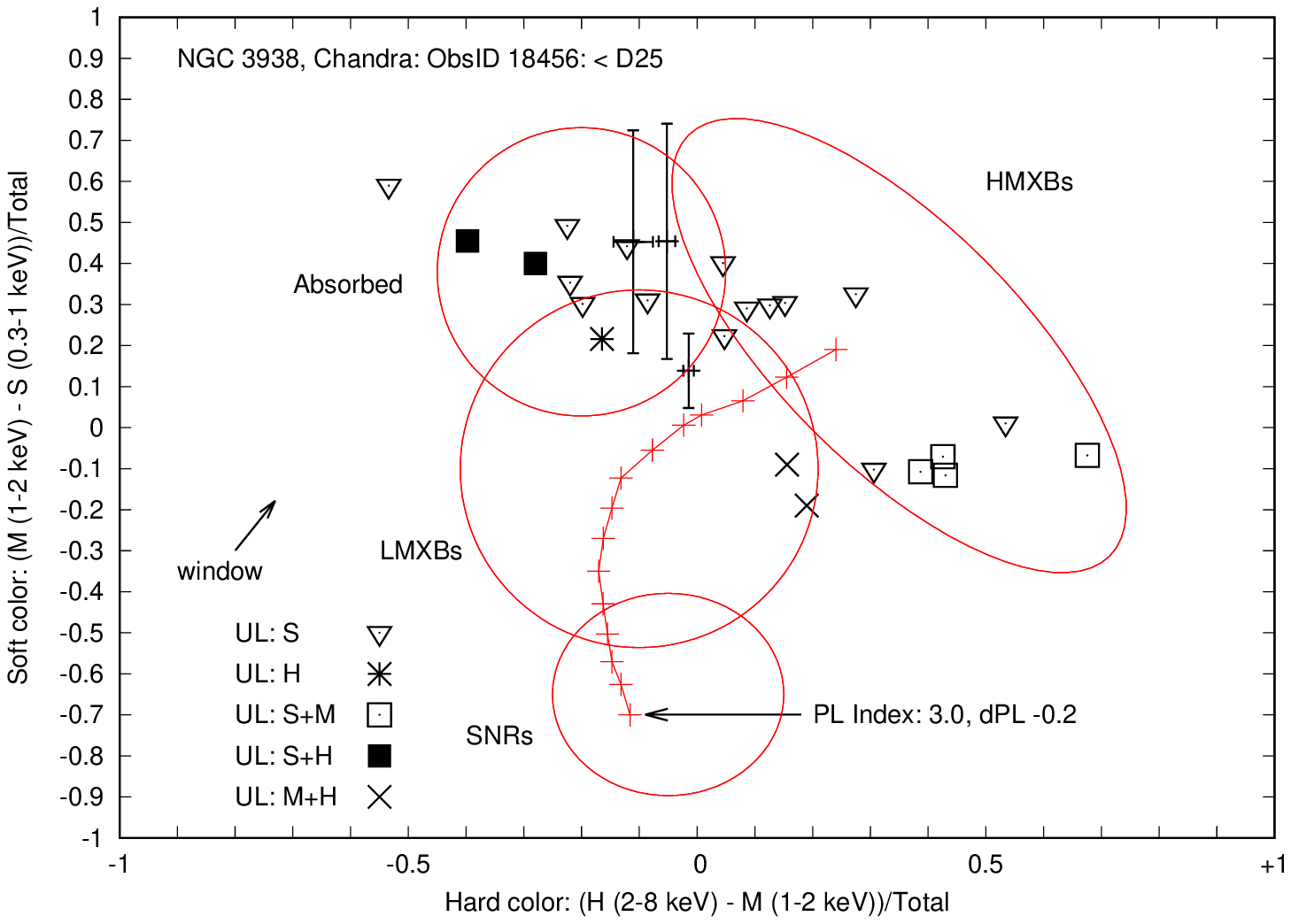}}}
 \label{colorcolorfig}
\end{figure}

Figures~\ref{colorcolorfig}(a) and (b) show the results for NGC 3938
for the individual epochs for sources {\it within} the D25 circle. 
We do not present a color-color plot for the merged data because of 
the optical blocking window build-up.

Of note in both color hardness plots is the complete lack of any
sources in the SNR circle, and relatively few in the LMXB circle.
As NGC 3938 is a star-forming galaxy, even if at a low rate,
it should exhibit some SNRs.  Further,
given that the galaxy is quite face-on to our perspective, we 
do not then get to claim `excess column density' within the 
galaxy as a straightforward explanation even though many of 
the sources are present in the `Absorbed' circle.  The 
likely explanation at this point, given that both observations 
were obtained somewhat late in {\it Chandra}'s lifetime, is 
the build-up of the absorbing contaminant on the ACIS optical 
blocking window.  

Even for the first epoch of data, that serves to reduce the
effective area of the telescope below $\sim$1 keV relative to
energies above 1 keV.  This would preferentially diminish
softer sources from detection with little change to medium and 
harder sources.  The question that remains is whether soft
sources such as supernova remnants, for example, are present
on the detected sources list at all.  To address that question
will likely require cross-wavelength analyses with pinpointed 
sources from other bands.

If we assume that much of the `motion' in the Epochs 1 to 
Soft color occurs because of the contaminant,
then that suggests that most of the detected sources in Epoch
1 are LMXBs.   That inference then suggests an additional
inference: NGC 3938 has not experienced a high star formation
rate in the past one-to-a-few Gyrs.   If that inference is
correct, that inference then goes against the historical 
presence of five supernovae in this galaxy (\S\ref{SpecSrcs}).

\begin{table}[h!]
 \centering
 \caption{Spectral Fit Parameters}
 \label{SpecFitTab}
 \begin{tabular}{rrllll}
         & ${\chi}^2$ & N$_{\rm H}$ & Brems (B, keV) or       & Calc'd Flux$^a$ & Calc'd Flux$^a$ \\
 SrcN    &  per dof   & ($10^{22}$ cm$^{-2}$) & Powerlaw Index (P)       & 0.5-2 keV   & 2-10 keV \\ \hline
  60     & 9.6/15     & $<$2.44 & P: 2.52${\pm}$2.31 & 3.6(-14)    & 5.4(-14) \\
  72-1   & 16.3/12    & $<$0.65 & P: 0.88${\pm}$1.18 & 4.6(-14)    & 1.7(-13) \\
  72-2   & 8.2/13     & $<$0.65 & P: 1.27${\pm}$1.10 & 4.5(-14)    & 1.6(-13) \\
  72-2   & 8.2/13     & $<$0.65 & B:  $<$15.7        & 5.3(-14)    & 1.2(-13) \\ \hline
\end{tabular}

$^a$Fluxes calculated using the best-fit model as listed; flux format:  n.n(-e) = n.n${\times}10^{-e}$.  Uncertainties on each
spectral data point were included in the fitting.  Uncertainties
on the fluxes are not included as the count rate-determined fluxes
provide a more statistically-defined value.

\end{table}

\section{Spectral Fits}\label{FitSpec}

There are at best two sources (60 and 72) that are sufficiently
bright to warrant the simplest model for spectral fitting. 
Source 60 lies inside the D25 circle and is only detected in
Epoch 1.  Source 72 is detected in both observations and lies
outside the D25 circle.  

We fit the spectra with either an absorbed power law or an
absorbed thermal bremsstrahlung spectrum plus a powerlaw for
the background.  All of the fits returned values with
significant uncertainties regardless of the model or fitting
approach we adopted.  We list the parameter values in 
Table~\ref{SpecFitTab} but do not include any figures of the
spectra or parameter contours to save space.  The spectra are
all peaked around 1 keV, both because of the drop in effective
area above that energy and because of the ACIS window build-up
below that energy. The result is that the fitted continuum
parameters are relatively unconstrained. The background region for 
these sources is determined by placing a circular region in the
portion of {\it Chandra's }I3 chip without any visible sources,
making it as large as possible.

Using our adopted distance of 22 Mpc, source 60 has a 0.5-2 keV flux
of ${\approx}2{\times}10^{39}$ erg s$^{-1}$.  If we reduce the
adopted distance to 17 Mpc, then the 0.5-2 keV flux drops to
${\approx}1.2{\times}10^{39}$ erg s$^{-1}$, so still within the ULX
definition.  

Source 72 lies outside of the D25 circle and corresponds with the
galaxy WISEA J114238.74+441247.0.  The NED-listed redshift is 0.227
which yields a recession velocity of 68300 km s$^{-1}$ relative to
the CMB or a Hubble distance of ${\sim}$1000 Mpc.  At that
distance, the 
observed fluxes are ${\sim}6{\times}10^{42}$ and 
${\sim}2{\times}10^{43}$ erg s$^{-1}$ for the 0.5-2 and 2-10 keV
bands, respectively.  Both values fall above the typical range of
normal galaxies and toward the lower end of active galaxies, 
suggesting a potential LINER or Seyfert AGN based solely on the
luminosity values.

\section{Timing}\label{TimDesc}
 
There are two time frames to be described: (i) variations {\it within}
an observation; and (ii) variations {\it between} the two epochs.
First, the variations within an observation and within the D25 radius
(sources outside D25 were also analyzed -- see the appendix.) were
analyzed using the {\tt CIAO.glvary} routine \citep{Gregory1992}.
This makes use of the aspect solution to project the effective area at
a given source's location so to account for pixel-to-pixel sensitivity
changes as well as effective area changes caused by chip edges.  The
{\tt CIAO} prescription for {\it
glvary}\footnote{https://cxc.cfa.harvard.edu/ciao/threads/variable/},
as described in the science thread, was followed.

For the first epoch, no sources within the D25 circle (and only one
source outside) were estimated to have a variability index of 2 or
more\footnote{An index $>$2 increasingly raises the probability that
the source is variable, starting at a probability of 50\%.  See the
previous footnote for a table linking variability index to
probabilities.}.

For the second epoch, only one source (77) had a variability index of
2 or more.  However, the source has two arguments against interpreting
it as a variable object: there were a total of ${\sim}$17 detected
counts and the object was positioned in a gap between ACIS CCDS I3 and
I1.  The presence of the source in the gap will almost certainly
introduce a variation in the detected count rate, even given the
Lissajous pattern {\it Chandra} uses to avoid sources falling
permanently between chip gaps.  We do not conclude that the source
varies.

\vspace{0.2in}

The variations {\it between} observations were analyzed with a
different approach.  For this problem, we are not interested in
small-scale changes, but in the total counts (or count rate) between
the two epochs.  Given that the aim-points of the two epoch were
within $3.2'$ of each other, and given the approximately symmetric
behavior of ACIS-I about the aimpoint, then the off-axis sources would
approximately experience the same off-axis behavior.  That means we
can compare the total counts for each source between the two epochs to
a reasonable accuracy.

The one item that must be addressed between the two epochs is the
increasing absorption accumulating on the ACIS optical blocking
window.  We used the CIAO proposal tool to calculate, for adopted {\it
bremsstrahlung} spectra of 1, 5, and 10 keV, the expected counts of
Epoch 2 (Cycle 18) targets given Epoch 1 (Cycle 15) count rates.  This
led to a `correction' factor to increase Epoch 2 counts for comparison
with the Epoch 1 counts.  The correction factor varied minimally for
different temperatures of the adopted spectrum.  We adopted an
Epoch~2~/~Epoch~1 correction factor of 1.23 to be applied to the
Epoch~2 source counts.  We ignore all sources that are only detected
in the merged data as well as sources that were not consistently on
the CCDs across both epochs.

With that correction factor in hand, we then generated
Table~\ref{N3938Var} which lists the combined sources detected in
Epochs 1 and 2, with their Epochs 1 and 2 counts, the corrected Epoch
2 counts, and a label describing whether a comparison of the Epoch 1
and corrected Epoch 2 counts were constant or varied within 3 standard
deviations.  With those requirements, five sources meet the `variable'
criterion: two are on the boundary between constant and variable,
while three are well beyond the `constant' criterion.  We caution the
reader that the window build-up introduces additional uncertainty
that only more timing data would address.  

\begin{table}
\caption{Variability Between Observation Epochs}
\label{N3938Var}
\scriptsize
\begin{tabular}{rrrrrrrrrrl}
     &   RA      &   Dec    & Ep-1  & Ep-1  &  Ep-1  & Flux & 3sig &  Ep-2 & Corr'd &       \\
SrcN & (J2000)   &  (J2000) &  Cts  &  Unc  &  Flux  & Unc  & Unc  &  Flux & Ep2-Fx & Vary?$^a$ \\ \hline
21 & 11:53:03.84 & 44:08:46 & ~33.7 & ~6.93 &  ~6.86 & 1.42 & 4.26 & ~4.25 & ~5.23 & No  \\
23 & 11:53:00.25 & 44:06:00 & ~85.4 & ~9.54 &  17.37 & 1.07 & 3.21 & 14.88 & 18.30 & No  \\
31 & 11:52:54.73 & 44:09:16 & ~10.1 & ~3.74 &  ~2.06 & 1.38 & 4.14 & ~2.73 & ~3.36 & No  \\
32 & 11:52:54.72 & 44:07:52 & ~87.7 & ~9.64 &  17.88 & 1.06 & 3.18 & 19.84 & 24.40 & Yes  \\
33 & 11:52:54.70 & 44:05:29 & ~25.0 & ~5.20 &  ~5.09 & 1.08 & 3.24 & ~4.08 & ~5.02 & No  \\
37 & 11:52:53.77 & 44:07:34 & ~20.0 & ~4.80 &  ~4.07 & 1.15 & 3.44 & ~5.15 & ~6.33 & No  \\
38 & 11:52:53.51 & 44:06:12 & ~41.0 & ~6.56 &  ~8.32 & 1.05 & 3.15 & ~7.61 & ~9.36 & No  \\
41 & 11:52:51.95 & 44:05:47 & ~63.0 & ~8.12 &  12.83 & 1.04 & 3.12 & ~6.47 & ~7.96 & Yes? \\
43 & 11:52:51.14 & 44:04:42 & ~21.2 & ~4.80 &  ~4.31 & 1.08 & 3.25 & ~8.26 & 10.16 & Yes? \\
46 & 11:52:50.58 & 44:06:28 & ~16.7 & ~4.24 &  ~3.39 & 1.08 & 3.23 & ~2.68 & ~3.30 & No  \\
47 & 11:52:50.52 & 44:04:36 & ~94.1 & ~9.95 &  19.09 & 1.05 & 3.16 & 23.81 & 29.29 & Yes \\
49 & 11:52:50.39 & 44:08:06 & 107.4 & 10.86 &  21.82 & 1.10 & 3.29 & 18.28 & 22.48 & No  \\
50 & 11:52:50.23 & 44:07:11 & ~26.1 & ~5.39 &  ~5.31 & 1.10 & 3.31 & ~6.84 & ~8.41 & No  \\
51 & 11:52:49.59 & 44:07:16 & ~36.0 & ~6.40 &  ~7.31 & 1.14 & 3.42 & ~4.80 & ~5.90 & No  \\
52 & 11:52:49.01 & 44:07:17 & ~15.2 & ~4.12 &  ~3.08 & 1.12 & 3.36 & ~3.73 & ~4.59 & No  \\
53 & 11:52:48.92 & 44:05:28 & ~20.3 & ~4.69 &  ~4.12 & 1.08 & 3.25 & ~4.04 & ~4.97 & No  \\
56 & 11:52:47.19 & 44:06:59 & ~16.2 & ~4.24 &  ~3.28 & 1.12 & 3.35 & ~2.03 & ~2.50 & No  \\
59 & 11:52:46.74 & 44:09:04 & ~13.4 & ~4.12 &  ~2.73 & 1.26 & 3.79 & ~3.33 & ~4.10 & No  \\
61 & 11:52:46.44 & 44:07:53 & ~46.8 & ~7.28 &  ~9.52 & 1.13 & 3.39 & 16.49 & 20.28 & Yes \\
63 & 11:52:46.26 & 44:06:28 & ~92.7 & ~9.80 &  18.79 & 1.04 & 3.12 & 14.76 & 18.15 & No  \\
65 & 11:52:44.47 & 44:09:14 & ~19.5 & ~5.00 &  ~3.97 & 1.29 & 3.86 & ~7.18 & ~8.83 & No  \\ \hline
\end{tabular}

Notes: SrcN = Source Number based on master source list; `Corr'd
Ep2-Fx' is the epoch 2 flux corrected for the build-up on the ACIS
optical blocking window from Epoch 1 to Epoch 2; `3sigUnc' = 3 times
flux uncertainty to determine whether source is significantly variable
or not. 

$^a$`Vary?': `No' or `Yes' based on the comparison of the 
Epoch 1 flux + 3sigUnc versus the corrected Epoch 2 flux.  Values
lying with 3${\sigma}$ of each other warrant a `No.'  Values 
$> 3{\sigma}$ convert to `Yes.'  Values near that boundary 
warrant `Yes?.'

\end{table}           

Finally, there are those sources that were not detected in both
epochs.  Sources within the D25 circle were within the fields-of-view
of the ACIS detector.  Furthermore, sources within the D25 circle fell
within the relatively flat `unit response' of the ACIS CCDs.
Consequently, we may infer that these sources underwent a
change-of-state between the two observations.  Those sources that were
detected in Epoch 1 but not in Epoch 2 are: 24, 27, 28, 39, 42, 45,
58, 60, 62, 64, 70, and 77.  Those sources detected in Epoch 2 but not
in Epoch 1 are: 26, 29, 30, 34, 54, and 74.   These results are noted
in the rightmost column of Table~\ref{MrgdSrc}.

In summary, then, NGC 3938 reveals sixteen X-ray sources that were
constant between the two observations, five sources detected to be
variable, and a further seventeen that were either on, then off or
off, then on, between the observations.  Consequently, a total of
twenty-two sources exhibited variability.

\section{Star Formation Rate and Metallicity of NGC 3938} \label{StFSect}

We now present estimates of the star formation rate (SFR) and the
metallicity of NGC 3938 based on optical and infrared flux
measurements made of the galaxy. We pattern the analysis
presented here after the analysis given in our previous work 
on the galaxy NGC 45 \citep{Pannuti2015}. In the present paper,
we implement an updated SFR relation derived by
\cite{Kennicutt2009}, but we use the
same metallicity relation derived by \cite{Lee2006} that we 
used in \cite{Pannuti2015}.

Based on the integrated H-${\alpha}$ and 24 micron luminosities of two samples of nearby galaxies, 
\cite{Kennicutt2009} derived the following relation for
the SFR of a galaxy:

$$\mbox{SFR (M$_{\odot}$ yr$^{-1}$)} = 
7.9{\times}10^{-42} 
[ L(H_{\alpha})_{obs} + 0.020 L(24)] ({\rm erg~s}^{-1}).$$

In this equation, $L(H_{\alpha})_{obs}$ is the integrated H-${\alpha}$
luminosity of the galaxy and L(24) is equal to ${\lambda} * L_{24}$,
where ${\lambda}$ is the wavelength of the observation and $L_{24}$ is
the integrated spectral luminosity of the galaxy (in units of ergs
s$^{-1}$ Hz$^{-1}$) at 24 microns. We determined $L(H_{\alpha})_{obs}$
for NGC 3938 as follows: based on the integrated H-${\alpha}$ flux of
$4.79{\times}10^{-12}$ ergs cm$^{-2}$ s$^{-1}$ from the galaxy as
measured by \cite{Kennicutt1983}, we computed a corresponding
integrated H-$\alpha$ luminosity of $2.77\times10^{41}$ ergs/s using
on our assumed distance to NGC 3938 of 22 Mpc. To compute L(24), we
adopted the integrated flux density at 24 microns of 1.09 Jy as
measured by \cite{Dale2007} using the Multiband Imaging Photometer for
Spitzer (MIPS) \citep{Rieke2004} aboard the {\it Spitzer Space
  Telescope} (SST) \citep{2004ApJS..154....1W}. From this measured
integrated flux density we computed an integrated spectral luminosity
for NGC 3938 at 24 microns of 6.31 $\times$ 10$^{29}$ ergs s$^{-1}$
Hz$^{-1}$ and a value of 7.89$\times$10$^{42}$ ergs s$^{-1}$ for
L(24).  Combining the relation derived by \cite{Kennicutt2009} and our
computed values for $L(H_{\alpha})_{obs}$ and L(24), we obtain a value
of 3.43 solar masses per year for the SFR of NGC 3938.  For
comparison, \cite{CalduPrimo2009} compute an average SFR of 0.33
$\pm$0.09 solar masses per year using data from five different
wavelength bands ranging from the ultraviolet through millimeter. We
cannot easily account for the discrepancy between our computed SFR and
the computed SFR presented by \cite{CalduPrimo2009}. Noting the large
uncertainty in the distance to NGC 3938 (see \S\ref{lumFunc}), if we
instead adopt a distance of 11 Mpc to NGC 3938, we compute an SFR of
0.86 solar masses per year. This value is a factor of ${\approx}$3
greater than the value determined by \cite{CalduPrimo2009}. We argue
that an elevated SFR for NGC 3938 makes sense in light of the high
historical supernova rate observed in this galaxy.

\par
To determine the metallicity of NGC 3938 (as expressed in terms
of log[O/H]), we use the following relation for metallicity as derived
by \cite{Lee2006}:

$$12~\mbox{log [O/H]} = 5.78 {\pm} 0.21 + (-0.122 {\pm} 0.012)) M_{4.5}.$$

In this relation, $M_{4.5}$ is the absolute magnitude of NGC 3938 at
the wavelength of 4.5 microns. We determined $M_{4.5}$ from the
integrated flux density of 0.21 Jy measured at this wavelength by 
\citet{Dale2007} using the Infrared Array Camera (IRAC) 
\citep{Fazio2004} aboard the SST. From
this measured flux density and adopting the standard IRAC 4.5 
micron zero-magnitude flux density \footnote{See 
https://irsa.ipac.caltech.edu/data/SPITZER/docs/irac/iracinstrumentha
ndbook/IRAC$\_$Instrument$\_$Handbook.pdf. of 179.7, we computed an 
integrated apparent magnitude of 7.33 at this wavelength for NGC 
3938. From this integrated apparent magnitude and our assumed
distance of 22 Mpc to NGC 3938, we computed a value of -24.38 
for $M_{4.5}$.
Using this value for $M_{4.5}$ and the relation derived by
\cite{Lee2006}, we compute a range of values from
8.17 to 9.18 for 12 + log [O/H]. To put this result in context, we 
compared
this computed metallicity with the metallicities of other galaxies of
the same Hubble type as presented by \cite{Zaritsky1994}. Those
authors found a range of metallicities from 8.4 to 9.2 
for Sc galaxies, therefore we conclude that our computed range of 
metallicity values are consistent with those of other Sc galaxies.
In addition, \cite{Calzetti2007} analyzed SST IRAC and MIPS 
observations of NGC 3938 (at the wavelengths of 8 $\mu$m and 
24$\mu$m) and estimated the metallicity 12 + log [O/H] of 
the galaxy to range from 8.35 to 9.07, which is consistent with our 
computed range of values.}

\begin{figure}[h!]
 \centering
 \caption{Luminosity function of the two observed epochs of NGC 3938,
  (data points with uncertainties) together with the `HMXB end' of the
  luminosity functions of NGC 628, NGC 4321, and NGC 5194 (black
  lines).  The NGC 3938 points are presented twice:  using a distance
  of 20 Mpc (black points) and a distance of 13 Mpc (red points).}
 \label{LumFFig}
 \scalebox{0.5}{\includegraphics{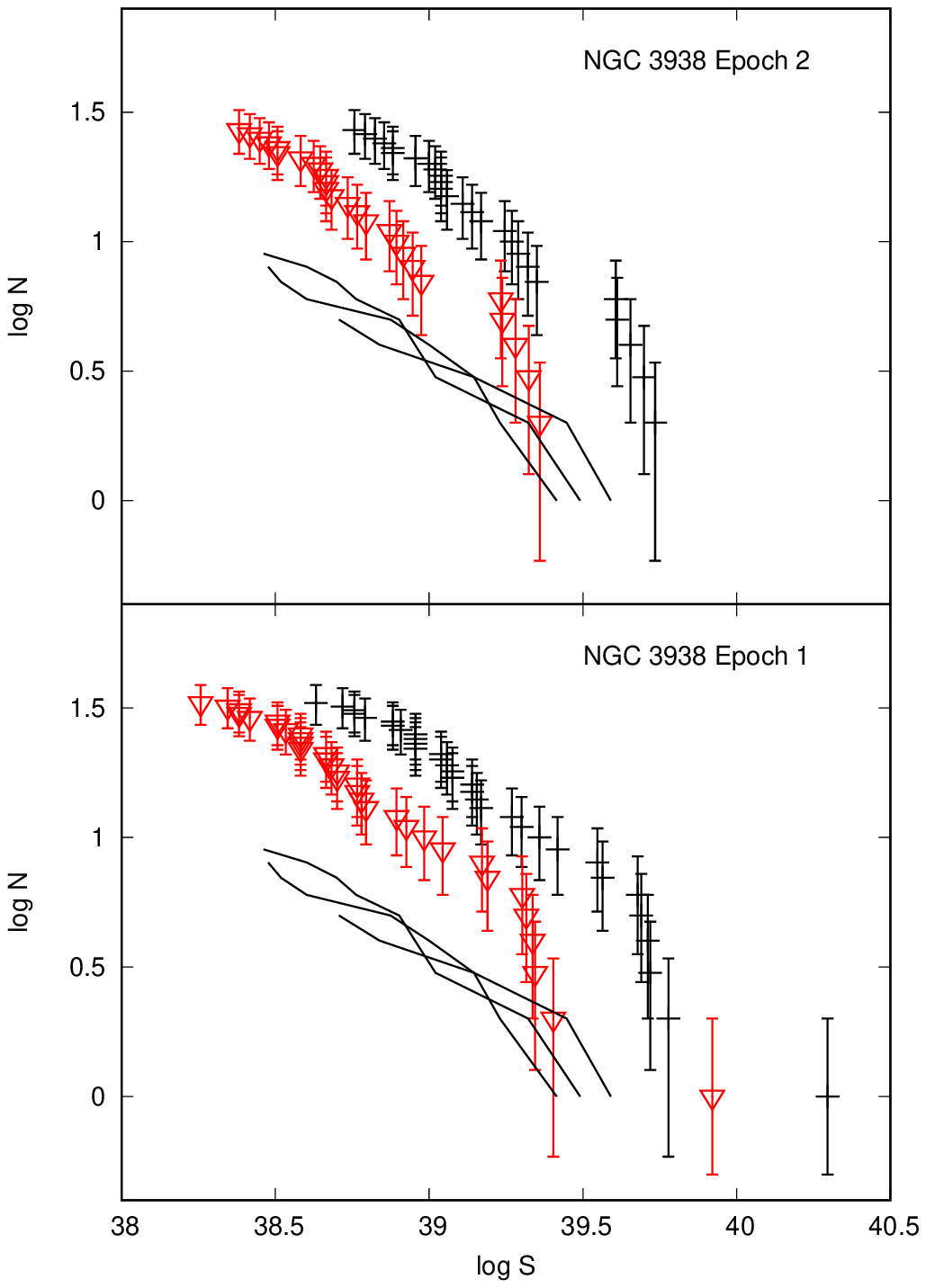}}
\end{figure}

\section{Luminosity function}\label{lumFunc}

Luminosity functions of X-ray binaries (XRBs) have received a
considerable amount of attention since the launch of {\it Chandra}:
e.g., \cite{Grimm2003}, \cite{Gilfanov2004}, \cite{Binder2017}, and
perhaps culminating with \cite{Lehmer2019}.  \cite{Lehmer2019}
analyzes the luminosity relations for 38 galaxies, including NGC 3938.
We do not attempt to re-create any of these analyses nor even fit
the luminosity function, dependent as it is upon the adopted distance. 

Our first task is the expected number of background sources within the
field-of-view.  We can look at this in two ways:  first, NGC 3938
essentially falls onto a single ACIS-I CCD, so the field-of-view is
64 sq arcmin.  Adopting the background estimator from \cite{Campana2001},
namely,
$$N(> L) = 360 (\frac{S}{2{\times}10^{-15}})^{-0.68}$$

{\noindent}sources per square degree for a measured flux of S in 
erg s$^{-1}$ cm$^{-2}$
in the 0.5-8 keV band then leads to five background sources.  We have
here assumed our minimum detectable threshold to be
${\sim}3{\times}10^{-15}$ erg s$^{-1}$ cm$^{-2}$.  If we drop the
minimum threshold to $2{\times}10^{-15}$, the number of sources rises
to six.  Alternatively, given that NGC 3938 occupies a smaller
fraction of the CCD, we can also adopt the D25 circle.  Using that for
our area, then 2-3 sources on the NGC 3938 list
(Table~\ref{MrgdSrc}) are expected to be background objects.

Our second task is the adopted distance to NGC 3938.  The
NED\footnote{NASA Extragalactic Database, ned.ipac.caltech.edu} lists
essentially two different distances: ${\approx}$5 Mpc and
${\approx}$17-22
Mpc.  Figure~\ref{LumFFig} shows the XRB luminosity function 
assuming 13 and 20 Mpc distances.  If we assume NGC 3938 matches the
Milky Way in size,
then its distance is closer to 16 Mpc.  If we adopt its redshift and
divide by an adopted H$_0$ of 70 km s$^{-1}$ Mpc$^{-1}$, its distance
should be ${\sim}$11 Mpc.  There is clearly a need to sort out the
distance to NGC 3938.  A distance of 10-12 Mpc would appear to match
our NGC 3938 luminosity functions of Lehmer et al.

If we consider variability, the luminosity function does not
differ significantly between the two epochs throughout the HMXB
range from ${\sim}10^{38}$ to ${\sim}10^{39}$ erg s$^{-1}$.  The
distributions do differ at the high end where a single object
is on in Epoch 1 and off
in Epoch 2 (source 60).  That result for NGC 3938 is in contrast
to the results of \cite{Binder2017} for NGC 300, which 
proposed that a majority of outbursting X-ray binaries occur at
sub-Eddington limits.  While two epochs is a minimal sample,
the difference may suggest a more active phase for NGC
300 and a less active phase for NGC 3938.
  
\section{Summary}

Much of the discussion of our results occurred throughout the paper as
we presented the data.  A brief summary of our results follows.

We analyzed two epochs of {\it Chandra} observations of the face-on
spiral NGC 3938.  We detect a total of 95 sources within the galaxy
from merging the data, with 66 and 48 sources detected for epochs 1
and 2, respectively.  We attempt to use color hardness to 
classify the detected sources within the D25 circle.  The contaminant
on the ACIS Optical Blocking Window hampers that effort as it 
`collapses' the color-color regional separation of the sources. 
Time variations are detected between the two
epochs, correcting for the increased absorption of the ACIS optical
blocking window, for about 1/4 of the sources in common between the
two epochs.  Only one source was detected to vary during an
observation, but its location in a chip gap renders its detected
variation unlikely.  Only one source inside the D25 circle was
particularly luminous -- so a possible variable ULX candidate.  
NGC 3938 is sufficiently distant that we could not probe below 
${\approx}10^{38}$ erg s$^{-1}$ in the luminosity function.  We 
also note a problem with the distance to NGC 3938 -- on the basis 
of the multiple-galaxy-defined luminosity functions
\citep{Lehmer2019}, the distance to NGC 3938 should fall in the 
10-12 Mpc band, not the ${\sim}$5 Mpc nor 17-22 Mpc bands that
dominate the NED values.

We also briefly describe the sources detected by {\it Chandra} that
lie {\it outside} of the D25 circle of NGC 3938.  Approximately sixty
percent of those sources have known counterparts.  Of the remaining
forty percent, half have one or two apparent counterparts within the
detection region, requiring relatively straightforward confirmation
of the X-ray source's counterpart.

\vspace{0.5in}
\begin{acknowledgments}

We thank the anonymous referee for comments that improved this paper.
This research has made use of the NASA/IPAC Infrared Science Archive
(IRSA) and the NASA/IPAC Extragalactic Database, both of which are
funded by the National Aeronautics and Space Administration and
operated by the California Institute of Technology. This research has
also made use of NASA's Astrophysics Data System.  The ADS is operated
by the Smithsonian Astrophysical Observatory under NASA Cooperative
Agreement 80NSSC21M0056.  We also used images obtained by the NASA/ESA
Hubble Space Telescope obtained from the Space Telescope Science
Institute, which is operated by the Association of Universities for
Research in Astronomy, Inc., under NASA contract NAS 5–26555. These
observations are associated with program 16239.  This research has
made use of data obtained from the Chandra Data Archive at the Chandra
X-ray Center.  This paper employs a list of Chandra data sets, 
obtained by the Chandra X-ray Observatory, contained in 
\dataset[DOI:10.25575/cdc.258]{https://doi.org/10.25574/cdc.258}.
\end{acknowledgments}

\vspace{5mm}
\facilities{CXO, ADS, IRSA, Spitzer, WISE, NED, MAST, HST}

\software{
          CIAO \citep{2006SPIE.6270E..60F}
          }

\begin{figure}[h!]
 \centering
 \caption{The {\it Chandra}-detected sources outside of the D25
   circle of NGC 3938.  The image is ${\sim}20'$ on a side.}
 \label{OutD25img}
 \scalebox{0.5}{\includegraphics{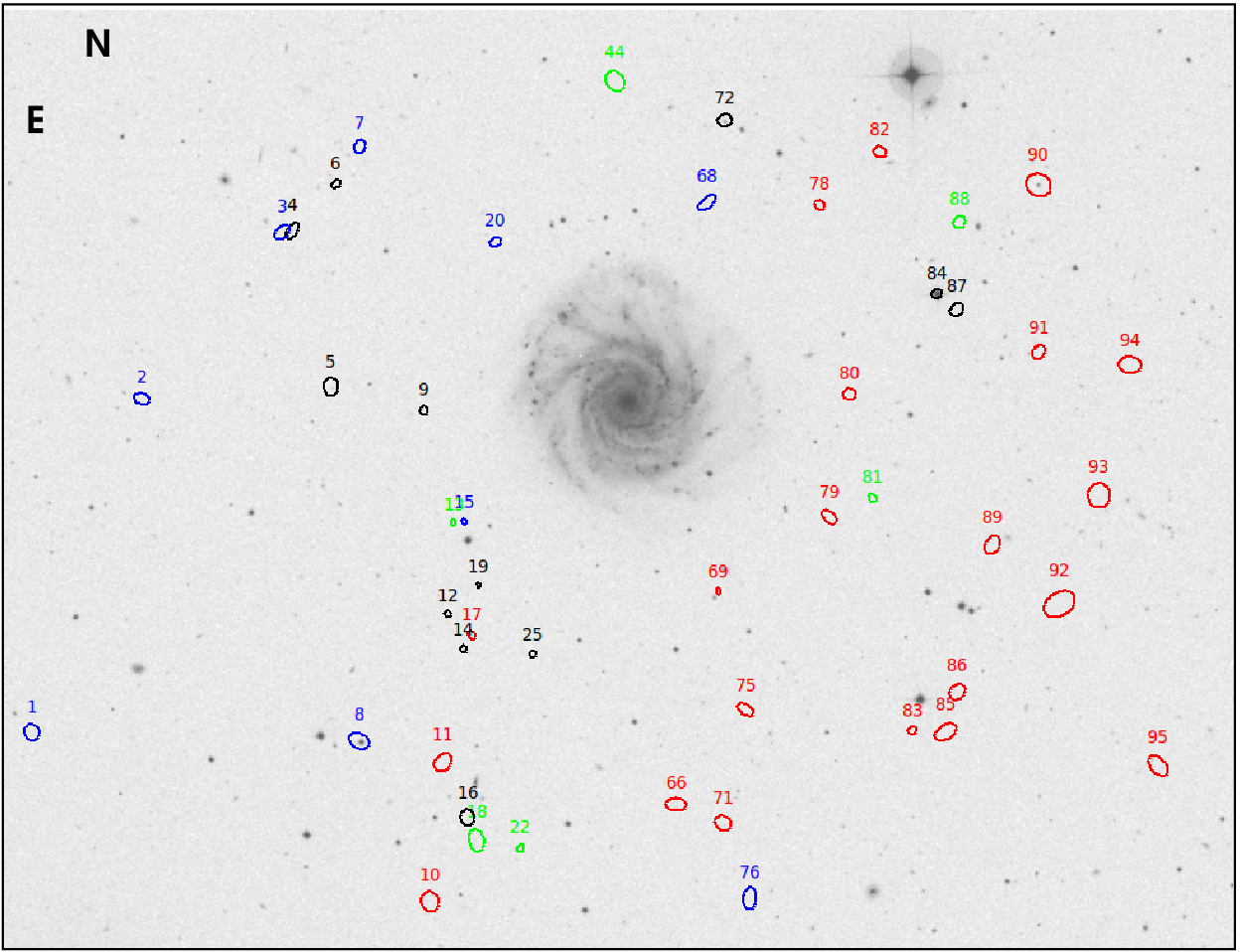}}
\end{figure}

\appendix

\section{Sources Detected Outside D25}\label{SrcOutD25}

\subsection{CSC correspondence} 

Figure~\ref{OutD25img} shows the 47 detected sources outside of
the D25 circle.  Of that 47, seven are not listed as a CSC
source (our merged numbers 1, 3, 4, 10, 15, 18, and 44).  Of
those, 1 and 10 lie on the edge of a CCD and could easily have
been dropped by imposing a minimum distance to the CCD edge.  
We elected not to impose such a minimum.  Two additional sources
(18, 44) are only detected in the merged data.  The CSC catalog
does not contain merged-epoch detections if the pointings between
multiple observations are greater than 1$'$.  There
remain three sources (3, 4, 15) for which we have no
explanation why they were not included on the CSC list.

\begin{table}[h!]
 \centering
 \label{CSCbelow}
 \caption{Sources in CSC, Outside D25 and Below Our Detection
 Threshold}
 \begin{tabular}{lll}
 CSC ID  &  CSC ID  & CSC ID \\ \hline
 {\it 2CXO J115220.8+440605} & {\it 2CXO J115218.7+440505} & {\it 2CXO J115227.7+440230} \\
 {\it 2CXO J115207.7+440431} & 2CXO J115205.6+440845 & {\it 2CXO J115203.6+440354} \\
 {\it 2CXO J115250.8+435811} & 2CXO J115218.3+441424 & {\it 2CXO J115153.1+440709} \\
 {\it 2CXO J115153.6+440503} & 2CXO J115217.2+435816 & {\it 2CXO J115212.0+435843} \\
 {\it 2CXO J115201.4+440012} & 2CXO J115348.0+440242 & 2CXO J115351.7+440417 \\
 {\it 2CXO J115203.3+435825}  \\ \hline
\end{tabular}

Sources in {\it italics} are listed as `marginal' in the CSC catalog.

\end{table}

Tackling the same question from the other direction finds 24
sources on the CSC list that were not detected by us.  Five of
those sources are automatically excluded by us -- the CSC team
elected to process {\it all} of the CCDs whereas we opted to 
process only those near the aimpoint.  For ACIS-I, that means 
we include CCDs I0 to I3, but not S3 and S4 which are typically
active during an ``ACIS-I'' observation.  For the analysis in 
this paper, those five sources lie outside the I0 to I3 CCDs.  
An additional sixteen CSC sources lie below our detection 
threshold.  Those sources are listed in Table~\ref{CSCbelow}.  
Of the sixteen, the CSC lists eleven as `marginal.'

That leaves just three sources: 2CXO J115246.1+440147; 2CXO
J115223.3+440342; and 2CXO J115209.7+441405.  The last one is
listed as `marginal' in the CSC catalog.  The first two appear to 
lie above our detection threshold.   However, we do not
detect them, yet they appear to match the {\it Chandra} PSF 
at their off-axis angles.   We have looked closely at the
detection parameters and can not provide a ready explanation 
for the two non-detections.  

\subsection{Identification of Sources Outside D25}\label{IdOutD25}

Returning to our detected sources that lie outside D25, if we adopt
the source classifications available from NED without question, we
then have the results shown in Table~\ref{SrcIDsOut} and listed
in the `Cntrprt' column of Table~\ref{MrgdSrc} (outside).   We
do not use the color-color approach as it was built for 
separation of X-ray binaries and not galaxies and AGN.

The `IR source' category would benefit from additional investigation
-- some sources are clearly separated into `star' or `galaxy' while
the IR sources are unknown at the present time.
Further, the `unknown' category could also be investigated: one can
overlay the {\it Chandra} region files for the undetected sources
onto
an IRAC image (eg, 3.6${\mu}$) and find matches, or at least
candidates, for {\it all} of the `unknown' X-ray sources (q.v.,
`Cntrprt' column in Table~\ref{MrgdSrc} (outside)).  However, 
tracking down or characterizing these identifications lies outside
the scope of this paper.

\begin{table}[h!]
 \centering
 \label{SrcIDsOut}
 \caption{Source Identification: {\it Out}side D25 (NED)}
 \begin{tabular}{lcl}
 Type   & Number   & Source Numbers \\  \hline
 star   &   5      & 2, 5, 16, 84, 93 \\
 galaxy &   13     & 4, 7, 9, 13, 17, 18, 19, 22, 25, 66, 71, 72, 91 \\
 IR source &  7    & 6, 20, 44, 69, 75, 78, 79, 86 \\
 QSO    &   1      & 14  \\
 Radio source &  1 & 87 \\
 No classification & 20 & 1, 3, 8, 10, 11, 12, 15, 68, 76, 78, 80 \\
                   &    & 81, 82, 83, 85, 88, 89, 90, 92, 94, 95 \\ \hline
\end{tabular}
\end{table}

\%end{figure}


\subsection{Timing Variations for Sources Outside D25}

We also examine the variation between Epochs 1 and 2 for sources
outside the D25 circle (Table~\ref{Outvary}).  We take the same
approach as we did for those sources inside D25 -- sources within an
epoch via {\tt CIAO.glvary} (next paragraphs) and sources across
epochs via the Epoch 1 vs Epoch 2 counts.

\begin{table}[h!]
 \centering
 \caption{Epoch 1 versus Epoch 2 for Sources Outside the D25 circle}
 \label{Outvary}
 \begin{tabular}{rrrrrrrrrrl}
     &    RA     &   Dec    & Ep-1  & Ep-1  &  Ep-1 & Flux & 3sig & Ep-2  & Corr'd &       \\
SrcN & (J2000)   &  (J2000) &  Cts  &  Unc  &  Flux & Unc  & Unc  &  Flux & Ep2-Fx & Vary?$^a$ \\ \hline
~4 & 11:53:26.52 & 44:10:35 & ~10.1 & ~4.00 & ~2.05 & 1.58 & 4.75 & ~4.03 & ~~4.96 & No  \\
~5 & 11:53:22.09 & 44:07:30 & ~55.8 & ~8.66 & 11.31 & 1.34 & 4.03 & ~5.68 & ~~6.99 & Yes? \\
~6 & 11:53:21.57 & 44:11:31 & ~43.9 & ~8.60 & ~8.93 & 1.69 & 5.06 & ~8.78 & ~10.80 & No  \\
~9 & 11:53:11.74 & 44:07:03 & ~25.9 & ~6.25 & ~5.27 & 1.51 & 4.52 & ~3.15 & ~~3.87 & Yes? \\ 
12 & 11:53:09.05 & 44:03:01 & ~16.1 & ~5.39 & ~3.26 & 1.80 & 5.40 & ~5.34 & ~~6.57 & Yes? \\
14 & 11:53:07.46 & 44:02:19 & ~84.3 & ~9.95 & 17.17 & 1.17 & 3.52 & 10.55 & ~12.98 & Yes? \\
16 & 11:53:07.05 & 43:58:59 & ~27.3 & ~6.00 & ~5.54 & 1.32 & 3.96 & ~6.97 & ~~8.57 & No  \\
19 & 11:53:05.81 & 44:03:35 & ~81.3 & ~9.27 & 16.56 & 1.06 & 3.17 & ~9.47 & ~11.65 & Yes? \\
25 & 11:52:59.72 & 44:02:13 & ~12.0 & ~4.00 & ~2.40 & 1.34 & 4.01 & ~7.75 & ~~9.53 & Yes? \\
72 & 11:52:38.72 & 44:12:47 & 297.7 & 18.47 & 60.50 & 1.15 & 3.44 & 89.03 & 109.51 & Yes \\
84 & 11:52:15.28 & 44:09:21 & ~26.7 & ~6.86 & ~5.32 & 1.76 & 5.29 & ~5.13 & ~~6.31 & No  \\
87 & 11:52:13.11 & 44:09:03 & 110.4 & 12.73 & 22.42 & 1.46 & 4.39 & 13.76 & ~16.92 & Yes? \\ \hline
 \end{tabular}

Notes: SrcN = SouRCe Number based on master source list; 
`Corr'd Ep2-Fx' is the EPoch 2 FluX CORRecteD for the build-up on 
the ACIS optical blocking window from Epoch 1 to Epoch 2; 
`3sigUnc' = 3 times the flux UNCertainty to determine whether the
source is significantly variable.

 $^a$Vary: to determine variability, the Epoch 1 flux and its
 uncertainty are compared to the corrected Epoch 2 flux.  Values lying
 within 3${\sigma}$ of each other warrant a `No'.  Values $> 3
 {\sigma}$ convert to `Yes'.  Values near the boundary warrant `Yes?'.

\end{table}

For the sources outside D25, there are two other categories to 
be assessed.  The first category consists of sources observed 
in Epoch 1 {\it or} Epoch 2 that were observed one time and 
did not fall within the detector's field-of-view during the
other epoch.  For Epoch 1, those sources are 10, 83, 85, 86, 
and 89 to 95; for epoch 2, the sources are 1 to 5, 8, and 44.
Of these two sets, only source 94 was detected by the {\tt
glvary} routine as variable within an observation with an 
index of 6 indicating near-certainty of variation.  Given 
that it was detected with ${\sim}$22 counts and given that 
it does not lie on a chip edge, it is quite possible that 
its variation is real.

The second category consists of those source that {\it could}
have been detected twice because both fields-of-view covered 
the source's position, but each source was only detected once. 
Within this category there are two sub-categories: (i) there 
was one detection of a somewhat weak source that was not 
detected a second time because of the reduced sensitivity 
owing to a greater off-axis observation or the increased 
contaminant on the optical blocking window; and (ii) the 
source was actually off during one of the observations.

For Epoch 1, the sources that were on but were not detected
during Epoch 2 are 11, 17, 66, 69, 71, 75, 78, 79, 80, and 
82.  The corresponding sources for Epoch 2 are sources 7, 
13, 20, 68, and 76.  We first eliminate the `near chip gap' 
or `larger off-axis angle' objects (Ep 1: 11, 17, 79, 82; 
Ep 2: 7, 68). 

Clearly, the second sub-category is more interesting than 
the first.  For Epoch 1, the sources for which the position
falls well within the detector for {\it both} epochs are 
66, 69, 71, 75, 78, and 80; the corresponding sources for 
Epoch 2 are 20 and 76.  All of these sources are relatively 
weak with total counts in the ${\sim}$17 to ${\sim}$32
range.  However, while weak, they all have suffcient counts 
that even with the ACIS window increase between the epochs,
detections would have been possible {\it if} each source had
remained constant.  Consequently, we deem all of the Epoch 1
sources as `on' during Epoch 1 and `off' during Epoch 2.  
The opposite situation applies to sources 20 and 68: `off'
during Epoch 1 but `on' during Epoch 2.

Source 76 was detected during Epoch 2, but not during Epoch 1. 
The source, however, happens to lie near the edge of {\it both}
observations.  We consequently deem it `off' during Epoch 1 and
`on' during Epoch 2, but it is possible that it suffers from
lying too close to a chip edge.

Finally, there are variations possible within an observation for the
sources outside D25 and not discussed above in the single-epoch
paragraph.  Most sources exhibit no variations.  {\tt Glvary} reports
three variables: sources 14, 19, and 88.  Of these, source 88 can be
eliminated quickly as it has ${\sim}$35 counts, is detected only in
the merged data, and lies near a chip edge for Epoch 2.  Sources 14
and 19 both lie near a chip gap, but have sufficiently large Epoch 1
counts that the drop to Epoch 2 suggests variability.  The
`within-Epoch' variability is likely caused by their locations near
the chip gap.

\section{Individual Epoch Tables} \label{SectAppB}

\subsection{Detected Source Tables}

The detected source tables are placed in this appendix solely to make
the article pages easier to read.  As stated in the analysis section,
`N' refers to the {\it merged} source number and is used for both epochs
of data.

\begin{table}[h!]
\centering
\caption{NGC 3938: Epoch 1 Detected Sources$^a$}
\label{N3938Ep1Srcs}
\tiny
\begin{tabular}{rrrrrrrrrrrrrr}
 N &      RA      &     Dec      & Counts &   Err & BkgCts & BkEr &  SrcRate &  SrcRErr &  BkgRate &  BkRErr  & Signif & Flux$^b$ & L$_X^c$ \\ \hline
\multicolumn{14}{c}{Inside D25 circle} \\
 21 & 11:53:03.84 & 44:08:46.21  &   33.7 &  6.93 &  14.3 & 0.09 &  6.79e-04 & 1.39e-04 & 2.87e-04 & 1.90e-06 &   6.9  & ~6.86 & ~3.9 \\
 23 & 11:53:00.25 & 44:05:59.87  &   85.4 &  9.54 &   5.6 & 0.05 &  1.72e-03 & 1.92e-04 & 1.12e-04 & 1.06e-06 &  24.3  & 17.37 & 10.0 \\
 24 & 11:53:00.14 & 44:07:00.33  &   25.7 &  5.48 &   4.3 & 0.05 &  5.17e-04 & 1.10e-04 & 8.72e-05 & 9.40e-07 &   7.9  & ~5.22 & ~3.0 \\
 27 & 11:52:57.15 & 44:08:08.41  &   65.7 &  8.43 &   5.3 & 0.05 &  1.32e-03 & 1.70e-04 & 1.07e-04 & 1.03e-06 &  19.0  & 13.33 & ~7.7 \\
 28 & 11:52:56.09 & 44:04:12.09  &    9.7 &  3.32 &   1.3 & 0.02 &  1.96e-04 & 6.67e-05 & 2.56e-05 & 5.02e-07 &   4.0  & ~1.98 & ~1.1 \\
 31 & 11:52:54.73 & 44:09:16.49  &   10.1 &  3.74 &   3.9 & 0.04 &  2.04e-04 & 7.53e-05 & 7.81e-05 & 8.88e-07 &   3.2  & ~2.06 & ~1.2 \\
 32 & 11:52:54.72 & 44:07:51.88  &   87.7 &  9.64 &   5.3 & 0.05 &  1.77e-03 & 1.94e-04 & 1.06e-04 & 1.04e-06 &  25.4  & 17.88 & 10.3 \\
 33 & 11:52:54.70 & 44:05:28.77  &   25.0 &  5.20 &   2.0 & 0.03 &  5.04e-04 & 1.05e-04 & 3.95e-05 & 6.22e-07 &   9.5  & ~5.09 & ~2.9 \\
 37 & 11:52:53.77 & 44:07:34.43  &   20.0 &  4.80 &   3.0 & 0.04 &  4.03e-04 & 9.65e-05 & 6.00e-05 & 7.59e-07 &   6.8  & ~4.07 & ~2.3 \\
 38 & 11:52:53.51 & 44:06:12.29  &   41.0 &  6.56 &   2.0 & 0.03 &  8.24e-04 & 1.32e-04 & 4.09e-05 & 6.28e-07 &  15.4  & ~8.32 & ~4.8 \\
 39 & 11:52:53.04 & 44:07:11.33  &   25.0 &  5.20 &   2.0 & 0.03 &  5.03e-04 & 1.05e-04 & 3.99e-05 & 6.29e-07 &   9.4  & ~5.08 & ~2.9 \\
 41 & 11:52:51.95 & 44:05:47.44  &   63.0 &  8.12 &   3.0 & 0.04 &  1.27e-03 & 1.63e-04 & 5.97e-05 & 7.81e-07 &  21.5  & 12.83 & ~7.4 \\
 42 & 11:52:51.60 & 44:03:43.92  &   16.3 &  4.24 &   1.7 & 0.03 &  3.28e-04 & 8.54e-05 & 3.37e-05 & 5.68e-07 &   6.4  & ~3.31 & ~1.9 \\
 43 & 11:52:51.14 & 44:04:41.94  &   21.2 &  4.80 &   1.8 & 0.03 &  4.27e-04 & 9.65e-05 & 3.53e-05 & 5.87e-07 &   8.2  & ~4.31 & ~2.5 \\
 45 & 11:52:50.80 & 44:06:53.31  &    8.1 &  3.16 &   1.9 & 0.03 &  1.63e-04 & 6.36e-05 & 3.87e-05 & 6.11e-07 &   3.1  & ~1.65 & ~0.9 \\
 46 & 11:52:50.58 & 44:06:27.82  &   16.7 &  4.24 &   1.3 & 0.02 &  3.36e-04 & 8.54e-05 & 2.63e-05 & 4.99e-07 &   6.9  & ~3.39 & ~1.9 \\
 47 & 11:52:50.52 & 44:04:35.66  &   94.1 &  9.95 &   4.9 & 0.05 &  1.89e-03 & 2.00e-04 & 9.86e-05 & 1.01e-06 &  27.9  & 19.09 & 11.0 \\
 49 & 11:52:50.39 & 44:08:06.18  &  107.4 & 10.86 &  10.6 & 0.07 &  2.16e-03 & 2.19e-04 & 2.13e-04 & 1.48e-06 &  24.6  & 21.82 & 12.6 \\
 50 & 11:52:50.23 & 44:07:11.25  &   26.1 &  5.39 &   2.9 & 0.04 &  5.26e-04 & 1.08e-04 & 5.79e-05 & 7.70e-07 &   9.0  & ~5.31 & ~3.1 \\
 51 & 11:52:49.59 & 44:07:16.27  &   36.0 &  6.40 &   5.0 & 0.05 &  7.24e-04 & 1.29e-04 & 1.01e-04 & 9.98e-07 &  10.6  & ~7.31 & ~4.2 \\
 52 & 11:52:49.01 & 44:07:17.40  &   15.2 &  4.12 &   1.8 & 0.03 &  3.05e-04 & 8.30e-05 & 3.68e-05 & 5.91e-07 &   5.8  & ~3.08 & ~1.7 \\
 53 & 11:52:48.92 & 44:05:28.44  &   20.3 &  4.69 &   1.7 & 0.03 &  4.08e-04 & 9.44e-05 & 3.47e-05 & 5.81e-07 &   7.9  & ~4.12 & ~2.4 \\ 
 56 & 11:52:47.19 & 44:06:58.65  &   16.2 &  4.24 &   1.8 & 0.03 &  3.25e-04 & 8.54e-05 & 3.69e-05 & 5.93e-07 &   6.2  & ~3.28 & ~1.9 \\
 58 & 11:52:46.86 & 44:10:07.77  &   16.2 &  5.00 &   8.8 & 0.07 &  3.26e-04 & 1.01e-04 & 1.78e-04 & 1.51e-06 &   4.0  & ~3.29 & ~1.9 \\
 59 & 11:52:46.74 & 44:09:03.59  &   13.4 &  4.12 &   3.6 & 0.04 &  2.70e-04 & 8.30e-05 & 7.23e-05 & 8.45e-07 &   4.3  & ~2.73 & ~1.6 \\
 60 & 11:52:46.65 & 44:06:48.45  &  354.5 & 19.13 &  11.5 & 0.08 &  7.13e-03 & 3.85e-04 & 2.31e-04 & 1.54e-06 &  78.9  & 71.91 & 41.5 \\
 61 & 11:52:46.44 & 44:07:53.24  &   46.8 &  7.28 &   6.2 & 0.06 &  9.43e-04 & 1.46e-04 & 1.24e-04 & 1.11e-06 &  12.9  & ~9.52 & ~5.5 \\
 62 & 11:52:46.33 & 44:03:52.53  &   10.0 &  3.32 &   1.0 & 0.02 &  2.02e-04 & 6.67e-05 & 1.98e-05 & 4.34e-07 &   4.3  & ~2.04 & ~1.2 \\
 63 & 11:52:46.26 & 44:06:27.71  &   92.7 &  9.80 &   3.3 & 0.04 &  1.86e-03 & 1.97e-04 & 6.71e-05 & 8.30e-07 &  30.7  & 18.79 & 10.8 \\
 64 & 11:52:45.64 & 44:05:26.68  &   13.8 &  3.87 &   1.2 & 0.02 &  2.79e-04 & 7.79e-05 & 2.32e-05 & 4.76e-07 &   5.8  & ~2.82 & ~1.6 \\
 65 & 11:52:44.47 & 44:09:14.20  &   19.5 &  5.00 &   5.5 & 0.05 &  3.93e-04 & 1.01e-04 & 1.10e-04 & 1.05e-06 &   5.6  & ~3.97 & ~2.3 \\
 70 & 11:52:38.95 & 44:06:02.49  &   21.5 &  4.80 &   1.5 & 0.03 &  4.32e-04 & 9.65e-05 & 3.07e-05 & 5.48e-07 &   8.6  & ~4.36 & ~2.5 \\
 77 & 11:52:34.44 & 44:05:16.27  &   10.7 &  3.46 &   1.3 & 0.03 &  2.16e-04 & 6.97e-05 & 2.59e-05 & 5.10e-07 &   4.4  & ~2.19 & ~1.3 \\
\multicolumn{14}{c}{Outside D25 circle} \\
 ~4 & 11:53:26.52 & 44:10:34.65  &   10.1 &  4.00 &   5.9 & 0.06 &  2.03e-04 & 8.05e-05 & 1.19e-04 & 1.22e-06 &   2.8  & ~2.05 & ~1.2 \\
 ~5 & 11:53:22.09 & 44:07:29.95  &   55.8 &  8.66 &  19.2 & 0.11 &  1.12e-03 & 1.74e-04 & 3.87e-04 & 2.23e-06 &  10.2  & 11.31 & ~6.5 \\
 ~6 & 11:53:21.57 & 44:11:31.38  &   43.9 &  8.60 &  30.1 & 0.14 &  8.84e-04 & 1.73e-04 & 6.05e-04 & 2.79e-06 &   6.7  & ~8.93 & ~5.1 \\
 ~9 & 11:53:11.74 & 44:07:03.09  &   25.9 &  6.25 &  13.1 & 0.09 &  5.22e-04 & 1.26e-04 & 2.63e-04 & 1.80e-06 &   5.5  & ~5.27 & ~3.0 \\
 10 & 11:53:11.11 & 43:57:19.53  &   73.6 &  9.85 &  23.4 & 0.12 &  1.48e-03 & 1.98e-04 & 4.70e-04 & 2.46e-06 &  12.5  & 14.95 & ~8.6 \\
 11 & 11:53:09.67 & 44:00:04.79  &   29.2 &  6.86 &  17.8 & 0.11 &  5.88e-04 & 1.38e-04 & 3.57e-04 & 2.15e-06 &   5.5  & ~5.94 & ~3.4 \\
 12 & 11:53:09.05 & 44:03:00.92  &   16.1 &  5.39 &  12.9 & 0.09 &  3.23e-04 & 1.08e-04 & 2.60e-04 & 1.79e-06 &   3.4  & ~3.26 & ~1.9 \\
 14 & 11:53:07.46 & 44:02:19.47  &   84.3 &  9.95 &  14.7 & 0.09 &  1.70e-03 & 2.00e-04 & 2.97e-04 & 1.91e-06 &  17.1  & 17.17 & ~9.9 \\
 16 & 11:53:07.05 & 43:58:58.78  &   27.3 &  6.00 &   8.7 & 0.07 &  5.49e-04 & 1.21e-04 & 1.75e-04 & 1.45e-06 &   6.7  & ~5.54 & ~3.1 \\
 17 & 11:53:06.57 & 44:02:36.21  &   39.8 &  7.68 &  19.2 & 0.11 &  8.00e-04 & 1.55e-04 & 3.87e-04 & 2.21e-06 &   7.3  & ~8.08 & ~4.7 \\
 19 & 11:53:05.81 & 44:03:34.88  &   81.3 &  9.27 &   4.7 & 0.05 &  1.64e-03 & 1.87e-04 & 9.55e-05 & 9.81e-07 &  24.3  & 16.56 & ~9.5 \\
 25 & 11:52:59.72 & 44:02:12.99  &   12.0 &  4.00 &   4.0 & 0.04 &  2.41e-04 & 8.05e-05 & 8.13e-05 & 8.95e-07 &   3.7  & ~2.43 & ~1.4 \\
 66 & 11:52:44.20 & 43:59:14.67  &   25.0 &  6.56 &  18.0 & 0.11 &  5.03e-04 & 1.32e-04 & 3.62e-04 & 2.13e-06 &   4.7  & ~5.08 & ~2.9 \\
 69 & 11:52:39.35 & 44:03:27.68  &   23.2 &  5.00 &   1.8 & 0.03 &  4.66e-04 & 1.01e-04 & 3.68e-05 & 6.01e-07 &   8.9  & ~4.70 & ~2.7 \\
 71 & 11:52:38.93 & 43:58:52.25  &   24.1 &  6.33 &  15.9 & 0.10 &  4.85e-04 & 1.27e-04 & 3.20e-04 & 2.02e-06 &   4.7  & ~4.90 & ~2.8 \\
 72 & 11:52:38.72 & 44:12:46.71  &  297.7 & 18.47 &  43.3 & 0.17 &  5.99e-03 & 3.72e-04 & 8.72e-04 & 3.40e-06 &  39.0  & 60.50 & 34.9 \\
 75 & 11:52:36.43 & 44:01:07.42  &   17.6 &  5.39 &  11.4 & 0.08 &  3.53e-04 & 1.08e-04 & 2.30e-04 & 1.68e-06 &   3.9  & ~3.56 & ~2.0 \\
 78 & 11:52:28.24 & 44:11:06.32  &   17.4 &  5.57 &  13.6 & 0.09 &  3.50e-04 & 1.12e-04 & 2.74e-04 & 1.85e-06 &   3.6  & ~3.53 & ~2.0 \\
 79 & 11:52:27.12 & 44:04:54.49  &   18.1 &  4.69 &   3.9 & 0.04 &  3.64e-04 & 9.44e-05 & 7.88e-05 & 8.97e-07 &   5.7  & ~3.68 & ~2.1 \\
 80 & 11:52:25.04 & 44:07:22.40  &   22.2 &  6.25 &  16.8 & 0.10 &  4.46e-04 & 1.26e-04 & 3.39e-04 & 2.05e-06 &   4.3  & ~4.51 & ~2.6 \\
 82 & 11:52:21.44 & 44:12:08.93  &   35.2 &  7.28 &  17.8 & 0.11 &  7.08e-04 & 1.47e-04 & 3.59e-04 & 2.13e-06 &   6.6  & ~7.15 & ~4.1 \\
 83 & 11:52:18.05 & 44:00:42.87  &   26.1 &  6.71 &  18.9 & 0.11 &  5.24e-04 & 1.35e-04 & 3.81e-04 & 2.19e-06 &   4.8  & ~5.29 & ~3.0 \\
 84 & 11:52:15.28 & 44:09:20.71  &   26.7 &  6.86 &  20.3 & 0.11 &  5.37e-04 & 1.38e-04 & 4.08e-04 & 2.27e-06 &   4.8  & ~5.42 & ~3.1 \\
 85 & 11:52:14.41 & 44:00:40.70  &   47.2 &  8.37 &  22.8 & 0.12 &  9.51e-04 & 1.68e-04 & 4.58e-04 & 2.41e-06 &   8.1  & ~9.60 & ~5.5 \\
 86 & 11:52:13.10 & 44:01:28.22  &   63.7 &  9.33 &  23.3 & 0.12 &  1.28e-03 & 1.88e-04 & 4.70e-04 & 2.43e-06 &  10.8  & 12.93 & ~7.5 \\
 87 & 11:52:13.11 & 44:09:02.54  &  110.4 & 12.73 &  51.6 & 0.18 &  2.22e-03 & 2.56e-04 & 1.04e-03 & 3.66e-06 &  13.4  & 22.42 & 12.9 \\
 89 & 11:52:09.24 & 44:04:22.25  &   28.5 &  6.78 &  17.5 & 0.10 &  5.73e-04 & 1.36e-04 & 3.52e-04 & 2.11e-06 &   5.4  & ~5.78 & ~3.3 \\
 90 & 11:52:04.12 & 44:08:10.54  &   26.1 &  7.14 &  24.9 & 0.13 &  5.25e-04 & 1.44e-04 & 5.01e-04 & 2.53e-06 &   4.3  & ~5.30 & ~3.0 \\
 91 & 11:52:04.06 & 44:11:29.88  &   44.4 &  9.38 &  43.6 & 0.17 &  8.94e-04 & 1.89e-04 & 8.77e-04 & 3.35e-06 &   5.8  & ~9.03 & ~5.2 \\
 92 & 11:52:01.68 & 44:03:13.27  &   84.5 & 12.45 &  70.5 & 0.21 &  1.70e-03 & 2.51e-04 & 1.42e-03 & 4.27e-06 &   9.0  & 17.17 & ~9.9 \\
 93 & 11:51:57.46 & 44:05:21.48  &   39.1 &  8.72 &  36.9 & 0.15 &  7.87e-04 & 1.75e-04 & 7.42e-04 & 3.08e-06 &   5.5  & ~7.95 & ~4.6 \\
 94 & 11:51:54.01 & 44:07:55.66  &   22.3 &  6.93 &  25.7 & 0.13 &  4.49e-04 & 1.39e-04 & 5.17e-04 & 2.57e-06 &   3.6  & ~4.53 & ~2.6 \\
 95 & 11:51:50.98 & 43:59:59.96  &   33.5 &  7.88 &  28.5 & 0.14 &  6.74e-04 & 1.58e-04 & 5.73e-04 & 2.74e-06 &   5.2  & ~6.81 & ~3.9 \\  \hline
\end{tabular}
\vspace{0.1in}

$^a$Column headings are defined as:  `N' = source number from
the {\it merged} source list; `RA', `Dec' = source position,
J2000, and sorted by decreasing RA; `Counts', `Err' = {\tt
wavdetect}-determined source counts and uncertainty; `BkgCts',
`BkgErr' = {\tt wavdetect}-determined background counts and
uncertainty; `SrcRate', `SrcRErr' = source count rate and
uncertainty rate = SrcCts / exposureTime; `BkgRate', `BkgRErr'
= background count rate and uncertainty rate and determined
in the same manner as the `SrcRate' and `SrcRErr'; `Signif' =
statistical significance as determined by {\tt wavdetect};
'SrcFlux' and 'L$_X$' = Source Flux and Luminosity.  The
count rates were converted to fluxes using WebPIMMS with an
adopted absorbed thermal bremsstrahlung model of 5 keV.  We 
do not include uncertainties because we do not know the 
source model.

$^b$Absorbed flux in the 0.5 - 8 keV band in units of
10$^{-15}$ erg s$^{-1}$ cm$^{-2}$.

$^c$Absorbed luminosity in the 0.5 - 8 keV band in units of
10$^{38}$ erg s$^{-1}$ assuming  a distance of 22 Mpc.  If the
distance is 17 Mpc, the luminosities will decrease by a
factor of 0.59.

\end{table}

\begin{table}[h!]
 \centering
 \caption{NGC 3938 Epoch 2 Detected Source List$^a$}
 \label{Ep2Srcs}
 \tiny
 \begin{tabular}{rrrrrrrrrrrrrr}
 N  &      RA     &     Dec     & Counts & Err  & BkgCts &  BkEr &  SrcRate &  SrcRErr &  BkgRate &  BkRErr  & Signif & Flux$^b$ & L$_X^c$ \\ \hline
\multicolumn{14}{c}{Inside D25 circle} \\
 21 & 11:53:03.92 & 44:08:46.46 &  15.6 &  4.12 &   1.4  &  0.03 & 3.43e-04 & 9.06e-05 & 3.04e-05 & 5.76e-07 &   6.3  & ~4.25 & ~2.4 \\
 23 & 11:53:00.25 & 44:06:00.43 &  54.5 &  7.48 &   1.5  &  0.03 & 1.20e-03 & 1.64e-04 & 3.34e-05 & 6.08e-07 &  21.7  & 14.88 & ~8.6 \\
 26 & 11:52:57.58 & 44:06:47.34 &  10.6 &  3.46 &   1.4  &  0.03 & 2.33e-04 & 7.61e-05 & 3.04e-05 & 5.54e-07 &   4.3  & ~2.89 & ~1.6 \\
 29 & 11:52:55.24 & 44:09:05.08 &  14.1 &  4.24 &   3.9  &  0.04 & 3.10e-04 & 9.32e-05 & 8.60e-05 & 9.83e-07 &   4.5  & ~3.84 & ~2.2 \\
 30 & 11:52:54.84 & 44:07:36.10 &   8.6 &  3.16 &   1.4  &  0.03 & 1.89e-04 & 6.95e-05 & 3.02e-05 & 5.57e-07 &   3.5  & ~2.34 & ~1.3 \\
 31 & 11:52:54.76 & 44:09:17.18 &  10.0 &  3.61 &   3.0  &  0.04 & 2.20e-04 & 7.92e-05 & 6.54e-05 & 8.40e-07 &   3.4  & ~2.73 & ~1.6 \\
 32 & 11:52:54.70 & 44:07:52.34 &  72.9 &  8.78 &   4.1  &  0.05 & 1.60e-03 & 1.93e-04 & 8.91e-05 & 9.94e-07 &  22.9  & 19.84 & 11.4 \\
 33 & 11:52:54.68 & 44:05:29.09 &  15.0 &  4.00 &   1.0  &  0.02 & 3.29e-04 & 8.79e-05 & 2.27e-05 & 4.90e-07 &   6.4  & ~4.08 & ~2.3 \\
 34 & 11:52:54.46 & 44:06:58.67 &  14.8 &  4.00 &   1.2  &  0.02 & 3.26e-04 & 8.79e-05 & 2.59e-05 & 5.19e-07 &   6.2  & ~4.04 & ~2.3 \\
 37 & 11:52:53.71 & 44:07:34.74 &  18.9 &  4.58 &   2.1  &  0.03 & 4.15e-04 & 1.01e-04 & 4.65e-05 & 7.07e-07 &   7.0  & ~5.15 & ~2.9 \\
 38 & 11:52:53.49 & 44:06:12.83 &  27.9 &  5.48 &   2.1  &  0.03 & 6.14e-04 & 1.20e-04 & 4.53e-05 & 6.99e-07 &  10.4  & ~7.61 & ~4.4 \\
 41 & 11:52:51.92 & 44:05:47.92 &  23.8 &  5.00 &   1.2  &  0.02 & 5.22e-04 & 1.10e-04 & 2.69e-05 & 5.39e-07 &   9.9  & ~6.47 & ~3.7 \\
 43 & 11:52:51.13 & 44:04:42.57 &  30.3 &  6.08 &   6.7  &  0.06 & 6.66e-04 & 1.34e-04 & 1.47e-04 & 1.27e-06 &   8.1  & ~8.26 & ~4.7 \\
 46 & 11:52:50.57 & 44:06:28.33 &   9.8 &  3.32 &   1.2  &  0.02 & 2.16e-04 & 7.29e-05 & 2.57e-05 & 5.14e-07 &   4.1  & ~2.68 & ~1.5 \\
 47 & 11:52:50.52 & 44:04:36.26 &  87.3 &  9.59 &   4.7  &  0.05 & 1.92e-03 & 2.11e-04 & 1.02e-04 & 1.07e-06 &  26.3  & 23.81 & 13.7 \\
 49 & 11:52:50.34 & 44:08:06.64 &  66.8 &  8.37 &   3.2  &  0.04 & 1.47e-03 & 1.84e-04 & 7.02e-05 & 8.81e-07 &  22.4  & 18.28 & 10.5 \\
 50 & 11:52:50.22 & 44:07:11.63 &  25.1 &  5.29 &   2.9  &  0.04 & 5.52e-04 & 1.16e-04 & 6.31e-05 & 8.33e-07 &   8.7  & ~6.84 & ~3.9 \\
 51 & 11:52:49.56 & 44:07:17.22 &  17.6 &  4.58 &   3.4  &  0.04 & 3.87e-04 & 1.01e-04 & 7.44e-05 & 8.96e-07 &   5.8  & ~4.80 & ~2.7 \\
 52 & 11:52:48.98 & 44:07:17.75 &  13.7 &  4.00 &   2.3  &  0.03 & 3.01e-04 & 8.79e-05 & 5.10e-05 & 7.49e-07 &   5.0  & ~3.73 & ~2.1 \\
 53 & 11:52:48.91 & 44:05:28.67 &  14.8 &  4.00 &   1.2  &  0.02 & 3.26e-04 & 8.79e-05 & 2.57e-05 & 5.18e-07 &   6.2  & ~4.04 & ~2.3 \\
 54 & 11:52:48.10 & 44:07:03.70 &   9.3 &  3.32 &   1.7  &  0.03 & 2.04e-04 & 7.29e-05 & 3.75e-05 & 6.23e-07 &   3.6  & ~2.53 & ~1.4 \\
 56 & 11:52:47.20 & 44:06:59.15 &   7.5 &  3.00 &   1.5  &  0.03 & 1.64e-04 & 6.59e-05 & 3.37e-05 & 6.05e-07 &   3.0  & ~2.03 & ~1.2 \\
 59 & 11:52:46.65 & 44:09:04.67 &  12.2 &  4.00 &   3.8  &  0.04 & 2.69e-04 & 8.79e-05 & 8.28e-05 & 9.49e-07 &   3.9  & ~3.33 & ~1.9 \\
 61 & 11:52:46.48 & 44:07:53.85 &  60.5 &  8.12 &   5.5  &  0.05 & 1.33e-03 & 1.78e-04 & 1.21e-04 & 1.16e-06 &  17.3  & 16.49 & ~9.5 \\
 63 & 11:52:46.24 & 44:06:28.30 &  54.3 &  7.68 &   4.7  &  0.05 & 1.19e-03 & 1.69e-04 & 1.04e-04 & 1.04e-06 &  16.3  & 14.76 & ~8.5 \\
 65 & 11:52:44.46 & 44:09:14.94 &  26.3 &  5.57 &   4.7  &  0.05 & 5.78e-04 & 1.22e-04 & 1.03e-04 & 1.05e-06 &   7.9  & ~7.17 & ~4.1 \\
 74 & 11:52:36.88 & 44:07:56.70 &  19.9 &  6.33 &  20.1  &  0.11 & 4.37e-04 & 1.39e-04 & 4.42e-04 & 2.51e-06 &   3.6  & ~5.42 & ~3.1 \\
\multicolumn{14}{c}{Outside D25 circle} \\
 ~1 & 11:53:54.95 & 44:00:39.43 &  17.8 &  5.57 &  13.2  &  0.09 & 3.91e-04 & 1.22e-04 & 2.90e-04 & 2.03e-06 &   3.8  & ~4.85 & ~2.8 \\
 ~2 & 11:53:42.85 & 44:07:15.65 &  42.7 &  7.94 &  20.3  &  0.11 & 9.37e-04 & 1.74e-04 & 4.47e-04 & 2.50e-06 &   7.6  & 11.16 & ~6.4 \\
 ~3 & 11:53:27.49 & 44:10:33.19 &  24.8 &  6.33 &  15.2  &  0.10 & 5.44e-04 & 1.39e-04 & 3.35e-04 & 2.16e-06 &   5.0  & ~6.75 & ~3.9 \\
 ~4 & 11:53:26.35 & 44:10:36.64 &  14.8 &  5.20 &  12.2  &  0.09 & 3.25e-04 & 1.14e-04 & 2.68e-04 & 1.93e-06 &   3.2  & ~4.03 & ~2.3 \\
 ~5 & 11:53:21.96 & 44:07:30.69 &  20.9 &  5.00 &   4.1  &  0.04 & 4.58e-04 & 1.10e-04 & 9.10e-05 & 9.88e-07 &   6.5  & ~5.68 & ~3.3 \\
 ~6 & 11:53:21.61 & 44:11:31.16 &  32.2 &  6.40 &   8.8  &  0.07 & 7.08e-04 & 1.41e-04 & 1.93e-04 & 1.61e-06 &   7.9  & ~8.78 & ~5.1 \\
 ~7 & 11:53:18.99 & 44:12:14.52 &  17.3 &  5.39 &  11.7  &  0.08 & 3.81e-04 & 1.18e-04 & 2.56e-04 & 1.87e-06 &   3.8  & ~4.72 & ~2.7 \\
 ~8 & 11:53:18.92 & 44:00:30.15 &  21.2 &  6.08 &  15.8  &  0.10 & 4.66e-04 & 1.34e-04 & 3.47e-04 & 2.19e-06 &   4.2  & ~5.78 & ~3.3 \\
 ~9 & 11:53:11.85 & 44:07:02.53 &  11.6 &  3.61 &   1.4  &  0.03 & 2.54e-04 & 7.92e-05 & 3.14e-05 & 5.70e-07 &   4.7  & ~3.15 & ~1.8 \\
 12 & 11:53:09.22 & 44:03:01.57 &  19.6 &  5.00 &   5.4  &  0.05 & 4.31e-04 & 1.10e-04 & 1.18e-04 & 1.13e-06 &   5.7  & ~5.34 & ~3.1 \\
 13 & 11:53:07.39 & 44:04:50.65 &  14.1 &  4.00 &   1.9  &  0.03 & 3.10e-04 & 8.79e-05 & 4.11e-05 & 6.58e-07 &   5.4  & ~3.84 & ~2.2 \\ 
 14 & 11:53:07.38 & 44:02:20.37 &  38.7 &  6.93 &   9.3  &  0.08 & 8.51e-04 & 1.52e-04 & 2.04e-04 & 1.65e-06 &   9.3  & 10.55 & ~6.1 \\
 16 & 11:53:07.03 & 43:59:00.44 &  25.6 &  6.86 &  21.4  &  0.12 & 5.62e-04 & 1.51e-04 & 4.71e-04 & 2.56e-06 &   4.5  & ~6.97 & ~4.0 \\
 19 & 11:53:05.80 & 44:03:34.66 &  34.8 &  6.16 &   3.2  &  0.04 & 7.64e-04 & 1.35e-04 & 7.06e-05 & 8.91e-07 &  11.6  & ~9.47 & ~5.5 \\
 20 & 11:53:04.03 & 44:10:22.25 &  25.5 &  5.48 &   4.5  &  0.05 & 5.61e-04 & 1.20e-04 & 9.82e-05 & 1.02e-06 &   7.8  & ~6.96 & ~4.0 \\
 25 & 11:52:59.85 & 44:02:13.49 &  28.5 &  6.48 &  13.5  &  0.09 & 6.25e-04 & 1.42e-04 & 2.97e-04 & 2.02e-06 &   6.0  & ~7.75 & ~4.5 \\
 68 & 11:52:40.71 & 44:11:09.34 &  15.0 &  5.10 &  11.0  &  0.08 & 3.29e-04 & 1.12e-04 & 2.42e-04 & 1.80e-06 &   3.4  & ~4.08 & ~2.3 \\
 72 & 11:52:38.75 & 44:12:47.34 & 326.9 & 19.29 &  45.1  &  0.17 & 7.18e-03 & 4.24e-04 & 9.91e-04 & 3.75e-06 &  42.1  & 89.03 & 51.4 \\
 76 & 11:52:35.96 & 43:57:25.09 &  32.4 &  7.55 &  24.6  &  0.13 & 7.12e-04 & 1.66e-04 & 5.40e-04 & 2.75e-06 &   5.4  & ~8.83 & ~5.1 \\
 84 & 11:52:15.20 & 44:09:20.64 &  18.8 &  5.83 &  15.2  &  0.10 & 4.14e-04 & 1.28e-04 & 3.33e-04 & 2.14e-06 &   3.8  & ~5.13 & ~2.9 \\
 87 & 11:52:12.97 & 44:09:01.63 &  50.6 &  8.95 &  29.4  &  0.14 & 1.11e-03 & 1.97e-04 & 6.47e-04 & 3.04e-06 &   7.8  & 13.76 & ~7.9 \\ \hline
\end{tabular}
\vspace{0.1in}

$^a$See footnote `a' to the Epoch 1 table.  The columns are
identically defined.

$^b$Absorbed flux in units of 10$^{-15}$ erg s$^{-1}$
cm$^{-2}$ in the 0.5 -- 8.0 keV band.

$^c$Absorbed luminosity in the 0.5 - 8 keV band in units of
10$^{38}$ erg s$^{-1}$ assuming 
a distance of 22 Mpc.  If the distance is 17 Mpc, the
luminosities will decrease by a factor of 0.59.

\end{table}

\subsection{Color-color tables}
The color-color tables are placed in this appendix solely to
make the article pages easier to read.   We divide both tables
into `within the D25 circle' and `outside D25'.  As stated in
the analysis section, `N' refers to the source number in the
{\it merged} source list and is used for both epochs of data.  

\begin{longrotatetable}
\begin{deluxetable*}{lrrrrrrrrrrrrrr}
\tablecaption{Color-Color Results By Source:  Epoch 1$^a$}
\label{Ep1ColTab}
\tabletypesize{\tiny}
\tablehead{
 \colhead{N} & \colhead{RA} & \colhead{Dec} & \colhead{BCts} & \colhead{BUnc} & \colhead{SCts} & \colhead{SUnc} &
 \colhead{MCts} & \colhead{MUnc} & \colhead{HCts} & \colhead{HUnc} & \colhead{M-S/B} & \colhead{UM-S/B} & \colhead{H-M/B} & \colhead{UH-S/B}
}
\startdata
   \multicolumn{15}{c}{Within D25 circle} \\
21 & 11:53:03.84 & 44:08:46.21 &   33.7 &  6.93 &   0.0 & 3.1  &  11.4 &  3.46 &  10.6 &  3.74 &  0.246 & 0.2623 & -0.024 & 0.0121 \\
23 & 11:53:00.25 & 44:05:59.87 &   85.4 &  9.54 &   6.8 & 2.65 &  36.7 &  6.08 &  42.4 &  6.63 &  0.350 & 0.1533 &  0.067 & 0.0169 \\
24 & 11:53:00.14 & 44:07:00.33 &   25.7 &  5.48 &   0.0 & 3.1  &  13.5 &  3.74 &  10.9 &  3.46 &  0.405 & 0.4287 & -0.101 & 0.0478 \\
27 & 11:52:57.15 & 44:08:08.41 &   65.7 &  8.43 &  11.1 & 3.46 &  44.4 &  6.71 &  10.3 &  3.46 &  0.507 & 0.1872 & -0.519 & 0.2025 \\
28 & 11:52:56.09 & 44:04:12.09 &    9.7 &  3.32 &   0.0 & 3.1  &   0.0 &  4.90 &   0.0 &  3.20 &  0.186 & 0.2700 & -0.175 & 0.2550 \\
31 & 11:52:54.73 & 44:09:16.49 &   10.1 &  3.74 &   0.0 & 3.1  &   4.7 &  2.24 &   0.0 &  3.20 &  0.158 & 0.1850 & -0.465 & 0.5435 \\
32 & 11:52:54.72 & 44:07:51.88 &   87.7 &  9.64 &  11.5 & 3.46 &  44.4 &  6.78 &  32.6 &  5.83 &  0.375 & 0.1331 & -0.135 & 0.0349 \\
33 & 11:52:54.70 & 44:05:28.77 &   25.0 &  5.20 &   0.0 & 3.1  &  15.7 &  4.00 &   7.6 &  2.83 &  0.504 & 0.5306 & -0.324 & 0.1610 \\
37 & 11:52:53.77 & 44:07:34.43 &   20.0 &  4.80 &   0.0 & 3.1  &   8.6 &  3.00 &  11.9 &  3.61 &  0.275 & 0.2986 &  0.165 & 0.0859 \\
38 & 11:52:53.51 & 44:06:12.29 &   41.0 &  6.56 &   0.0 & 3.1  &  15.7 &  4.00 &  22.2 &  4.80 &  0.307 & 0.3209 &  0.159 & 0.0587 \\
39 & 11:52:53.04 & 44:07:11.33 &   25.0 &  5.20 &   0.0 & 3.1  &  16.6 &  4.12 &   0.0 &  3.20 &  0.540 & 0.5676 & -0.664 & 0.6979 \\
41 & 11:52:51.95 & 44:05:47.44 &   63.0 &  8.12 &   5.9 & 2.45 &  36.3 &  6.08 &  21.0 &  4.69 &  0.483 & 0.2248 & -0.243 & 0.0747 \\
42 & 11:52:51.60 & 44:03:43.92 &   16.3 &  4.24 &   0.0 & 3.1  &   6.9 &  2.65 &   3.6 &  2.00 &  0.233 & 0.2570 & -0.202 & 0.1465 \\
43 & 11:52:51.14 & 44:04:41.94 &   21.2 &  4.80 &   0.0 & 3.1  &   9.8 &  3.16 &   7.6 &  2.83 &  0.316 & 0.3397 & -0.104 & 0.0563 \\
45 & 11:52:50.80 & 44:06:53.31 &    8.1 &  3.16 &   0.0 & 3.1  &   0.0 &  4.90 &   6.1 &  2.65 &  0.222 & 0.3260 &  0.753 & 0.8721 \\
46 & 11:52:50.58 & 44:06:27.82 &   16.7 &  4.24 &   0.0 & 3.1  &   7.8 &  2.83 &   8.7 &  3.00 &  0.281 & 0.3078 &  0.054 & 0.0302 \\
47 & 11:52:50.52 & 44:04:35.66 &   94.1 &  9.95 &   0.0 & 3.1  &  37.7 &  6.25 &  53.9 &  7.48 &  0.368 & 0.3747 &  0.172 & 0.0414 \\
49 & 11:52:50.39 & 44:08:06.18 &  107.4 & 10.86 &   7.2 & 2.83 &  52.2 &  7.28 &  42.4 &  6.71 &  0.419 & 0.1798 & -0.091 & 0.0213 \\
50 & 11:52:50.23 & 44:07:11.25 &   26.1 &  5.39 &   7.6 & 2.83 &  14.3 &  3.87 &   5.6 &  2.45 &  0.257 & 0.1295 & -0.333 & 0.1848 \\
51 & 11:52:49.59 & 44:07:16.27 &   36.0 &  6.00 &   2.7 & 1.73 &  13.4 &  3.74 &   7.1 &  2.83 &  0.297 & 0.2143 & -0.175 & 0.0907 \\
52 & 11:52:49.01 & 44:07:17.40 &   15.2 &  4.12 &   0.0 & 3.1  &   9.8 &  3.16 &   4.5 &  2.24 &  0.441 & 0.4783 & -0.349 & 0.2274 \\
53 & 11:52:48.92 & 44:05:28.44 &   20.3 &  4.69 &   0.0 & 3.1  &   0.0 &  4.90 &  17.5 &  4.24 &  0.089 & 0.1271 &  0.862 & 0.9091 \\
56 & 11:52:47.19 & 44:06:58.65 &   16.2 &  4.24 &   5.8 & 2.45 &   8.7 &  3.00 &   2.6 &  1.73 &  0.179 & 0.1083 & -0.377 & 0.2989 \\
58 & 11:52:46.86 & 44:10:07.77 &   16.2 &  5.00 &   0.0 & 3.1  &   6.6 &  2.65 &   0.0 &  3.20 &  0.216 & 0.2422 & -0.407 & 0.4567 \\
59 & 11:52:46.74 & 44:09:03.59 &   13.4 &  4.12 &   0.0 & 3.1  &   5.7 &  2.45 &   0.0 &  3.20 &  0.194 & 0.2195 & -0.425 & 0.4811 \\
60 & 11:52:46.65 & 44:06:48.45 &  354.5 & 19.13 &  43.0 & 6.63 & 181.2 & 13.60 & 132.1 & 11.70 &  0.390 & 0.0701 & -0.139 & 0.0177 \\
61 & 11:52:46.44 & 44:07:53.24 &   46.8 &  7.28 &   3.7 & 2.00 &  27.2 &  5.29 &  13.8 &  3.87 &  0.502 & 0.2988 & -0.286 & 0.1074 \\
62 & 11:52:46.33 & 44:03:52.53 &   10.0 &  3.32 &   4.9 & 2.24 &   4.9 &  2.24 &   0.0 &  3.20 &  0.000 & 0.0000 & -0.170 & 0.1953 \\
63 & 11:52:46.26 & 44:06:27.71 &   92.7 &  9.80 &   9.7 & 3.16 &  49.2 &  7.07 &  34.0 &  5.92 &  0.426 & 0.1583 & -0.164 & 0.0409 \\
64 & 11:52:45.64 & 44:05:26.68 &   13.8 &  3.87 &   2.9 & 1.73 &   7.9 &  2.83 &   2.7 &  1.73 &  0.362 & 0.2718 & -0.377 & 0.2961 \\
65 & 11:52:44.47 & 44:09:14.20 &   19.5 &  5.00 &   0.0 & 3.1  &   7.8 &  2.83 &   9.5 &  3.32 &  0.241 & 0.2637 &  0.087 & 0.0493 \\
70 & 11:52:38.95 & 44:06:02.49 &   21.5 &  4.80 &   0.0 & 3.1  &   8.9 &  3.00 &  10.5 &  3.32 &  0.270 & 0.2910 &  0.074 & 0.0382 \\
77 & 11:52:34.44 & 44:05:16.27 &   10.7 &  3.46 &   0.0 & 3.1  &   4.9 &  2.24 &   5.6 &  2.45 &  0.168 & 0.1928 &  0.065 & 0.0465 \\ \hline
   \multicolumn{15}{c}{Epoch 1: Outside D25 circle} \\
 4 & 11:53:26.52 & 44:10:34.65 &   10.1 &  4.00 &   0.0 & 3.1  &   0.0 &  4.90 &   0.0 &  3.20 &  0.178 & 0.2617 & -0.168 & 0.2472 \\
 5 & 11:53:22.09 & 44:07:29.95 &   55.8 &  8.66 &   0.0 & 3.1  &  22.2 &  4.90 &  27.4 &  5.92 &  0.342 & 0.3545 &  0.093 & 0.0322 \\
 6 & 11:53:21.57 & 44:11:31.38 &   43.9 &  8.60 &   0.0 & 3.1  &  19.1 &  4.80 &  25.8 &  6.08 &  0.364 & 0.3825 &  0.153 & 0.0605 \\
 9 & 11:53:11.74 & 44:07:03.09 &   25.9 &  6.25 &   0.0 & 3.1  &   6.4 &  2.83 &  12.0 &  4.00 &  0.127 & 0.1427 &  0.216 & 0.1306 \\
10 & 11:53:11.11 & 43:57:19.53 &   73.6 &  9.85 &   0.0 & 3.1  &  32.0 &  6.16 &  32.1 &  6.32 &  0.393 & 0.4033 &  0.001 & 0.0004 \\
11 & 11:53:09.67 & 44:00:04.79 &   29.2 &  6.86 &   0.0 & 3.1  &  14.0 &  4.00 &  16.5 &  4.90 &  0.373 & 0.3980 &  0.086 & 0.0406 \\
12 & 11:53:09.05 & 44:03:00.92 &   16.1 &  5.39 &   0.0 & 3.1  &  10.2 &  3.32 &   0.0 &  3.20 &  0.441 & 0.4867 & -0.634 & 0.6992 \\
14 & 11:53:07.46 & 44:02:19.47 &   84.3 &  9.95 &  13.0 & 3.74 &  41.3 &  6.63 &  30.5 &  5.83 &  0.336 & 0.1175 & -0.128 & 0.0354 \\
16 & 11:53:07.05 & 43:58:58.78 &   27.3 &  6.00 &   0.0 & 3.1  &  14.8 &  4.00 &   8.7 &  3.32 &  0.429 & 0.4538 & -0.223 & 0.1155 \\
17 & 11:53:06.57 & 44:02:36.21 &   39.8 &  7.68 &   0.0 & 3.1  &   0.0 &  4.90 &  28.6 &  5.74 &  0.045 & 0.0646 &  0.719 & 0.7459 \\
19 & 11:53:05.81 & 44:03:34.88 &   81.3 &  9.27 &   2.6 & 1.73 &  32.5 &  5.74 &  47.6 &  7.00 &  0.368 & 0.2566 &  0.186 & 0.0477 \\
25 & 11:52:59.72 & 44:02:12.99 &   12.0 &  4.00 &   0.0 & 3.1  &   0.0 &  4.90 &   0.0 &  3.20 &  0.150 & 0.2179 & -0.142 & 0.2058 \\
66 & 11:52:44.20 & 43:59:14.67 &   25.0 &  6.56 &   0.0 & 3.1  &   0.0 &  4.90 &  12.3 &  4.00 &  0.072 & 0.1036 &  0.492 & 0.5332 \\
69 & 11:52:39.35 & 44:03:27.68 &   37.4 &  6.71 &   5.9 & 2.45 &  11.7 &  3.46 &   6.1 &  2.65 &  0.155 & 0.0838 & -0.150 & 0.0831 \\
71 & 11:52:38.93 & 43:58:52.25 &   24.1 &  6.33 &   0.0 & 3.1  &   0.0 &  4.90 &  25.0 &  5.66 &  0.075 & 0.1074 &  1.037 & 1.0979 \\
72 & 11:52:38.72 & 44:12:46.71 &  297.7 & 18.47 &  31.1 & 6.08 & 136.0 & 12.04 & 129.9 & 12.25 &  0.352 & 0.0787 & -0.020 & 0.0029 \\
75 & 11:52:36.43 & 44:01:07.42 &   17.6 &  5.39 &   0.0 & 3.1  &   6.2 &  2.65 &   8.0 &  3.32 &  0.176 & 0.1990 &  0.102 & 0.0685 \\
78 & 11:52:28.24 & 44:11:06.32 &   17.4 &  5.57 &   0.0 & 3.1  &   0.0 &  4.90 &   0.0 &  3.20 &  0.103 & 0.1500 & -0.098 & 0.1417 \\
79 & 11:52:27.12 & 44:04:54.49 &   18.1 &  4.69 &   0.0 & 3.1  &   8.9 &  3.00 &   7.7 &  3.00 &  0.320 & 0.3482 & -0.066 & 0.0382 \\
80 & 11:52:25.04 & 44:07:22.40 &   22.2 &  6.25 &   0.0 & 3.1  &  12.4 &  3.61 &   0.0 &  3.20 &  0.419 & 0.4520 & -0.559 & 0.6026 \\
82 & 11:52:21.44 & 44:12:08.93 &   35.2 &  7.28 &   0.0 & 3.1  &  13.2 &  4.00 &  19.6 &  5.39 &  0.287 & 0.3056 &  0.182 & 0.0834 \\
83 & 11:52:18.05 & 44:00:42.87 &   26.1 &  6.71 &   0.0 & 3.1  &   0.0 &  4.90 &  14.7 &  4.58 &  0.069 & 0.0991 &  0.563 & 0.6074 \\
84 & 11:52:15.28 & 44:09:20.71 &   26.7 &  6.86 &  11.7 & 3.61 &  12.5 &  3.74 &   0.0 &  3.20 &  0.030 & 0.0150 & -0.348 & 0.3744 \\
85 & 11:52:14.41 & 44:00:40.70 &   47.2 &  8.37 &   0.0 & 3.1  &  13.0 &  3.87 &  31.9 &  6.56 &  0.210 & 0.2220 &  0.400 & 0.1613 \\
86 & 11:52:13.11 & 44:01:28.22 &   63.7 &  9.33 &   0.0 & 3.1  &  32.1 &  6.00 &  26.1 &  5.83 &  0.455 & 0.4679 & -0.094 & 0.0307 \\
87 & 11:52:13.11 & 44:09:02.54 &  110.4 & 12.73 &   0.0 & 3.1  &  47.1 &  7.35 &  56.0 &  8.37 &  0.399 & 0.4060 &  0.081 & 0.0197 \\
89 & 11:52:09.24 & 44:04:22.25 &   28.5 &  6.78 &   0.0 & 3.1  &  17.8 &  4.47 &   0.0 &  3.20 &  0.516 & 0.5458 & -0.625 & 0.6609 \\
90 & 11:52:04.12 & 44:08:10.54 &   26.1 &  7.14 &   0.0 & 3.1  &   9.8 &  3.46 &   0.0 &  3.20 &  0.257 & 0.2811 & -0.375 & 0.4112 \\
91 & 11:52:04.06 & 44:11:29.88 &   44.4 &  9.38 &   9.2 & 3.46 &  30.2 &  6.71 &   0.0 &  3.20 &  0.473 & 0.2295 & -0.608 & 0.6360 \\
92 & 11:52:01.68 & 44:03:13.27 &   84.5 & 12.45 &   0.0 & 3.1  &  41.7 &  6.93 &  29.9 &  7.21 &  0.457 & 0.4679 & -0.140 & 0.0458 \\
93 & 11:51:57.46 & 44:05:21.48 &   39.1 &  8.72 &   0.0 & 3.1  &   0.0 &  4.90 &  39.1 &  8.72 &  0.046 & 0.0659 &  1.000 & 1.0486 \\
94 & 11:51:54.01 & 44:07:55.66 &   22.3 &  6.93 &   0.0 & 3.1  &   7.1 &  3.00 &   0.0 &  3.20 &  0.179 & 0.2025 & -0.318 & 0.3595 \\
95 & 11:51:50.98 & 43:59:59.96 &   33.5 &  7.88 &   0.0 & 3.1  &  28.9 &  6.71 &   0.0 &  3.20 &  0.770 & 0.8111 & -0.863 & 0.9086 \\ 
\enddata
\vspace{0.1in}

$^a$Column headings for this and the next table are defined as
follows:  `N' = source number from the {\it merged} source list;
`RA', `Dec' = source position, J2000, in descending RA order;
`BCts' = Broad band CounTS; `BUnc' = Broad band counts
UNCertainty; similarly for S, M, and H = Soft, Medium, Hard
bands; `M-S/B' = (M - S)/B; UM-S/B = Uncertain of (M - S)/B; and
similarly for `H-M/B' and 'UH-M/B'.  The equation columns were
compressed (smaller font, fewer characters) to allow the tables
to fit on the page.

\end{deluxetable*}
\end{longrotatetable}

\begin{longrotatetable}
\begin{deluxetable*}{lrrrrrrrrrrrrrr}
\tablecaption{Color-Color Results By Source:  Epoch 2$^a$}
\label{Ep2ColTab}
\tabletypesize{\tiny}
\tablehead{
 \colhead{N} & \colhead{RA} & \colhead{Dec} & \colhead{BCts} & \colhead{BUnc} & \colhead{SCts} & \colhead{SUnc} &
 \colhead{MCts} & \colhead{MUnc} & \colhead{HCts} & \colhead{HUnc} & \colhead{M-S/B} & \colhead{UM-S/B} & \colhead{H-M/B} & \colhead{UH-S/B}
}
\startdata
   \multicolumn{15}{c}{Within D25 circle} \\
21 & 11:53:03.92 & 44:08:46.46 &  15.6 &  4.12 &  0.0 & 3.60 &  10.7 &  3.32 &   0.0 &  4.50 &   0.4551 & 0.4915 & -0.3974 & 0.4292 \\
23 & 11:53:00.25 & 44:06:00.43 &  54.5 &  7.48 &  0.0 & 3.60 &  27.7 &  5.29 &  21.1 &  4.69 &   0.4422 & 0.4543 & -0.1211 & 0.0392 \\
26 & 11:52:57.58 & 44:06:47.34 &  10.6 &  3.46 &  0.0 & 3.60 &   6.8 &  2.65 &   4.7 &  2.24 &   0.3019 & 0.3387 & -0.1981 & 0.1380 \\
29 & 11:52:55.24 & 44:09:05.08 &  14.1 &  4.24 &  0.0 & 3.60 &   0.0 &  2.60 &   8.6 &  3.16 &  -0.0709 & 0.1025 &  0.4255 & 0.4711 \\
30 & 11:52:54.84 & 44:07:36.10 &   8.6 &  3.16 &  0.0 & 3.60 &   0.0 &  2.60 &   6.3 &  2.65 &  -0.1163 & 0.1699 &  0.4302 & 0.4928 \\
31 & 11:52:54.76 & 44:09:17.18 &  10.0 &  3.61 &  4.5 & 2.24 &   0.0 &  2.60 &   0.0 &  4.50 &  -0.1900 & 0.2230 &  0.1900 & 0.2773 \\
32 & 11:52:54.70 & 44:07:52.34 &  72.9 &  8.78 &  2.9 & 1.73 &  36.0 &  6.08 &  32.2 &  5.83 &   0.4540 & 0.2868 & -0.0521 & 0.0144 \\
33 & 11:52:54.68 & 44:05:29.09 &  15.0 &  4.00 &  0.0 & 3.60 &   8.9 &  3.00 &   5.6 &  2.45 &   0.3533 & 0.3846 & -0.2200 & 0.1349 \\
34 & 11:52:54.46 & 44:06:58.67 &  14.8 &  4.00 &  0.0 & 3.60 &   6.9 &  2.65 &   7.6 &  2.83 &   0.2230 & 0.2463 &  0.0473 & 0.0283 \\
37 & 11:52:53.71 & 44:07:34.74 &  18.9 &  4.58 &  0.0 & 3.60 &   3.8 &  2.00 &  13.9 &  3.87 &   0.0106 & 0.0122 &  0.5344 & 0.3435 \\
38 & 11:52:53.49 & 44:06:12.83 &  27.9 &  5.48 &  0.0 & 3.60 &  11.7 &  3.46 &  14.1 &  3.87 &   0.2903 & 0.3081 &  0.0860 & 0.0386 \\
41 & 11:52:51.92 & 44:05:47.92 &  23.8 &  5.00 &  0.0 & 3.60 &  10.7 &  3.32 &  13.7 &  3.74 &   0.2983 & 0.3186 &  0.1261 & 0.0584 \\
43 & 11:52:51.13 & 44:04:42.57 &  30.3 &  6.08 &  0.0 & 3.60 &  13.0 &  3.74 &  10.4 &  3.46 &   0.3102 & 0.3288 & -0.0858 & 0.0415 \\
46 & 11:52:50.57 & 44:06:28.33 &   9.8 &  3.32 &  0.0 & 3.60 &   2.6 &  1.73 &   5.6 &  2.45 &  -0.1020 & 0.1273 &  0.3061 & 0.2649 \\
47 & 11:52:50.52 & 44:04:36.26 &  87.3 &  9.59 &  0.0 & 3.60 &  32.0 &  5.74 &  56.0 &  7.68 &   0.3253 & 0.3324 &  0.2749 & 0.0690 \\
49 & 11:52:50.34 & 44:08:06.64 &  66.8 &  8.37 &  0.0 & 3.60 &  30.4 &  5.57 &  33.4 &  5.92 &   0.4012 & 0.4110 &  0.0449 & 0.0128 \\
50 & 11:52:50.22 & 44:07:11.63 &  25.1 &  5.29 &  0.0 & 3.60 &  18.4 &  4.36 &   5.0 &  2.45 &   0.5896 & 0.6186 & -0.5339 & 0.3116 \\
51 & 11:52:49.56 & 44:07:17.22 &  17.6 &  4.58 &  0.0 & 3.60 &   7.4 &  2.83 &   0.0 &  4.50 &   0.2159 & 0.2379 & -0.1648 & 0.1815 \\
52 & 11:52:48.98 & 44:07:17.75 &  13.7 &  4.00 &  5.5 & 2.45 &   7.4 &  2.83 &   7.2 &  2.83 &   0.1387 & 0.0909 & -0.0146 & 0.0091 \\
53 & 11:52:48.91 & 44:05:28.67 &  14.8 &  4.00 &  0.0 & 3.60 &   0.0 &  2.60 &  12.6 &  3.61 &  -0.0676 & 0.0973 &  0.6757 & 0.7262 \\
54 & 11:52:48.10 & 44:07:03.70 &   9.3 &  3.32 &  0.0 & 3.60 &   0.0 &  2.60 &   6.2 &  2.65 &  -0.1075 & 0.1568 &  0.3871 & 0.4431 \\
56 & 11:52:47.20 & 44:06:59.15 &   7.5 &  3.00 &  0.0 & 3.60 &   6.6 &  2.65 &   0.0 &  4.50 &   0.4000 & 0.4598 & -0.2800 & 0.3218 \\
59 & 11:52:46.65 & 44:09:04.67 &  12.2 &  4.00 &  3.7 & 2.00 &   0.0 &  2.60 &   0.0 &  4.50 &  -0.0902 & 0.1067 &  0.1557 & 0.2261 \\
61 & 11:52:46.48 & 44:07:53.85 &  60.5 &  8.12 &  3.6 & 2.00 &  31.0 &  5.66 &  24.3 &  5.10 &   0.4529 & 0.2717 & -0.1107 & 0.0342 \\
63 & 11:52:46.24 & 44:06:28.30 &  54.3 &  7.68 &  0.0 & 3.60 &  30.3 &  5.57 &  18.1 &  4.47 &   0.4917 & 0.5048 & -0.2247 & 0.0761 \\
65 & 11:52:44.46 & 44:09:14.94 &  26.3 &  5.57 &  0.0 & 3.60 &  11.6 &  3.46 &  15.6 &  4.24 &   0.3042 & 0.3239 &  0.1521 & 0.0693 \\
74 & 11:52:36.88 & 44:07:56.70 &  19.9 &  6.33 &  0.0 & 3.60 &   0.0 &  2.60 &   0.0 &  4.50 &  -0.0503 & 0.0728 &  0.0955 & 0.1384 \\  \hline
\multicolumn{15}{c}{Epoch 2: Outside D25 circle} \\
 1 & 11:53:54.95 & 44:00:39.43 &  17.8 &  5.57 &  0.0 & 3.60 &  11.7 &  3.87 &   0.0 &  4.50 &   0.4551 & 0.5000 & -0.4045 & 0.4445 \\
 2 & 11:53:42.85 & 44:07:15.65 &  42.7 &  7.94 &  0.0 & 3.60 &  21.1 &  5.00 &  18.5 &  5.20 &   0.4098 & 0.4280 & -0.0609 & 0.0251 \\
 3 & 11:53:27.49 & 44:10:33.19 &  24.8 &  6.33 &  0.0 & 3.60 &   7.5 &  3.00 &  16.1 &  5.10 &   0.1573 & 0.1741 &  0.3468 & 0.1978 \\
 4 & 11:53:26.35 & 44:10:36.64 &  14.8 &  5.20 &  0.0 & 3.60 &   8.6 &  3.16 &   0.0 &  4.50 &   0.3378 & 0.3790 & -0.2770 & 0.3108 \\
 5 & 11:53:21.96 & 44:07:30.69 &  20.9 &  5.00 &  0.0 & 3.60 &   8.5 &  3.00 &   8.8 &  3.16 &   0.2344 & 0.2549 &  0.0144 & 0.0080 \\
 6 & 11:53:21.61 & 44:11:31.16 &  32.2 &  6.40 &  0.0 & 3.60 &  20.2 &  4.58 &   0.0 &  4.50 &   0.5155 & 0.5385 & -0.4876 & 0.5093 \\
 7 & 11:53:18.99 & 44:12:14.52 &  17.3 &  5.39 &  0.0 & 3.60 &  11.7 &  3.61 &   0.0 &  4.50 &   0.4682 & 0.5112 & -0.4162 & 0.4544 \\
 8 & 11:53:18.92 & 44:00:30.15 &  21.2 &  6.08 &  0.0 & 3.60 &   0.0 &  2.60 &   0.0 &  4.50 &  -0.0472 & 0.0681 &  0.0896 & 0.1293 \\
 9 & 11:53:11.85 & 44:07:02.53 &  11.6 &  3.61 &  0.0 & 3.60 &   6.9 &  2.65 &   0.0 &  4.50 &   0.2845 & 0.3173 & -0.2069 & 0.2308 \\
12 & 11:53:09.22 & 44:03:01.57 &  19.6 &  5.00 &  0.0 & 3.60 &  10.2 &  3.32 &   9.3 &  3.46 &   0.3367 & 0.3644 & -0.0459 & 0.0255 \\
14 & 11:53:07.38 & 44:02:20.37 &  38.7 &  6.93 &  0.0 & 3.60 &  17.2 &  4.36 &  15.6 &  4.36 &   0.3514 & 0.3680 & -0.0413 & 0.0173 \\
15 & 11:53:07.39 & 44:04:50.65 &  14.1 &  4.00 &  0.0 & 3.60 &   5.8 &  2.45 &   7.2 &  2.83 &   0.1560 & 0.1751 &  0.0993 & 0.0638 \\
16 & 11:53:07.03 & 43:59:00.44 &  25.6 &  6.86 &  0.0 & 3.60 &  18.0 &  4.58 &   0.0 &  4.50 &   0.5625 & 0.5997 & -0.5273 & 0.5622 \\
19 & 11:53:05.80 & 44:03:34.66 &  34.8 &  6.16 &  0.0 & 3.60 &  14.1 &  3.87 &  20.5 &  4.69 &   0.3017 & 0.3174 &  0.1839 & 0.0733 \\
20 & 11:53:04.03 & 44:10:22.25 &  25.5 &  5.48 &  0.0 & 3.60 &  11.4 &  3.46 &   9.5 &  3.46 &   0.3059 & 0.3263 & -0.0745 & 0.0388 \\
25 & 11:52:59.85 & 44:02:13.49 &  28.5 &  6.48 &  0.0 & 3.60 &  14.6 &  4.00 &  12.6 &  4.36 &   0.3860 & 0.4097 & -0.0702 & 0.0348 \\
68 & 11:52:40.71 & 44:11:09.34 &  15 5.&  1.00 &  0.0 & 3.60 &   6.5 &  2.65 &  10.3 &  3.87 &   0.1933 & 0.2189 &  0.2533 & 0.1648 \\
72 & 11:52:38.75 & 44:12:47.34 & 326.9 & 19.29 & 27.1 & 5.39 & 149.2 & 12.53 & 151.5 & 13.12 &   0.3735 & 0.0836 &  0.0070 & 0.0009 \\
76 & 11:52:35.96 & 43:57:25.09 &  32.4 &  7.55 &  0.0 & 3.60 &  16.6 &  4.80 &  13.0 &  4.58 &   0.4012 & 0.4280 & -0.1111 & 0.0569 \\
84 & 11:52:15.20 & 44:09:20.64 &  18.8 &  5.83 & 12.2 & 3.74 &   0.0 &  2.60 &   0.0 &  4.50 &  -0.5106 & 0.5571 &  0.1011 & 0.1463 \\
87 & 11:52:12.97 & 44:09:01.63 &  50.6 &  8.95 &  0.0 & 3.60 &  11.6 &  3.87 &  38.7 &  7.68 &   0.1581 & 0.1690 &  0.5356 & 0.2285 \\
\enddata
\vspace{0.1in}

$^a$See Table~\ref{Ep1ColTab} for the column definitions which
are identical to the columns in this table.
\end{deluxetable*}
\end{longrotatetable}

\bibliography{sample631}{}

\begin{thebibliography}{}
%
%
\bibitem[Bekhti et al.(2016)]{Bekhti2016}
Bekhti et al. 2016, A\&A, 594, A116

\bibitem[Bertin \& Arnouts(1996)]{1996AAS..117..393B} Bertin, E. \&
  Arnouts, S. 1996 A\&AS, 117, 393

\bibitem[Bianchi(2014)]{Bianchi2014} Bianchi, L. 2014, Ap\&SS,
    354, 103
    
\bibitem[Binder et al.(2017)]{Binder2017} Binder, B.; Gross, J.;
  Williams, B. F. et al. 2017, ApJ, 834, 128
  
\bibitem[Brandt et al.(2001)]{Brandt2001} Brandt, W. N. et al. 2001, 
   AJ, 122, 2810

\bibitem[Buhidar \& Schlegel(2017)]{Buhidar2017} Buhidar, K. \& 
   Schlegel, E. M. 2017, BAAS, 22914412

\bibitem[Cald\'u-Primo et al.(2009)]{CalduPrimo2009} Cald\'u-Primo, A.,
  Cruz-Gonz\'alez, I.; \& Morisset, C. 2009, A\&A, 493, 33 

\bibitem[Calzetti et al.(2007)]{Calzetti2007} Calzetti, D., 
Kennicutt, R.~C., Engelbracht, C.~W., et al.\ 2007, \apj, 666, 870

\bibitem[Campana et al.(2001)]{Campana2001} Campana, S., Moretti, A., 
  Lazatti, D., \& Tagliaferri, G. 2001, ApJ, 560, L19

\bibitem[Chandar et al.(2020)]{Chandar2020} Chandar, R.; Johns, P.; Mok, A.;
  Prestwich, A.; Gallo, E.; \& Hunt, Q. 2020, ApJ, 890, 150

\bibitem[Charles \& Seward(1995)]{CS1995} Charles, P.A. \& Seward, F.D.
  1995, Exploring the X-ray Universe (Cambridge: Cambridge Univ Press)
  
\bibitem[Dale et al.(2007)]{Dale2007} Dale, D. A., Gil de Paz, A.;
et al. 2007, ApJ, 655, 863

\bibitem[Evans et al.(2010)]{Evans2010} Evans, I. N., Primini, 
F. A., Glotfelty, K. J. et al. 2010, ApJS, 189, 37

\bibitem[Fabbiano \& Elvis(2019)]{Fabbiano2019} Fabbiano, G. \& Elvis, M.
  2019, ApJ, 884, 163.
  
\bibitem[Fabbiano, Kim, \& Trinchieri(1992)]{Fabbiano1992} Fabbiano, G.;
  Kim, D.-W.; \& Trinchieri, G. 1992, ApJS, 80, 531

\bibitem[Fazio et al.(2004)]{Fazio2004} Fazio, G. G., Hora,
  J. L., Allen, L. E. et al. 2003 ApJS, 154, 10

\bibitem[Freeman et al.(2002)]{2002ApJS..138..185F} Freeman, P. E.,
  Kashyap, V.; Rosner, R.; \& Lamb, D. Q. 2002 ApJS, 138, 185

\bibitem[Fruscione et al.(2006)]{2006SPIE.6270E..60F} Fruscione, A.,
   et al.\ 2006, \procspie, 6270

\bibitem[Gaetz et al.(2000)]{Gaetz2000}, Gaetz, T. J.; Jerius, D.; Edgar, R. J.;
  Van Speybroeck, L. P.; Schwartz, D. A.; Markevitch, M. L.; Taylor, S. C.;
  \& Schulz, N. S. 2000, SPIE, 4012, 41
  
\bibitem[Garmire et al.(2003)]{2003SPIE.4851...28G} Garmire, G.; Bautz, M. W.;
  Ford, P. G.; Nousek, J. A.; \& Ricker, Jr., G. R. 2003  SPIE, 4851, 28

\bibitem[Gilfanov(2004)]{Gilfanov2004} Gilfanov, M. 2004, MNRAS, 349, 146

\bibitem[Gregory \& Loredo(1992)]{Gregory1992} Gregory, P. C. \& 
   Loredo, T. J. 1992, ApJ, 398, 146

\bibitem[Grimm et al.(2003)]{Grimm2003} Grimm, H.-J., Gilfanov, M. \&
  Sunyaev, R.  2003, MNRAS, 339, 793

\bibitem[Hamilton \& Sarazin(1984)]{HS1984} Hamilton, A.J.S. \& 
  Sarazin, C. L. 1984, ApJ, 284, 601
  
\bibitem[Ho et al.(1997)]{Ho1997} Ho, L., Filippenko, A. V., \&
   Sargent, W. L. W. (1997), ApJS, 112, 315

\bibitem[Hunt et al.(2021)]{Hunt2021} Hunt, Q.; Gallo, E.; Chandar, R.;
  Mulia, P. J.; Mok, A.; Prestwich, A.; \& Liu, S. 2021, ApJ, 912, 31

\bibitem[Hunt et al.(2023a)]{Hunt2023a} Hunt, Q.; Gallo, E.; Chandar, R.;
  Mok, A.; \& Prestwich, A. 2023, ApJ, 947, 31

\bibitem[Hunt et al.(2023b)]{Hunt2023b} Hunt, Q.; Chandar, R.; Gallo, E.;
  Floyd, M.; Maccarone, T. J.; \& Thilker, D. A. 2023, ApJ, 953, 126

\bibitem[JimenezVicente et al.(1999)]{JimenezVicente1999}
   Jimenez-Vicente, J.; Battaner, E.,; Rozas, M.; Castenada, H.; \&
   Poncel, C. 1999, A\&A, 342, 417

\bibitem[Kennicutt et al.(2009)]{Kennicutt2009} Kennicutt, R. C. Jr.; 
  Hao, C.-N.; Calzetti, D.; Moustakas, J.; Dale, D. A.; Bendo, G.;
  Engelbracht, C. W.; Johnson, B. D.; \& Lee, J. C. 2009, ApJ, 703, 1672

\bibitem[Kennicutt \& Kent(1983)]{Kennicutt1983} Kennicutt, R. C. \&
  Kent, S. M. 1983, AJ, 88, 1094

\bibitem[Kim et al.(2007)]{2007ApJS..169..401K} Kim, D.-W.; et al.
  2007, ApJS, 169, 401

\bibitem[Lee et al.(2006)]{Lee2006} Lee, H., Skillman, E.~D., Cannon, J.~M., et al.\ 2006, \apj, 647, 970

\bibitem[Lehmer et al.(2021)]{Lehmer2021} Lehmer, B.; Eufrasio, R. T.;
  Basu-Zych, A. et al. 2021, ApJ, 907, 17

\bibitem[Lehmer et al.(2019)]{Lehmer2019} Lehmer, B.; Eufrasio, R. T.;
  Tzanavaris, P. et al. 2019, ApJS, 243, 3

\bibitem[Long(2017)]{Long2017} Long, K. 2020, in Handbook of
Supernovae, Alsabti, A. W. \& Murdin, P. (eds) (SpringerLink) (also
arXiv 1712.05331)

\bibitem[Marchesi et al.(2020)]{Marchesi2020} Marchesi, S., Gilli, R.,
   Lanzuisi, G. et al. 2020, A\&A, 642, A184

\bibitem[Marin et al.(2023)]{Marin2023} Marin, F. et al. 2023, Nature,
    619, 41
   
\bibitem[MAST(2021)]{MAST2021} MAST Portal Guide, eds. R. A. Shaw, B.
  Cherinka, P. Forshay (ver 2; Baltimore: STScI)

\bibitem[Mineo et al.(2012)]{Mineo2012} Mineo, S., Gilfanov, M. \&
   Sunyaev, R. 2012, MNRAS, 419, 2095
   
\bibitem[Mushotzky et al.(2000)]{Mushotzky2000} Mushotzky, R. F.,
  Cowie, L. L., Barger, A. J., \& Arnaud, K. A. 2000, Nature, 404, 459

\bibitem[Pannuti et al.(2015)]{Pannuti2015} Pannuti, T. G., Swartz, D. A.,
  Laine, S., Schlegel, E. M., Lacey, C. K., Moffitt, W. P., Sharma, B., 
  Lackey-Stewart, A. M., Kosakowski, A. R., Filipovi\'c, M. D., \& Payne, J. L.
  2015, AJ, 150, a91

\bibitem[Pellegrini et al.(2000)]{Pellegrini2000} Pellegrini, S., Cappi, M.
   et al. 2000, A\&A, 353, 447

\bibitem[Portegies Zwart et al.(2007)]{PortegiesZwart2007} Portegies Zwart,
  S. F., McMillan, S. L. W., \& Makino, J. 2007, MNRAS, 374, 95
   
\bibitem[Poznanski et al.(2009)]{Poznanski2009} Poznanski, D.; Butler, N.
  et al. 2009, ApJ, 694, 1067

\bibitem[Prestwich(2004)]{Prestwich2004} Prestwich, A. 2004, RMxAC, 20, 3

\bibitem[Prestwich et al.(2003)]{2003ApJ...595..719P} Prestwich, A.
  et al. 2003, ApJ, 595, 719

\bibitem[Rieke et al.(2004)]{Rieke2004} Rieke, G. H.; Young, E. T.;
  Engelbracht, C. W. et al. 2004, ApJS, 154, 25

\bibitem[Rodriguez et al.(2014)]{Rodriguez2014} Rodriguez, O., Clochiatti,
   A., \& Hamuy, M. 2014, AJ, 148, 107

\bibitem[Rosati et al.(2002)]{Rosati2002} Rosati, P. et al. 2002, 
  ApJ, 566, 667

\bibitem[Sasaki(2020)]{Sasaki2020} Sasaki, M. 2020, AN, 341, 156

\bibitem[Satyapal(2008)]{Satyapal2008} Satyapal, S.; Vega, D.; Dudik,
  R. P.; Abel, N. P.  \& Heckman, T. 2008, ApJ, 677, 926

\bibitem[Schlegel \& Pannuti(2003)]{SchlegelPannuti2003} Schlegel, E. M. \&
   Pannuti, T. G. 2003, AJ, 125, 3025

\bibitem[Soria \& Wu(2003)]{SoriaWu2003} Soria, R. \& Wu, K. 2003,
   A\&A, 410, 53

\bibitem[Swartz et al.(2003)]{Swartz2003} Swartz, D.; Ghosh, K. K.;
  McCullough, M. L. et al. 2003, ApJS, 144, 213

\bibitem[Thornley(2004)]{Thornley2004} Thornley, M. D. 2004 in
  Dense Interstellar Medium in Galaxies, ed. S. Pfalzner et al.
  (Berlin:  Springer), p109
  
\bibitem[Townsley et al.(2003)]{Townsley2003} Townsley, L. K.,
  Feigelson, E. D., Montmerle, T. et al. 2003, ApJ, 593, 874
  
\bibitem[Tully \& Shaya(1984)]{TullyShaya1984} Tully, R. B. \& Shaya,
   E. J.. 1984, ApJ, 281, 31

\bibitem[Van Dyk et al.(2017)]{VanDyk2017} Van Dyk, S.; Filippenko,
    A., Fox, O. D. et al. 2017, ATEL 10485

\bibitem[van Speybroeck et al.(1997)]{vanSpeybroeck1997} van Speybroeck, L. P.;
   Jerius, D.; Edgar, R. J.; Gaetz, T. J.; Zhao, P.; \& Reid, P. B. 1997
   SPIE, 3113, 89
   
\bibitem[Weisskopf et al.(2002)]{2002PASP..114....1W} Weisskopf,
  M. C., Brinkman, B., Canizares, C.; Garmire, G.; Murray, S.; \& Van
  Speybroeck, L. P. 2002 PASP, 114, 1

\bibitem[Werner et al.(2004)]{2004ApJS..154....1W} Werner, M. W.; Roellig, T. L.
  Low, F. S. et al. 2004 ApJS, 154, 1

\bibitem[Zampieri(2006)]{Zampieri2006} Zampieri, L. 2006, MSAIS, 9, 378

\bibitem[Zaritsky et al.(1994)]{Zaritsky1994} Zaritsky, D.; Kennicutt, R. C.
  Jr.; Huchra, J. P. 1994, ApJ 420, 87

\end{thebibliography}
\bibliographystyle{aasjournal}
{}

\end{document}